\documentclass[aps, reprint,10pt,twocolumn,superscriptaddress]{revtex4-1}
\usepackage{amsmath}
\usepackage{latexsym}
\usepackage{amssymb}
\usepackage{graphics,epstopdf}
\usepackage{hyperref}
\hypersetup{
    colorlinks = true,
    linkcolor =blue,
	citecolor=blue, 
	urlcolor=blue 
}

\newcommand{\ed}{\end{document}}
\newcommand{\beq}{\begin{equation}}
\newcommand{\eeq}{\end{equation}}

\usepackage{mathtools}
\usepackage{natbib}

\begin{document}
\title{Sub-diffusive phases in open clean long-range systems}
\author{Archak Purkayastha}
\email{archak.p@tcd.ie}
\affiliation{Department of Physics, Trinity College Dublin, Dublin, Ireland}

\author{Madhumita Saha}
\email{madhumita.saha@acads.iiserpune.ac.in} 
\affiliation{Department of Physics, Indian Institute of Science Education and Research Pune, Dr. Homi Bhabha Road, Ward No. 8, NCL Colony, Pashan, Pune, Maharashtra 411008, India}

\author{Bijay Kumar Agarwalla}
\email{bijay@iiserpune.ac.in}
\affiliation{Department of Physics, Indian Institute of Science Education and Research Pune, Dr. Homi Bhabha Road, Ward No. 8, NCL Colony, Pashan, Pune, Maharashtra 411008, India}

\date{\today} 
\begin{abstract}
{We show that a one-dimensional ordered fermionic lattice system with power-law-decaying hopping, when connected to two baths at its two ends with different chemical potentials at zero temperature, features two phases showing sub-diffusive scaling of conductance with system size. These phases have no analogues in the isolated system (i.e, in absence of the baths) where the transport is perfectly ballistic. In the open system scenario, interestingly, there occurs two chemical-potential-driven sub-diffusive to ballistic phase transitions at zero temperature. We discuss how these phase transitions, to our knowledge, are different from all the known non-equilibrium quantum phase transitions.  We provide a clear understanding of the microscopic origin of these phases and argue that the sub-diffusive phases are robust against the presence of arbitrary number-conserving many-body interactions in the system. { These phases showing sub-diffusive scaling of conductance with system size in a two-terminal set-up are therefore universal properties of all ordered one-dimensional number-conserving fermionic systems with power-law-decaying hopping at zero temperature.  }
}
\end{abstract}

\maketitle

{\it Introduction ---}
Normal metals (conductors) have their own well-defined conductivity at a given temperature. For a metal wire of cross-sectional area $A$, length $N$, conductivity $\sigma$, and connected to two-terminals at its two ends, the conductance (i.e, inverse of resistance) $G$  is given as $G=\sigma A/N$. Importantly, since conductivity is independent of the dimension of the metal used, if the length of the wire is changed keeping the cross-sectional area fixed, the conductance scales inversely with $N$, i.e., $G \sim N^{-1}$. This corresponds to normal diffusive transport. 
In absence of this behavior, conductivity no longer remains as a property of the material, but rather depends on the dimension of the wire in a non-trivial way. Two simple examples of these are perfect insulators (no transport) with $G \sim e^{-\lambda N}$ and perfect conductors (ballistic transport) with $G$ independent of $N$.

It has been well-established that transport behavior may deviate from the ones described above,  especially for low-dimensional systems \cite{Dhar_2008,Ilievski_2018,Bertini_2021,Landi_2021}. Such transport behavior, where $G~\sim~N^{-\delta}$, with $0<\delta\neq 1$, is often called anomalous. Rapid miniaturization of devices has taken technology to limits where realizing such low-dimensional systems have become a real possibility, and thus understanding their transport properties has become imperative \cite{Quantum_device1,Quantum_device2,Quantum_device3,
thermal_machines_review}.   One of the most intriguing behavior among the anomalous transport is the so-called sub-diffusive transport, which corresponds to $\delta>1$ (as opposed to super-diffusive transport for $0<\delta<1$). In this case, even though the conductivity of the wire goes to zero as $N\rightarrow \infty$, for any finite length, its conductance is exponentially larger than what one would expect for a perfect insulator. For finite-size systems, relevant in quantum technology, quantum chemistry and mesoscopic physics, this particular feature can make a significant difference. Sub-diffusive behavior is often observed for systems residing at the critical regions separating localization-delocalization transitions \cite{Griffith0,Griffith1,Griffith2,Griffith3,DeRoeck_2020,taylor_2021,
Archak1, Archak2, Archak3,Sutradhar_2019, subdiffusion_interaction,Fractal1,Fractal2,Fibonacci0,Fibonacci1, Fibonacci2,Fibonacci3, Fibonacci4,Fibonacci5}. It is usually associated with the presence of correlated or uncorrelated disorder in the system, although a complete microscopic understanding is largely missing. To the contrary, in this work, we reveal, and microscopically explain,  a completely different way in which sub-diffusive scaling of conductance with system length can occur in a large class of systems, even in complete absence of disorder.

\begin{figure}
\includegraphics[width=0.95\columnwidth]{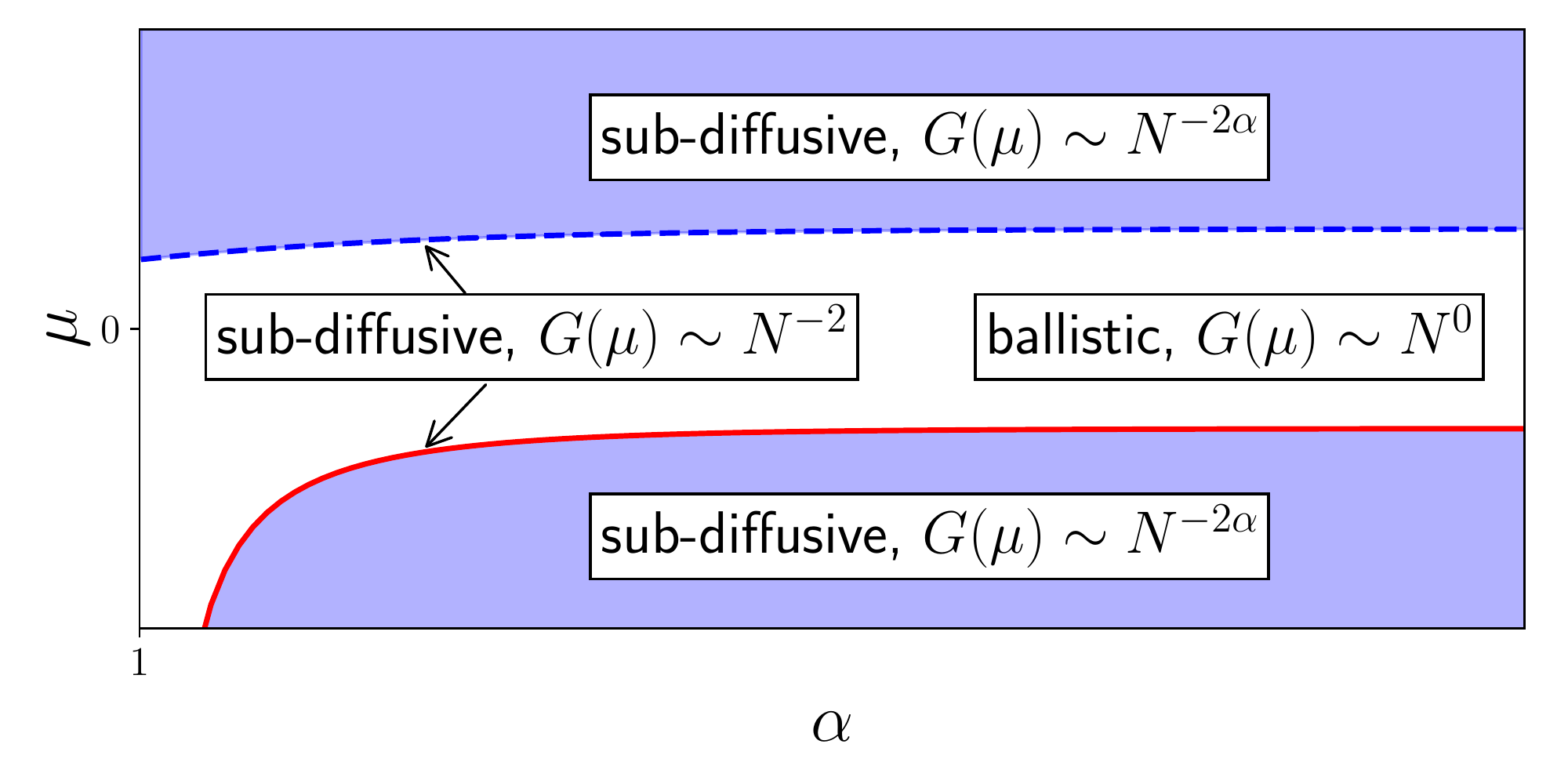} 
\caption{Non-equilibrium phase diagram obtained from system size ($N$) scaling of zero-temperature conductance $G(\mu)$ as a function of chemical potential $\mu$ and long-range hopping exponent $\alpha$ for a one-dimensional open clean long range system (Eq.~(\ref{Hs})).  
The critical lines correspond to the system band edges $\mu=2\eta(\alpha)$ (blue dashed line)  and $\mu=-2\zeta(\alpha)$ (red line), where $\eta(\alpha)$ is Dirichlet-eta function and $\zeta(\alpha)$ is the Riemann-zeta function.}
\label{fig:phase-diagram} 
\end{figure} 

Specifically, we show that, a one-dimensional fermionic wire with long-range power-law-decaying hopping connected to two-terminals at the two ends, surprisingly features two phases at zero temperature, showing sub-diffusive scaling of conductance with $N$ in absence of any disorder.
These unique sub-diffusive phases arise due to an interplay between the long-range hopping and the dissipation governed by the two terminals, and have no analog in absence of either. 
We observe two chemical-potential-driven  dissipative quantum phase transitions between phases featuring sub-diffusive and ballistic transport (see Fig.~(\ref{fig:phase-diagram})). We discuss how these phase transitions are different from all the previously well-known dissipative quantum phase transitions. Furthermore, we provide a clear microscopic understanding of the sub-diffusive transport by connecting the corresponding scaling exponent to a non-analyticity in the dispersion relation. We also argue that these sub-diffusive phases are immune to the presence of many-body interactions in the system as long as the net number of particles within the system is conserved. The sub-diffusive phases are therefore { universal} properties of   all clean number-conserving one-dimensional fermionic systems with power-law-decaying hopping in the two-terminal set-up.

Low-dimensional long-range lattice systems have been realized in various controlled experimental platforms \cite{expt_Rydberg_atoms1,expt_Rydberg_atoms2,expt_Rydberg_atoms3,
expt_Rydberg_atoms4,expt_trapped_ions1,
expt_trapped_ions2,expt_trapped_ions3,expt_trapped_ions4,expt_trapped_ions5,
expt_trapped_ions7,expt_time_crystal2,expt_trapped_ions6,
Experiment_transport,expt_polar_molecules1,
expt_polar_molecules3,expt_dipolar_gas1,expt_nuclear_spins,
expt_polar_molecules2,
expt_time_crystal1,superconducting_qubit_long_range_experiment}, and have been reported to show exotic physics like time-crystals \cite{expt_time_crystal1,expt_time_crystal2}, prethermalization \cite{expt_trapped_ions3}, dynamical phase transitions \cite{expt_nuclear_spins,expt_trapped_ions2,expt_trapped_ions7,
expt_atoms_in_trap}, environment assisted transport \cite{Experiment_transport} etc.  
This has lead to a large number studies in quantum transport which is so far limited mostly to isolated (non-dissipative) systems in presence and absence of disorder \cite{Archak4,long_range_transport,Archak5,Kloss_2019,Kloss_2020,Kawa_2020,
Schneider_2021,Prasad_2021,Ranjan_Tanay}, and a very recent study on dissipative spin chain at infinite temperature \cite{Katzer_2020}. But, interestingly, the physics of dissipative long-range fermionic systems at low temperatures, a class of which reveals the unique universal physics described here, has remained entirely unexplored previously.

{\it The clean long-range hopping model ---} We consider the following one-dimensional lattice model of fermions with long-range hopping decaying as a power-law
\begin{align}
\label{Hs}
\hat{\mathcal{H}}_S = -\sum_{m=1}^N \left(\sum_{r=1}^{N-m} \frac{1}{m^\alpha} \left(\hat{c}_r^\dagger \hat{c}_{r+m} + \hat{c}_{r+m}^\dagger \hat{c}_{r}\right)\right),
\end{align}
where $\hat{c}_r$ is the fermionic annihilation operator at the $r$th site of the system.  Interestingly, this long-range model Hamiltonian has recently been realized using Floquet engineering technique in superconducting qubits \cite{superconducting_qubit_long_range_experiment}. The above system Hamiltonian can be written as, $\hat{\mathcal{H}}_S = \sum_{\ell m=1}^N \mathbf{H}_{\ell m} \hat{c}_\ell^\dagger \hat{c}_m$, where  the matrix $\mathbf{H}$ is a Toeplitz matrix with elements given by
$
\mathbf{H}_{\ell m} = \frac{1}{|\ell-m|^\alpha},~~\forall~\ell \neq m,
$
and $\mathbf{H}_{\ell \ell}=0$.  The eigenspectrum of this matrix, which correspond to the single-particle eigenvalues and eigenvectors of $\hat{\mathcal{H}}_S$, are difficult to find analytically for arbitrary $N$. But, in the thermodynamic limit, $N\rightarrow \infty$, the single particle eigenvalues can be obtained via a Fourier transform, and correspond to the dispersion relation \cite{supp},  
$
\varepsilon(k,\alpha)= -2\sum_{m=1}^\infty \frac{ \cos(m k)}{m^\alpha} . 
$ 
The infinite series summation in the dispersion relation is absolutely convergent for all $k$ if  $\alpha>1$. It is in this case that the thermodynamic limit ($N\rightarrow \infty$) is well-defined. We will therefore always consider $\alpha>1$. It can be numerically verified that the eigenvalues of $\mathbf{H}$ tend to this dispersion relation in the large $N$ limit and the corresponding single-particle eigenvectors of the system are completely delocalized. This property indicates that there should be ballistic transport in the system \cite{Archak5}. 
On the contrary, as we will show below, in the open system scenario, there is a surprising sub-diffusive to ballistic phase transition as a function of chemical potential for all $\alpha>1$ at zero temperature. We note that the band-edges of the dispersion relation correspond to $\varepsilon(0,\alpha)=-2\zeta(\alpha)$, where $\zeta(\alpha)=\sum_{m=1}^\infty \frac{1}{m^\alpha}$ is the Riemann-zeta function, and $\varepsilon(\pm \pi,\alpha)= 2\eta(\alpha) $ with  $\eta(\alpha)=\sum_{m=1}^\infty \frac{(-1)^{m-1}}{m^\alpha}$ being 
the Dirichlet-eta function.

{\it Open system conductance at zero temperature---}
To calculate the conductance in open quantum system setting, we consider the two terminal transport set-up where the system is connected to two baths at its two ends, i.e., the first and the $N$th sites. Such open system set-up is exactly what is used for realizing  autonomous (continuous) quantum heat engines,  refrigerators, thermoelectric generators etc \cite{Kosloff_Levy_2014,Benenti_2017, Landi_2021}.   
The left (right) bath is modelled by a non-interacting Hamiltonian with an infinite number of modes $\hat{\mathcal{H}}_{B_1}=\sum_{r=1}^\infty \Omega_{r1} \hat{B}_{r1}^\dagger\hat{B}_{r1} $ ($\hat{\mathcal{H}}_{B_N}=\sum_{r=1}^\infty \Omega_{rN} \hat{B}_{rN}^\dagger\hat{B}_{rN}$), where $\hat{B}_{r1}$ ($\hat{B}_{rN}$) is the fermionic annhilation operator of the $r$th mode of the left (right) bath. The baths are connected to the system with the system-bath coupling Hamiltonian $\hat{\mathcal{H}}_{SB}=\sum_{\ell=1,N}\sum_{r=1}^\infty (\kappa_{r\ell} \hat{c}_\ell^\dagger\hat{B}_{r\ell} +~\kappa_{r\ell}^* \hat{B}_{r\ell}^\dagger\hat{c}_\ell)$.   Initially, the baths are assumed to be at their own thermal states with their own temperatures and chemical potentials ($\mu_1,\mu_N$), while the system's initial state is arbitrary. We are specifically interested in the non-equilibrium steady state (NESS) in the zero temperature limit and linear response regime, $\mu_1~=~\mu,~\mu_N~=~\mu\!-\!\Delta\mu,~\Delta\mu \rightarrow~0$.

It is possible to obtain the exact NESS properties of the system using the { non-equilibrium Green's function} (NEGF) approach \cite{Jauho_book,di_Ventra_book,Dhar_2006}. The { retarded} NEGF for such a set-up is given by $\mathbf{G}^{+}(\omega)=~ \left[\omega\mathbb{I}-\mathbf{H}-\Sigma^{(1)}(\omega)-\Sigma^{(N)}(\omega)\right]^{-1}$, where $\mathbb{I}$ is the $N$-dimensional identity matrix, and $\Sigma^{(1)}(\omega)$ ($\Sigma^{(N)}(\omega)$) is the self-energy matrix due to the left (right) bath. The only non-zero element in the $N\times N$ left (right) bath self-energy matrix is the top left (bottom right) corner element,   $\Sigma^{(\ell)}_{\ell \ell}(\omega)= -i\frac{\mathfrak{J}_\ell(\omega)}{2}-\mathcal{P}\int \frac{d\omega^\prime}{2\pi}\frac{\mathfrak{J}_\ell(\omega^\prime)}{\omega-\omega^\prime}$, $\ell=\{1,N\}$ \cite{supp}. Here $\mathcal{P}$ denotes principal value, and $\mathfrak{J}_\ell(\omega)$ is the bath spectral function, defined as $\mathfrak{J}_\ell(\omega)~=~2\pi \sum_{r=1}^\infty | \kappa_{r \ell}|^2 \delta(\omega - \Omega_{r \ell})$. The zero temperature particle conductance $G(\mu)$ is given in terms of the NEGF as,
\begin{align}
\label{conductance}
G(\mu) = \lim_{\Delta \mu \rightarrow 0} \frac{I}{\Delta \mu} = \frac{1}{2\pi}\mathcal{T}(\mu)=\frac{\mathfrak{J}_1(\mu)\mathfrak{J}_N(\mu) |\mathbf{G}^{+}_{1N}(\mu)|^2}{2\pi},
\end{align} 
where { $\mathbf{G}^{+}_{1N}(\mu)$ denotes the $(1, N)$th element of the matrix $\mathbf{G}^{+}(\mu)$}, $\mathcal{T}(\omega)$ is the transmission function and  $I~=~\int_{\mu_1}^{\mu_N}~\frac{d \omega}{2\pi}~\mathcal{T}(\omega)$ is the NESS particle current.  The scaling of conductance with system-size $N$ is used to classify transport properties as described in the introduction. Note that, for anomalous transport, this classification of different transport regimes, which is standard in an open system setting, may not lead the to same results as its corresponding isolated system counterpart where the regimes are classified via time scaling of spread of correlations \cite{Archak3}.

\begin{figure}
\includegraphics[width=\columnwidth]{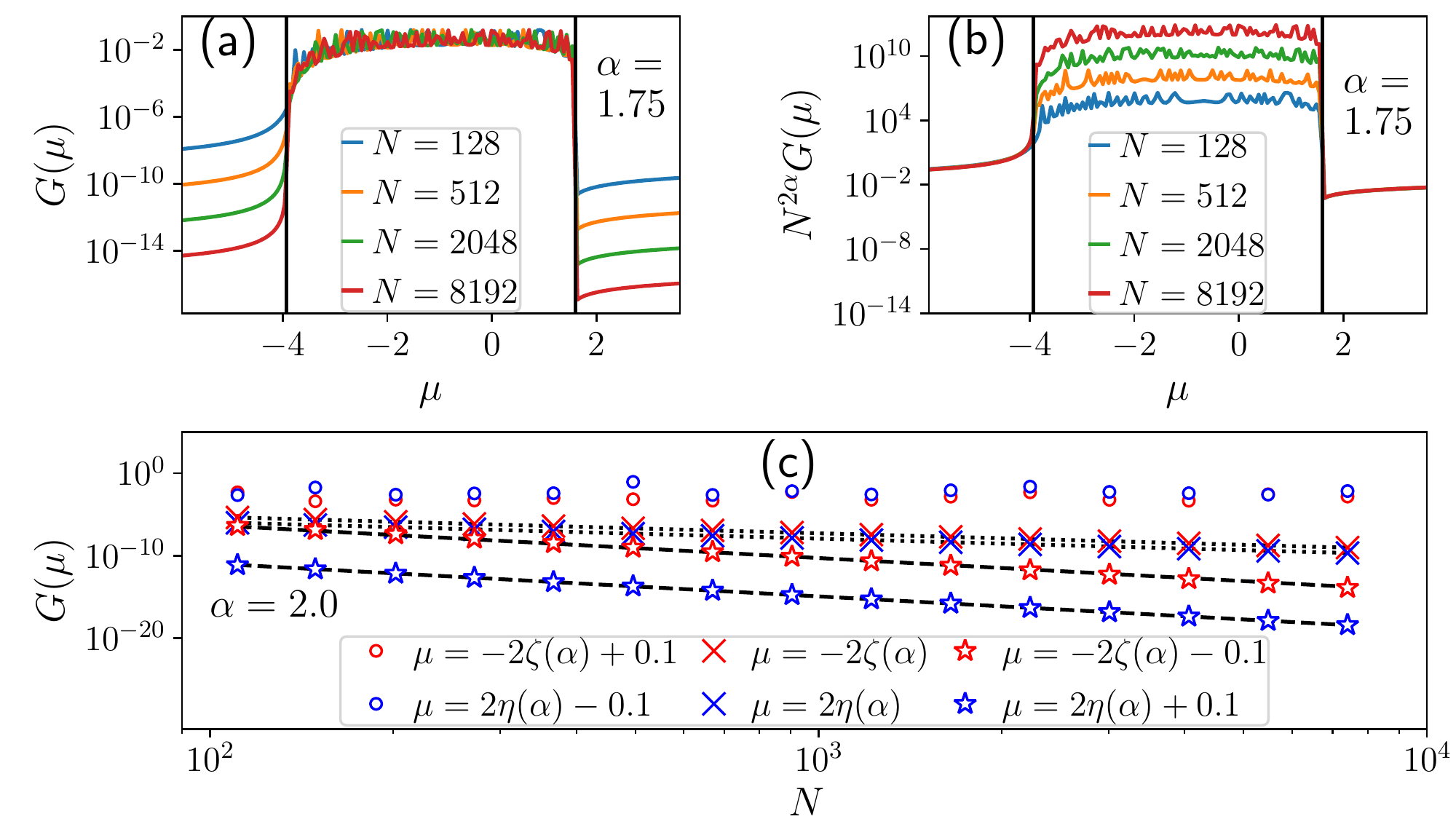} 
\caption{{\bf (a)}  Zero-temperature conductance $G(\mu)$ as a function of chemical potential $\mu$, at a chosen value of $\alpha=1.75$ for various system sizes $N$. The two vertical lines correspond to band-edges $\mu=-2\,\zeta(\alpha)$ and $\mu=2\,\eta(\alpha)$. {\bf (b)}~The same plot as in (a) but with the y-axis scaled by $N^{2\alpha}$. {\bf (c)}~The scaling of $G(\mu)$ with system size at various values of $\mu$, for $\alpha=2.0$. The black dashed lines are fits of $N^{-2\alpha}$. The black dotted lines are fits of $N^{-2}$.   For the plots, the bath spectral functions are chosen to be $\mathfrak{J}_1(\omega)=\mathfrak{J}_N(\omega)=\Gamma \sqrt{1-\left(\frac{\omega}{\Lambda}\right)^2}$, with $\Lambda=8$, $\Gamma=10$. All energy scales are in units of nearest neighbour hopping strength. }
\label{fig:conductance_scaling} 
\end{figure}

{\it Chemical-potential-driven sub-diffusive to ballistic phase transition ---}
We numerically  calculate the exact transmission function, and study the conductance scaling with system size. Our central result is as follows. For $1<\alpha<\infty$,
\begin{align}
\label{main_result}
&G(\mu) \sim N^{-2\alpha},~~\forall~\mu<-2\zeta(\alpha),~\mu>2\eta(\alpha), \nonumber \\
&G(\mu) \sim N^{-2}, ~~\textrm{at}~\mu=-2\zeta(\alpha),2\eta(\alpha), \\
&G(\mu) \sim N^0, ~~\forall~-2\zeta(\alpha)<\mu<2\eta(\alpha), \nonumber
\end{align}
where, as mentioned before, $\zeta(\alpha)$ is the Riemann-zeta function and $\eta(\alpha)$ is the Dirichlet-eta function.
In other words, when the chemical potential $\mu$ is within the band of the system, the transport is ballistic, as expected. But, surprisingly, when $\mu$ lies outside the band of the system, the transport is sub-diffusive, with an exponent of $2\alpha$. Moreover, when $\mu$ is located exactly at the band-edges, the transport is again sub-diffusive but with an $\alpha$ independent exponent. Representative plots showing the above behavior are given  in Fig.~\ref{fig:conductance_scaling}. Fig.~\ref{fig:conductance_scaling}(a) shows the behavior $G(\mu)$ with $\mu$ for various system sizes at a chosen value of $\alpha$ ($\alpha=1.75$). Clearly,  within the band, i.e., $-2\,\zeta(\alpha)<\mu<2\,\eta(\alpha)$, there is no scaling of $G(\mu)$ with $N$, confirming perfect ballistic behavior, whereas outside that regime $G(\mu)$ scales with system size. Fig.~\ref{fig:conductance_scaling}(b) shows the same result as in Fig.~\ref{fig:conductance_scaling}(a) with the y-axis now scaled by $N^{2\alpha}$. All data points outside the band of the system   collapse perfectly, thereby confirming the sub-diffusive scaling.  Likewise, the $\alpha$ independent scaling at the band-edges can also be checked (not shown in the figure for $\alpha=1.75$) numerically.  Interestingly, this behavior is seen at all values of $\alpha>1$. Fig~\ref{fig:conductance_scaling}(c) shows conductance scaling with system size at values close to the system band-edges for a different value of $\alpha$ ($\alpha=2)$. The behavior consistent with Eq.(\ref{main_result}) is clearly observed here. 

{\it Origin of the sub-diffusive phases---}
The origin of these surprising sub-diffusive phases for chemical potentials outside the band of the system can be traced to the non-analyticity property of the dispersion relation at its minimum value at $k=0$. From Eqs.~(\ref{conductance}), it is evident that the system size scaling of conductance originates from that of $\mathbf{G}^{+}_{1N}(\mu)$. Since the baths are attached only to the first and the last sites, we conjecture that, for large $N$, system size scaling of  $\mathbf{G}^{+}_{1N}(\mu)$ will be same as $\mathbf{g}_{1N}^{+}(\mu)$, where $\mathbf{g}^{+}(\mu)=[(\mu -i\epsilon ) \mathbb{I}-\mathbf{H}]^{-1}$ is the { retarded Green's} function of the system in absence of the baths. That is, $\mathbf{G}^{+}_{1N}(\mu)\propto \mathbf{g}^{+}_{1N}(\mu)$ with the proportionality constant being independent of $N$. Also, since the system is clean (ordered), in the $N\rightarrow\infty$ limit, one can obtain the bare { retarded Green's} function via a Fourier transform, $\mathbf{g}_{pq}^+(\mu)=\lim_{\epsilon \rightarrow 0}\int dk~g^+(k,\mu)e^{-ik|p-q|}$, where $g^+(k,\mu)=[\mu-\varepsilon(k,\alpha)-i\epsilon]^{-1}$. Combining all of these, we have, for large $N$,
\begin{align}
\label{conjecture}
\mathbf{G}_{1N}^{+}(\mu) \propto \lim_{\epsilon\rightarrow 0} \int_{-\pi}^{\pi} dk \frac{e^{-ikN}}{\mu-\varepsilon(k,\alpha)-i\epsilon}.
\end{align}
The above heuristic expression, in combination with Eq.(\ref{conductance}), relates the scaling of conductance with system size of an open system with  spectral properties of the isolated system in the thermodynamic limit.  

The major contribution to the above integral comes from the singularities of the integrand. It can be checked that $k=0$ is always a singular point because $\varepsilon(k,\alpha)$ is non-analytic at $k=0$,
$
\lim_{k\rightarrow 0} \frac{\partial^p \varepsilon(k,\alpha)}{\partial k^p} \rightarrow \infty,~~\forall~p>\alpha-1.
$ To capture the effect of this non-analyticity, we derive a non-trivial non-analytic small $k$ expansion of $\varepsilon(k,\alpha)$ for non-integer $\alpha>1$ \cite{supp},
$
\varepsilon(k,\alpha)\simeq-2\left[ \zeta(\alpha)- a_1 |k|^{\alpha-1}- a_2 k^2 \right],~~\forall~~|k|\ll 1
$
where $a_1$ and $a_2$ are real numbers. The presence of $|k|^{\alpha-1}$ makes the above expression explicitly non-analytic, clearly distinguishing it from a standard Taylor expansion. While evaluating the integral in Eq.(\ref{conjecture}) via contour integration,  the non-integer value of $\alpha$ leads to a branch whose contribution to Eq.(\ref{conjecture}) can be shown to scale with system-size as $N^{-\alpha}$ \cite{supp}.  Though these results are obtained for non-integer values of $\alpha$,  integer values of $\alpha$ can be included  by assuming an arbitrarily small fractional part.

Now, when $\mu$ is within the band of the system,  $-2\,\zeta(\alpha)<\mu<2\,\eta(\alpha)$, there are additional poles on the real-line. It can be checked easily that such poles can at best generate an oscillatory behavior with $N$ and thus cannot provide a scaling with $N$. These poles within the band therefore gives the leading behavior $\mathbf{G}^{+}_{1N}(\mu)\sim N^0$ implying ballistic transport. When $\mu$ is below the band of the system, i.e., $\mu<-2\,\zeta(\alpha)$, the additional poles on the real-line do not exist, and  the main contribution to the integral comes from the non-analytic point $k\rightarrow 0$.  As mentioned above, in this case, the contour integration generates a scaling of the form $\mathbf{G}^{+}_{1N}(\mu)\sim N^{-\alpha}~~\forall~\mu<-2\zeta(\alpha)$, leading to a sub-diffusive exponent of $2\alpha$ from Eq.(\ref{conductance}). On the other hand, when $\mu>2\eta(\alpha)$, one may argue that the main contribution to the integral comes from $k\sim\pm \pi$ corresponding to the upper band edge, $\varepsilon(\pm \pi,\alpha)=2\eta(\alpha)$, where the denominator in Eq.(\ref{conjecture}) would be minimum. However, an expansion about this point, $\varepsilon(k\pm \pi,\alpha)~\simeq~2\eta(\alpha)~-~2a_2(1~-~2^{3-\alpha})k^2 $, $|k|\ll 1$, shows that, unlike the lower band edge at $k=0$, this point is analytic, and its contribution to the integral in Eq.(\ref{conjecture}) decays exponentially with $N$ \cite{supp}. Consequently, for large enough $N$, the leading contribution once again stems from the singularity at $k=0$, and giving $\mathbf{G}^{+}_{1N}(\mu)\sim N^{-\alpha}~~\forall~\mu>2\eta(\alpha)$, leading to the same sub-diffusive exponent. However, interestingly, since the denominator in Eq.(\ref{conjecture}) is now large, the value of $\mathbf{G}^{+}_{1N}(\mu)$, and therefore the conductance, for $\mu>2\eta(\alpha)$ is much smaller than that for $\mu<-2\zeta(\alpha)$, even though the system size scaling is the same. This is clearly seen in all the plots of Fig.~\ref{fig:conductance_scaling}. 

A more careful analysis is required at the critical points  $\mu=-2\zeta(\alpha), 2\eta(\alpha)$. At any finite $N$, the critical $\mu$ values always lie slightly outside the system band, but the minimum and maximum eigenvalues of $\mathbf{H}$ approach these values with increase in $N$.  We find that it becomes difficult to use Eq.(\ref{conjecture}) to capture this behavior. Nevertheless, the conjecture $\mathbf{G}^{+}_{1N}(\mu)\propto \mathbf{g}^{+}_{1N}(\mu)$ still holds, and it can be directly numerically checked  for finite $N$ that $\mathbf{g}^{+}_{1N}(-2\zeta(\alpha)),~\mathbf{g}^{+}_{1N}(2\eta(\alpha)) \sim N^{-1}$, independent of $\alpha$ \cite{supp}. This therefore clearly gives the origin of $N^{-2}$ scaling of conductance at $\mu=-2\zeta(\alpha), 2\eta(\alpha)$.

{\it {A different type dissipative quantum phase transition} ---}
Since $G(\mu)\propto |\mathbf{G}^{+}_{1N}(\mu)|^2$, a non-analytic change in $G(\mu)$ corresponds to a non-analytic change in NESS, thereby pointing to a  dissipative  quantum phase transition. This type of dissipative quantum phase transition, to our knowledge, has not been discussed before. In existing examples of dissipative phase transitions in the literature that we know of (for example, \cite{non_eq_phase_transition_expt1,non_eq_phase_transition_expt2,
non_eq_phase_transition_expt3,non_eq_phase_transition_expt4,
Jo_2021,Gamayun_2021,Zamora_2020,Jo_2019,Carollo_2019,Minganti_2018,
Marcuzzi_2016, Dagvadorj_2015,Nagy_2015,Bastidas_2012,Prosen_2008}) the phase transition occurs on changing either a parameter in the system Hamiltonian, or the strength of the system-bath couplings. In contrast, here, the phase transition occurs as a function of the chemical potentials of the baths. These are not Hamiltonian parameters, either of the system or of the baths, but rather are the thermodynamic parameters fixed by the initial state of the baths. These control the zero temperature noise that originates from the baths, which, in turn control the NESS. 

 As is clear from the above results, this phase transition stems from the non-analyticity of the dispersion relation of the system in the thermodynamic limit. It is therefore a property of the system in the large $N$ limit, and is completely independent of details of the baths, as long as there is a unique NESS. In fact to guarantee a unique NESS,  only two properties of the bath spectral functions are required: (a) the spectral functions for both the baths must be continuous, (b) the band of the baths must encompass the band of the system \cite{Dhar_2006}. Notably, the strength of system-bath coupling, while determining the value of conductance, does not affect the system-size scaling of conductance. This is evident from validity of the conjecture $\mathbf{G}^{+}_{1N}(\mu)\propto \mathbf{g}^{+}_{1N}(\mu)$, and can also be verified numerically \cite{supp}. { However, the presence of the baths are crucial to allow sub-diffusive transport at the chemical potentials beyond the system bandwidth. 
The isolated system at such chemical potentials would either be completely empty or completely filled, thereby having no possibility of transport.} Thus, the sub-diffusive behavior observed here has no isolated system analogue.

This phase transition is clearly a quantum phase transition, as it occurs strictly at zero temperature. At any finite temperature, at all values of chemical potentials of the baths, calculation of current or conductance will have finite contribution from energies within the system energy bands. At low temperatures, for chemical potentials outside the system band, this contribution will be small, but as system size is increased, will eventually be the leading contribution. So, at finite but low temperatures, for chemical potentials outside the system band, there will be a crossover from the sub-diffusive to the ballistic behavior as a function of system size. Thus, like standard quantum phase transitions, this phase transition gives rise to a finite size crossover at finite but low temperatures.

It is important to note that the standard Lindblad equation approaches in local and global (eigenbasis) forms \cite{Breuer_book,Landi_2021, Walls1970,Wichterich_2007,Rivas_2010,barranco_2014,Levy2014,Archak6,
Trushechkin_2016,
Eastham_2016,Hofer_2017,Gonzalez_2017,Mitchison_2018,
Cattaneo_2019,Hartmann_2020_1,konopik_2020local,
Scali_2021} cannot capture these sub-diffusive phases. This is because such approaches, by construction, neglect contributions coming from bath energies which are away from system energy scales. Such an equation would therefore wrongly predict zero conductance for chemical potentials outside the system band at all system sizes \cite{supp}. Whether more refined quantum master equation approaches \cite{ule,Kleinherbers_2020,Davidovic_2020,mozgunov2020,mccauley2020,
kirvsanskas2018}, including the Redfield equation \cite{Breuer_book,Landi_2021,Hartmann_2020_1,Archak6}, can capture the sub-diffusive behavior remains to be seen and requires further investigation. 

{\it {Universality} of the sub-diffusive phases---}
When $\mu\leq -2\zeta(\alpha)$, it is intuitive and can be numerically checked \cite{supp} that there is a sub-extensive number of particles in the system. If a number-conserving many-body interaction term (i.e, higher than quadratic term, for example,  $\hat{\mathcal{H}}_{\rm int}~=~\sum_{\ell,m} V_{\ell m} \hat{c}_\ell^\dagger\hat{c}_\ell \hat{c}_m^\dagger \hat{c}_m$, $\hat{\mathcal{H}}_S \rightarrow \hat{\mathcal{H}}_S+\hat{\mathcal{H}}_{\rm int}$) is now switched on, at large enough $N$, due to extremely low particle density in the system,  it will play a negligible role for $\mu\leq-2\zeta(\alpha)$. Similar argument, in terms of holes rather than particles, can be made for $\mu\geq 2\eta(\alpha)$. { Thus, the sub-diffusive phases for chemical potential outside the system band, as well as the critical points, are robust against presence of arbitrary number-conserving many-body interactions in the system. Therefore they are universal. 

For $\mu<-2\zeta(\alpha)$ ($\mu>2\eta(\alpha)$), the intuitive picture that emerges is that few particles (holes) tunnel into the system from one bath due to quantum fluctuations, and then hop into the other bath with essentially a single long-range hop of amplitude $\sim|N^{-\alpha}|^2$. This is consistent with the scaling $G(\mu)\sim N^{-2\alpha}$. However, such a simple picture does not explain the $G(\mu)\sim N^{-2}$ scaling at the critical points $\mu=-2\zeta(\alpha), 2\eta(\alpha)$.   Moreover, the scaling at the critical points is super-universal since it is also independent of $\alpha$, which controls the effective range of hopping.}   Conversely, for $-2\zeta(\alpha)<\mu<2\eta(\alpha)$, there will be a finite particle density in the system and the many-body interactions can have a non-trivial effect which can change the nature of the transport, making this regime non-universal. 

Direct demonstration of above statements in presence of many-body interactions is currently beyond the state-of-the-art numerical techniques. { But, interestingly, long-range magnetization-conserving spin Hamiltonians with power-law-decaying interactions have been realized in several experimental platforms \cite{expt_trapped_ions6,
Experiment_transport,expt_polar_molecules1,
expt_polar_molecules3,expt_dipolar_gas1,expt_nuclear_spins}. These can be mapped via Jordan-Wigner transformation into number-conserving fermionic Hamiltonians with power-law-decaying hopping and many-body interactions \cite{supp}. This makes experimental verification of the universal sub-diffusive phases plausible.} The effect of uncorrelated or correlated disorder \cite{Ranjan_Tanay} on such sub-diffusive phases remains to be seen.

{\it Acknowledgement---}
BKA acknowledges the MATRICS grant MTR/2020/000472 from SERB, Government of India and the Shastri Indo-Canadian Institute for providing financial support for this research work in the form of a Shastri Institutional Collaborative Research Grant (SICRG). MS acknowledge financial support through National Postdoctoral Fellowship (NPDF), SERB file no.  PDF/2020/000992. AP acknowledges funding from European Unions Horizon 2020 research and innovation programme under the
Marie Sklodowska-Curie grant agreement No. 890884.

\bibliography{ref_open_long_range}

\begin{thebibliography}{101}%
\makeatletter
\providecommand \@ifxundefined [1]{%
 \@ifx{#1\undefined}
}%
\providecommand \@ifnum [1]{%
 \ifnum #1\expandafter \@firstoftwo
 \else \expandafter \@secondoftwo
 \fi
}%
\providecommand \@ifx [1]{%
 \ifx #1\expandafter \@firstoftwo
 \else \expandafter \@secondoftwo
 \fi
}%
\providecommand \natexlab [1]{#1}%
\providecommand \enquote  [1]{``#1''}%
\providecommand \bibnamefont  [1]{#1}%
\providecommand \bibfnamefont [1]{#1}%
\providecommand \citenamefont [1]{#1}%
\providecommand \href@noop [0]{\@secondoftwo}%
\providecommand \href [0]{\begingroup \@sanitize@url \@href}%
\providecommand \@href[1]{\@@startlink{#1}\@@href}%
\providecommand \@@href[1]{\endgroup#1\@@endlink}%
\providecommand \@sanitize@url [0]{\catcode `\\12\catcode `\$12\catcode
  `\&12\catcode `\#12\catcode `\^12\catcode `\_12\catcode `\%12\relax}%
\providecommand \@@startlink[1]{}%
\providecommand \@@endlink[0]{}%
\providecommand \url  [0]{\begingroup\@sanitize@url \@url }%
\providecommand \@url [1]{\endgroup\@href {#1}{\urlprefix }}%
\providecommand \urlprefix  [0]{URL }%
\providecommand \Eprint [0]{\href }%
\providecommand \doibase [0]{http://dx.doi.org/}%
\providecommand \selectlanguage [0]{\@gobble}%
\providecommand \bibinfo  [0]{\@secondoftwo}%
\providecommand \bibfield  [0]{\@secondoftwo}%
\providecommand \translation [1]{[#1]}%
\providecommand \BibitemOpen [0]{}%
\providecommand \bibitemStop [0]{}%
\providecommand \bibitemNoStop [0]{.\EOS\space}%
\providecommand \EOS [0]{\spacefactor3000\relax}%
\providecommand \BibitemShut  [1]{\csname bibitem#1\endcsname}%
\let\auto@bib@innerbib\@empty
\bibitem [{\citenamefont {Dhar}(2008)}]{Dhar_2008}%
  \BibitemOpen
  \bibfield  {author} {\bibinfo {author} {\bibfnamefont {A.}~\bibnamefont
  {Dhar}},\ }\href {\doibase 10.1080/00018730802538522} {\bibfield  {journal}
  {\bibinfo  {journal} {Advances in Physics}\ }\textbf {\bibinfo {volume}
  {57}},\ \bibinfo {pages} {457–537} (\bibinfo {year} {2008})}\BibitemShut
  {NoStop}%
\bibitem [{\citenamefont {Ilievski}\ \emph {et~al.}(2018)\citenamefont
  {Ilievski}, \citenamefont {De~Nardis}, \citenamefont {Medenjak},\ and\
  \citenamefont {Prosen}}]{Ilievski_2018}%
  \BibitemOpen
  \bibfield  {author} {\bibinfo {author} {\bibfnamefont {E.}~\bibnamefont
  {Ilievski}}, \bibinfo {author} {\bibfnamefont {J.}~\bibnamefont {De~Nardis}},
  \bibinfo {author} {\bibfnamefont {M.}~\bibnamefont {Medenjak}}, \ and\
  \bibinfo {author} {\bibfnamefont {T.}~\bibnamefont {Prosen}},\ }\href
  {\doibase 10.1103/PhysRevLett.121.230602} {\bibfield  {journal} {\bibinfo
  {journal} {Phys. Rev. Lett.}\ }\textbf {\bibinfo {volume} {121}},\ \bibinfo
  {pages} {230602} (\bibinfo {year} {2018})}\BibitemShut {NoStop}%
\bibitem [{\citenamefont {Bertini}\ \emph {et~al.}(2021)\citenamefont
  {Bertini}, \citenamefont {Heidrich-Meisner}, \citenamefont {Karrasch},
  \citenamefont {Prosen}, \citenamefont {Steinigeweg},\ and\ \citenamefont
  {\ifmmode \check{Z}\else \v{Z}\fi{}nidari\ifmmode~\check{c}\else
  \v{c}\fi{}}}]{Bertini_2021}%
  \BibitemOpen
  \bibfield  {author} {\bibinfo {author} {\bibfnamefont {B.}~\bibnamefont
  {Bertini}}, \bibinfo {author} {\bibfnamefont {F.}~\bibnamefont
  {Heidrich-Meisner}}, \bibinfo {author} {\bibfnamefont {C.}~\bibnamefont
  {Karrasch}}, \bibinfo {author} {\bibfnamefont {T.}~\bibnamefont {Prosen}},
  \bibinfo {author} {\bibfnamefont {R.}~\bibnamefont {Steinigeweg}}, \ and\
  \bibinfo {author} {\bibfnamefont {M.}~\bibnamefont {\ifmmode \check{Z}\else
  \v{Z}\fi{}nidari\ifmmode~\check{c}\else \v{c}\fi{}}},\ }\href {\doibase
  10.1103/RevModPhys.93.025003} {\bibfield  {journal} {\bibinfo  {journal}
  {Rev. Mod. Phys.}\ }\textbf {\bibinfo {volume} {93}},\ \bibinfo {pages}
  {025003} (\bibinfo {year} {2021})}\BibitemShut {NoStop}%
\bibitem [{\citenamefont {Landi}\ \emph {et~al.}(2021)\citenamefont {Landi},
  \citenamefont {Poletti},\ and\ \citenamefont {Schaller}}]{Landi_2021}%
  \BibitemOpen
  \bibfield  {author} {\bibinfo {author} {\bibfnamefont {G.~T.}\ \bibnamefont
  {Landi}}, \bibinfo {author} {\bibfnamefont {D.}~\bibnamefont {Poletti}}, \
  and\ \bibinfo {author} {\bibfnamefont {G.}~\bibnamefont {Schaller}},\
  }\href@noop {} {\  (\bibinfo {year} {2021})},\ \Eprint
  {http://arxiv.org/abs/2104.14350} {arXiv:2104.14350 [quant-ph]} \BibitemShut
  {NoStop}%
\bibitem [{\citenamefont {Smith}(1996)}]{Quantum_device1}%
  \BibitemOpen
  \bibfield  {author} {\bibinfo {author} {\bibfnamefont {C.~G.}\ \bibnamefont
  {Smith}},\ }\href {\doibase 10.1088/0034-4885/59/2/003} {\bibfield  {journal}
  {\bibinfo  {journal} {Reports on Progress in Physics}\ }\textbf {\bibinfo
  {volume} {59}},\ \bibinfo {pages} {235} (\bibinfo {year} {1996})}\BibitemShut
  {NoStop}%
\bibitem [{\citenamefont {Fleischmann}\ and\ \citenamefont
  {Geisel}(2002)}]{Quantum_device2}%
  \BibitemOpen
  \bibfield  {author} {\bibinfo {author} {\bibfnamefont {R.}~\bibnamefont
  {Fleischmann}}\ and\ \bibinfo {author} {\bibfnamefont {T.}~\bibnamefont
  {Geisel}},\ }\href {\doibase 10.1103/PhysRevLett.89.016804} {\bibfield
  {journal} {\bibinfo  {journal} {Phys. Rev. Lett.}\ }\textbf {\bibinfo
  {volume} {89}},\ \bibinfo {pages} {016804} (\bibinfo {year}
  {2002})}\BibitemShut {NoStop}%
\bibitem [{\citenamefont {Zagoskin}(2011)}]{Quantum_device3}%
  \BibitemOpen
  \bibfield  {author} {\bibinfo {author} {\bibfnamefont {A.~M.}\ \bibnamefont
  {Zagoskin}},\ }\href@noop {} {\bibfield  {journal} {\bibinfo  {journal}
  {Cambridge University Press}\ } (\bibinfo {year} {2011})}\BibitemShut
  {NoStop}%
\bibitem [{\citenamefont {Benenti}\ \emph
  {et~al.}(2017{\natexlab{a}})\citenamefont {Benenti}, \citenamefont {Casati},
  \citenamefont {Saito},\ and\ \citenamefont
  {Whitney}}]{thermal_machines_review}%
  \BibitemOpen
  \bibfield  {author} {\bibinfo {author} {\bibfnamefont {G.}~\bibnamefont
  {Benenti}}, \bibinfo {author} {\bibfnamefont {G.}~\bibnamefont {Casati}},
  \bibinfo {author} {\bibfnamefont {K.}~\bibnamefont {Saito}}, \ and\ \bibinfo
  {author} {\bibfnamefont {R.~S.}\ \bibnamefont {Whitney}},\ }\href {\doibase
  https://doi.org/10.1016/j.physrep.2017.05.008} {\bibfield  {journal}
  {\bibinfo  {journal} {Physics Reports}\ }\textbf {\bibinfo {volume} {694}},\
  \bibinfo {pages} {1 } (\bibinfo {year} {2017}{\natexlab{a}})}\BibitemShut
  {NoStop}%
\bibitem [{\citenamefont {Luitz}\ \emph {et~al.}(2016)\citenamefont {Luitz},
  \citenamefont {Laflorencie},\ and\ \citenamefont {Alet}}]{Griffith0}%
  \BibitemOpen
  \bibfield  {author} {\bibinfo {author} {\bibfnamefont {D.~J.}\ \bibnamefont
  {Luitz}}, \bibinfo {author} {\bibfnamefont {N.}~\bibnamefont {Laflorencie}},
  \ and\ \bibinfo {author} {\bibfnamefont {F.}~\bibnamefont {Alet}},\ }\href
  {\doibase 10.1103/PhysRevB.93.060201} {\bibfield  {journal} {\bibinfo
  {journal} {Phys. Rev. B}\ }\textbf {\bibinfo {volume} {93}},\ \bibinfo
  {pages} {060201} (\bibinfo {year} {2016})}\BibitemShut {NoStop}%
\bibitem [{\citenamefont {Potter}\ \emph {et~al.}(2015)\citenamefont {Potter},
  \citenamefont {Vasseur},\ and\ \citenamefont {Parameswaran}}]{Griffith1}%
  \BibitemOpen
  \bibfield  {author} {\bibinfo {author} {\bibfnamefont {A.~C.}\ \bibnamefont
  {Potter}}, \bibinfo {author} {\bibfnamefont {R.}~\bibnamefont {Vasseur}}, \
  and\ \bibinfo {author} {\bibfnamefont {S.~A.}\ \bibnamefont {Parameswaran}},\
  }\href {\doibase 10.1103/PhysRevX.5.031033} {\bibfield  {journal} {\bibinfo
  {journal} {Phys. Rev. X}\ }\textbf {\bibinfo {volume} {5}},\ \bibinfo {pages}
  {031033} (\bibinfo {year} {2015})}\BibitemShut {NoStop}%
\bibitem [{\citenamefont {Agarwal}\ \emph {et~al.}(2015)\citenamefont
  {Agarwal}, \citenamefont {Gopalakrishnan}, \citenamefont {Knap},
  \citenamefont {M\"uller},\ and\ \citenamefont {Demler}}]{Griffith2}%
  \BibitemOpen
  \bibfield  {author} {\bibinfo {author} {\bibfnamefont {K.}~\bibnamefont
  {Agarwal}}, \bibinfo {author} {\bibfnamefont {S.}~\bibnamefont
  {Gopalakrishnan}}, \bibinfo {author} {\bibfnamefont {M.}~\bibnamefont
  {Knap}}, \bibinfo {author} {\bibfnamefont {M.}~\bibnamefont {M\"uller}}, \
  and\ \bibinfo {author} {\bibfnamefont {E.}~\bibnamefont {Demler}},\ }\href
  {\doibase 10.1103/PhysRevLett.114.160401} {\bibfield  {journal} {\bibinfo
  {journal} {Phys. Rev. Lett.}\ }\textbf {\bibinfo {volume} {114}},\ \bibinfo
  {pages} {160401} (\bibinfo {year} {2015})}\BibitemShut {NoStop}%
\bibitem [{\citenamefont {Vosk}\ \emph {et~al.}(2015)\citenamefont {Vosk},
  \citenamefont {Huse},\ and\ \citenamefont {Altman}}]{Griffith3}%
  \BibitemOpen
  \bibfield  {author} {\bibinfo {author} {\bibfnamefont {R.}~\bibnamefont
  {Vosk}}, \bibinfo {author} {\bibfnamefont {D.~A.}\ \bibnamefont {Huse}}, \
  and\ \bibinfo {author} {\bibfnamefont {E.}~\bibnamefont {Altman}},\ }\href
  {\doibase 10.1103/PhysRevX.5.031032} {\bibfield  {journal} {\bibinfo
  {journal} {Phys. Rev. X}\ }\textbf {\bibinfo {volume} {5}},\ \bibinfo {pages}
  {031032} (\bibinfo {year} {2015})}\BibitemShut {NoStop}%
\bibitem [{\citenamefont {De~Roeck}\ \emph {et~al.}(2020)\citenamefont
  {De~Roeck}, \citenamefont {Huveneers},\ and\ \citenamefont
  {Olla}}]{DeRoeck_2020}%
  \BibitemOpen
  \bibfield  {author} {\bibinfo {author} {\bibfnamefont {W.}~\bibnamefont
  {De~Roeck}}, \bibinfo {author} {\bibfnamefont {F.}~\bibnamefont {Huveneers}},
  \ and\ \bibinfo {author} {\bibfnamefont {S.}~\bibnamefont {Olla}},\ }\href
  {\doibase 10.1007/s10955-020-02496-1} {\bibfield  {journal} {\bibinfo
  {journal} {Journal of Statistical Physics}\ }\textbf {\bibinfo {volume}
  {180}},\ \bibinfo {pages} {678} (\bibinfo {year} {2020})}\BibitemShut
  {NoStop}%
\bibitem [{\citenamefont {Taylor}\ and\ \citenamefont
  {Scardicchio}(2021)}]{taylor_2021}%
  \BibitemOpen
  \bibfield  {author} {\bibinfo {author} {\bibfnamefont {S.~R.}\ \bibnamefont
  {Taylor}}\ and\ \bibinfo {author} {\bibfnamefont {A.}~\bibnamefont
  {Scardicchio}},\ }\href {\doibase 10.1103/PhysRevB.103.184202} {\bibfield
  {journal} {\bibinfo  {journal} {Phys. Rev. B}\ }\textbf {\bibinfo {volume}
  {103}},\ \bibinfo {pages} {184202} (\bibinfo {year} {2021})}\BibitemShut
  {NoStop}%
\bibitem [{\citenamefont {Purkayastha}\ \emph {et~al.}(2018)\citenamefont
  {Purkayastha}, \citenamefont {Sanyal}, \citenamefont {Dhar},\ and\
  \citenamefont {Kulkarni}}]{Archak1}%
  \BibitemOpen
  \bibfield  {author} {\bibinfo {author} {\bibfnamefont {A.}~\bibnamefont
  {Purkayastha}}, \bibinfo {author} {\bibfnamefont {S.}~\bibnamefont {Sanyal}},
  \bibinfo {author} {\bibfnamefont {A.}~\bibnamefont {Dhar}}, \ and\ \bibinfo
  {author} {\bibfnamefont {M.}~\bibnamefont {Kulkarni}},\ }\href {\doibase
  10.1103/PhysRevB.97.174206} {\bibfield  {journal} {\bibinfo  {journal} {Phys.
  Rev. B}\ }\textbf {\bibinfo {volume} {97}},\ \bibinfo {pages} {174206}
  (\bibinfo {year} {2018})}\BibitemShut {NoStop}%
\bibitem [{\citenamefont {Purkayastha}\ \emph {et~al.}(2017)\citenamefont
  {Purkayastha}, \citenamefont {Dhar},\ and\ \citenamefont
  {Kulkarni}}]{Archak2}%
  \BibitemOpen
  \bibfield  {author} {\bibinfo {author} {\bibfnamefont {A.}~\bibnamefont
  {Purkayastha}}, \bibinfo {author} {\bibfnamefont {A.}~\bibnamefont {Dhar}}, \
  and\ \bibinfo {author} {\bibfnamefont {M.}~\bibnamefont {Kulkarni}},\ }\href
  {\doibase 10.1103/PhysRevB.96.180204} {\bibfield  {journal} {\bibinfo
  {journal} {Phys. Rev. B}\ }\textbf {\bibinfo {volume} {96}},\ \bibinfo
  {pages} {180204} (\bibinfo {year} {2017})}\BibitemShut {NoStop}%
\bibitem [{\citenamefont {Purkayastha}(2019)}]{Archak3}%
  \BibitemOpen
  \bibfield  {author} {\bibinfo {author} {\bibfnamefont {A.}~\bibnamefont
  {Purkayastha}},\ }\href {\doibase 10.1088/1742-5468/ab02f4} {\bibfield
  {journal} {\bibinfo  {journal} {Journal of Statistical Mechanics: Theory and
  Experiment}\ }\textbf {\bibinfo {volume} {2019}},\ \bibinfo {pages} {043101}
  (\bibinfo {year} {2019})}\BibitemShut {NoStop}%
\bibitem [{\citenamefont {Sutradhar}\ \emph {et~al.}(2019)\citenamefont
  {Sutradhar}, \citenamefont {Mukerjee}, \citenamefont {Pandit},\ and\
  \citenamefont {Banerjee}}]{Sutradhar_2019}%
  \BibitemOpen
  \bibfield  {author} {\bibinfo {author} {\bibfnamefont {J.}~\bibnamefont
  {Sutradhar}}, \bibinfo {author} {\bibfnamefont {S.}~\bibnamefont {Mukerjee}},
  \bibinfo {author} {\bibfnamefont {R.}~\bibnamefont {Pandit}}, \ and\ \bibinfo
  {author} {\bibfnamefont {S.}~\bibnamefont {Banerjee}},\ }\href {\doibase
  10.1103/PhysRevB.99.224204} {\bibfield  {journal} {\bibinfo  {journal} {Phys.
  Rev. B}\ }\textbf {\bibinfo {volume} {99}},\ \bibinfo {pages} {224204}
  (\bibinfo {year} {2019})}\BibitemShut {NoStop}%
\bibitem [{\citenamefont {Lev}\ \emph {et~al.}(2017)\citenamefont {Lev},
  \citenamefont {Kennes}, \citenamefont {Klöckner}, \citenamefont {Reichman},\
  and\ \citenamefont {Karrasch}}]{subdiffusion_interaction}%
  \BibitemOpen
  \bibfield  {author} {\bibinfo {author} {\bibfnamefont {Y.~B.}\ \bibnamefont
  {Lev}}, \bibinfo {author} {\bibfnamefont {D.~M.}\ \bibnamefont {Kennes}},
  \bibinfo {author} {\bibfnamefont {C.}~\bibnamefont {Klöckner}}, \bibinfo
  {author} {\bibfnamefont {D.~R.}\ \bibnamefont {Reichman}}, \ and\ \bibinfo
  {author} {\bibfnamefont {C.}~\bibnamefont {Karrasch}},\ }\href {\doibase
  10.1209/0295-5075/119/37003} {\bibfield  {journal} {\bibinfo  {journal}
  {{EPL} (Europhysics Letters)}\ }\textbf {\bibinfo {volume} {119}},\ \bibinfo
  {pages} {37003} (\bibinfo {year} {2017})}\BibitemShut {NoStop}%
\bibitem [{\citenamefont {Kohmoto}\ \emph {et~al.}(1987)\citenamefont
  {Kohmoto}, \citenamefont {Sutherland},\ and\ \citenamefont
  {Tang}}]{Fractal1}%
  \BibitemOpen
  \bibfield  {author} {\bibinfo {author} {\bibfnamefont {M.}~\bibnamefont
  {Kohmoto}}, \bibinfo {author} {\bibfnamefont {B.}~\bibnamefont {Sutherland}},
  \ and\ \bibinfo {author} {\bibfnamefont {C.}~\bibnamefont {Tang}},\ }\href
  {\doibase 10.1103/PhysRevB.35.1020} {\bibfield  {journal} {\bibinfo
  {journal} {Phys. Rev. B}\ }\textbf {\bibinfo {volume} {35}},\ \bibinfo
  {pages} {1020} (\bibinfo {year} {1987})}\BibitemShut {NoStop}%
\bibitem [{\citenamefont {Zhong}\ and\ \citenamefont
  {Mosseri}(1995)}]{Fractal2}%
  \BibitemOpen
  \bibfield  {author} {\bibinfo {author} {\bibfnamefont {J.~X.}\ \bibnamefont
  {Zhong}}\ and\ \bibinfo {author} {\bibfnamefont {R.}~\bibnamefont
  {Mosseri}},\ }\href {\doibase 10.1088/0953-8984/7/44/008} {\bibfield
  {journal} {\bibinfo  {journal} {Journal of Physics: Condensed Matter}\
  }\textbf {\bibinfo {volume} {7}},\ \bibinfo {pages} {8383} (\bibinfo {year}
  {1995})}\BibitemShut {NoStop}%
\bibitem [{\citenamefont {Hiramoto}\ and\ \citenamefont
  {Abe}(1988)}]{Fibonacci0}%
  \BibitemOpen
  \bibfield  {author} {\bibinfo {author} {\bibfnamefont {H.}~\bibnamefont
  {Hiramoto}}\ and\ \bibinfo {author} {\bibfnamefont {S.}~\bibnamefont {Abe}},\
  }\href {\doibase 10.1143/JPSJ.57.230} {\bibfield  {journal} {\bibinfo
  {journal} {Journal of the Physical Society of Japan}\ }\textbf {\bibinfo
  {volume} {57}},\ \bibinfo {pages} {230} (\bibinfo {year} {1988})}\BibitemShut
  {NoStop}%
\bibitem [{\citenamefont {Zhong}\ \emph {et~al.}(2001)\citenamefont {Zhong},
  \citenamefont {Diener}, \citenamefont {Steck}, \citenamefont {Oskay},
  \citenamefont {Raizen}, \citenamefont {Plummer}, \citenamefont {Zhang},\ and\
  \citenamefont {Niu}}]{Fibonacci1}%
  \BibitemOpen
  \bibfield  {author} {\bibinfo {author} {\bibfnamefont {J.}~\bibnamefont
  {Zhong}}, \bibinfo {author} {\bibfnamefont {R.~B.}\ \bibnamefont {Diener}},
  \bibinfo {author} {\bibfnamefont {D.~A.}\ \bibnamefont {Steck}}, \bibinfo
  {author} {\bibfnamefont {W.~H.}\ \bibnamefont {Oskay}}, \bibinfo {author}
  {\bibfnamefont {M.~G.}\ \bibnamefont {Raizen}}, \bibinfo {author}
  {\bibfnamefont {E.~W.}\ \bibnamefont {Plummer}}, \bibinfo {author}
  {\bibfnamefont {Z.}~\bibnamefont {Zhang}}, \ and\ \bibinfo {author}
  {\bibfnamefont {Q.}~\bibnamefont {Niu}},\ }\href {\doibase
  10.1103/PhysRevLett.86.2485} {\bibfield  {journal} {\bibinfo  {journal}
  {Phys. Rev. Lett.}\ }\textbf {\bibinfo {volume} {86}},\ \bibinfo {pages}
  {2485} (\bibinfo {year} {2001})}\BibitemShut {NoStop}%
\bibitem [{\citenamefont {Varma}\ and\ \citenamefont {\ifmmode \check{Z}\else
  \v{Z}\fi{}nidari\ifmmode~\check{c}\else \v{c}\fi{}}(2019)}]{Fibonacci2}%
  \BibitemOpen
  \bibfield  {author} {\bibinfo {author} {\bibfnamefont {V.~K.}\ \bibnamefont
  {Varma}}\ and\ \bibinfo {author} {\bibfnamefont {M.}~\bibnamefont {\ifmmode
  \check{Z}\else \v{Z}\fi{}nidari\ifmmode~\check{c}\else \v{c}\fi{}}},\ }\href
  {\doibase 10.1103/PhysRevB.100.085105} {\bibfield  {journal} {\bibinfo
  {journal} {Phys. Rev. B}\ }\textbf {\bibinfo {volume} {100}},\ \bibinfo
  {pages} {085105} (\bibinfo {year} {2019})}\BibitemShut {NoStop}%
\bibitem [{\citenamefont {Jagannathan}(2021)}]{Fibonacci3}%
  \BibitemOpen
  \bibfield  {author} {\bibinfo {author} {\bibfnamefont {A.}~\bibnamefont
  {Jagannathan}},\ }\href@noop {} {\  (\bibinfo {year} {2021})},\ \Eprint
  {http://arxiv.org/abs/2012.14744} {arXiv:2012.14744 [cond-mat.stat-mech]}
  \BibitemShut {NoStop}%
\bibitem [{\citenamefont {Chiaracane}\ \emph {et~al.}(2021)\citenamefont
  {Chiaracane}, \citenamefont {Pietracaprina}, \citenamefont {Purkayastha},\
  and\ \citenamefont {Goold}}]{Fibonacci4}%
  \BibitemOpen
  \bibfield  {author} {\bibinfo {author} {\bibfnamefont {C.}~\bibnamefont
  {Chiaracane}}, \bibinfo {author} {\bibfnamefont {F.}~\bibnamefont
  {Pietracaprina}}, \bibinfo {author} {\bibfnamefont {A.}~\bibnamefont
  {Purkayastha}}, \ and\ \bibinfo {author} {\bibfnamefont {J.}~\bibnamefont
  {Goold}},\ }\href {\doibase 10.1103/PhysRevB.103.184205} {\bibfield
  {journal} {\bibinfo  {journal} {Phys. Rev. B}\ }\textbf {\bibinfo {volume}
  {103}},\ \bibinfo {pages} {184205} (\bibinfo {year} {2021})}\BibitemShut
  {NoStop}%
\bibitem [{\citenamefont {Settino}\ \emph {et~al.}(2020)\citenamefont
  {Settino}, \citenamefont {Talarico}, \citenamefont {Cosco}, \citenamefont
  {Plastina}, \citenamefont {Maniscalco},\ and\ \citenamefont
  {Lo~Gullo}}]{Fibonacci5}%
  \BibitemOpen
  \bibfield  {author} {\bibinfo {author} {\bibfnamefont {J.}~\bibnamefont
  {Settino}}, \bibinfo {author} {\bibfnamefont {N.~W.}\ \bibnamefont
  {Talarico}}, \bibinfo {author} {\bibfnamefont {F.}~\bibnamefont {Cosco}},
  \bibinfo {author} {\bibfnamefont {F.}~\bibnamefont {Plastina}}, \bibinfo
  {author} {\bibfnamefont {S.}~\bibnamefont {Maniscalco}}, \ and\ \bibinfo
  {author} {\bibfnamefont {N.}~\bibnamefont {Lo~Gullo}},\ }\href {\doibase
  10.1103/PhysRevB.101.144303} {\bibfield  {journal} {\bibinfo  {journal}
  {Phys. Rev. B}\ }\textbf {\bibinfo {volume} {101}},\ \bibinfo {pages}
  {144303} (\bibinfo {year} {2020})}\BibitemShut {NoStop}%
\bibitem [{\citenamefont {Ryabtsev}\ \emph {et~al.}(2010)\citenamefont
  {Ryabtsev}, \citenamefont {Tretyakov}, \citenamefont {Beterov},\ and\
  \citenamefont {Entin}}]{expt_Rydberg_atoms1}%
  \BibitemOpen
  \bibfield  {author} {\bibinfo {author} {\bibfnamefont {I.~I.}\ \bibnamefont
  {Ryabtsev}}, \bibinfo {author} {\bibfnamefont {D.~B.}\ \bibnamefont
  {Tretyakov}}, \bibinfo {author} {\bibfnamefont {I.~I.}\ \bibnamefont
  {Beterov}}, \ and\ \bibinfo {author} {\bibfnamefont {V.~M.}\ \bibnamefont
  {Entin}},\ }\href {\doibase 10.1103/PhysRevLett.104.073003} {\bibfield
  {journal} {\bibinfo  {journal} {Phys. Rev. Lett.}\ }\textbf {\bibinfo
  {volume} {104}},\ \bibinfo {pages} {073003} (\bibinfo {year}
  {2010})}\BibitemShut {NoStop}%
\bibitem [{\citenamefont {B\'eguin}\ \emph {et~al.}(2013)\citenamefont
  {B\'eguin}, \citenamefont {Vernier}, \citenamefont {Chicireanu},
  \citenamefont {Lahaye},\ and\ \citenamefont
  {Browaeys}}]{expt_Rydberg_atoms2}%
  \BibitemOpen
  \bibfield  {author} {\bibinfo {author} {\bibfnamefont {L.}~\bibnamefont
  {B\'eguin}}, \bibinfo {author} {\bibfnamefont {A.}~\bibnamefont {Vernier}},
  \bibinfo {author} {\bibfnamefont {R.}~\bibnamefont {Chicireanu}}, \bibinfo
  {author} {\bibfnamefont {T.}~\bibnamefont {Lahaye}}, \ and\ \bibinfo {author}
  {\bibfnamefont {A.}~\bibnamefont {Browaeys}},\ }\href {\doibase
  10.1103/PhysRevLett.110.263201} {\bibfield  {journal} {\bibinfo  {journal}
  {Phys. Rev. Lett.}\ }\textbf {\bibinfo {volume} {110}},\ \bibinfo {pages}
  {263201} (\bibinfo {year} {2013})}\BibitemShut {NoStop}%
\bibitem [{\citenamefont {Browaeys}\ \emph {et~al.}(2016)\citenamefont
  {Browaeys}, \citenamefont {Barredo},\ and\ \citenamefont
  {Lahaye}}]{expt_Rydberg_atoms3}%
  \BibitemOpen
  \bibfield  {author} {\bibinfo {author} {\bibfnamefont {A.}~\bibnamefont
  {Browaeys}}, \bibinfo {author} {\bibfnamefont {D.}~\bibnamefont {Barredo}}, \
  and\ \bibinfo {author} {\bibfnamefont {T.}~\bibnamefont {Lahaye}},\ }\href
  {\doibase 10.1088/0953-4075/49/15/152001} {\bibfield  {journal} {\bibinfo
  {journal} {Journal of Physics B: Atomic, Molecular and Optical Physics}\
  }\textbf {\bibinfo {volume} {49}},\ \bibinfo {pages} {152001} (\bibinfo
  {year} {2016})}\BibitemShut {NoStop}%
\bibitem [{\citenamefont {Guardado-Sanchez}\ \emph {et~al.}(2021)\citenamefont
  {Guardado-Sanchez}, \citenamefont {Spar}, \citenamefont {Schauss},
  \citenamefont {Belyansky}, \citenamefont {Young}, \citenamefont {Bienias},
  \citenamefont {Gorshkov}, \citenamefont {Iadecola},\ and\ \citenamefont
  {Bakr}}]{expt_Rydberg_atoms4}%
  \BibitemOpen
  \bibfield  {author} {\bibinfo {author} {\bibfnamefont {E.}~\bibnamefont
  {Guardado-Sanchez}}, \bibinfo {author} {\bibfnamefont {B.~M.}\ \bibnamefont
  {Spar}}, \bibinfo {author} {\bibfnamefont {P.}~\bibnamefont {Schauss}},
  \bibinfo {author} {\bibfnamefont {R.}~\bibnamefont {Belyansky}}, \bibinfo
  {author} {\bibfnamefont {J.~T.}\ \bibnamefont {Young}}, \bibinfo {author}
  {\bibfnamefont {P.}~\bibnamefont {Bienias}}, \bibinfo {author} {\bibfnamefont
  {A.~V.}\ \bibnamefont {Gorshkov}}, \bibinfo {author} {\bibfnamefont
  {T.}~\bibnamefont {Iadecola}}, \ and\ \bibinfo {author} {\bibfnamefont
  {W.~S.}\ \bibnamefont {Bakr}},\ }\href {\doibase 10.1103/PhysRevX.11.021036}
  {\bibfield  {journal} {\bibinfo  {journal} {Phys. Rev. X}\ }\textbf {\bibinfo
  {volume} {11}},\ \bibinfo {pages} {021036} (\bibinfo {year}
  {2021})}\BibitemShut {NoStop}%
\bibitem [{\citenamefont {Korenblit}\ \emph {et~al.}(2012)\citenamefont
  {Korenblit}, \citenamefont {Kafri}, \citenamefont {Campbell}, \citenamefont
  {Islam}, \citenamefont {Edwards}, \citenamefont {Gong}, \citenamefont {Lin},
  \citenamefont {Duan}, \citenamefont {Kim}, \citenamefont {Kim},\ and\
  \citenamefont {Monroe}}]{expt_trapped_ions1}%
  \BibitemOpen
  \bibfield  {author} {\bibinfo {author} {\bibfnamefont {S.}~\bibnamefont
  {Korenblit}}, \bibinfo {author} {\bibfnamefont {D.}~\bibnamefont {Kafri}},
  \bibinfo {author} {\bibfnamefont {W.~C.}\ \bibnamefont {Campbell}}, \bibinfo
  {author} {\bibfnamefont {R.}~\bibnamefont {Islam}}, \bibinfo {author}
  {\bibfnamefont {E.~E.}\ \bibnamefont {Edwards}}, \bibinfo {author}
  {\bibfnamefont {Z.-X.}\ \bibnamefont {Gong}}, \bibinfo {author}
  {\bibfnamefont {G.-D.}\ \bibnamefont {Lin}}, \bibinfo {author} {\bibfnamefont
  {L.-M.}\ \bibnamefont {Duan}}, \bibinfo {author} {\bibfnamefont
  {J.}~\bibnamefont {Kim}}, \bibinfo {author} {\bibfnamefont {K.}~\bibnamefont
  {Kim}}, \ and\ \bibinfo {author} {\bibfnamefont {C.}~\bibnamefont {Monroe}},\
  }\href {\doibase 10.1088/1367-2630/14/9/095024} {\bibfield  {journal}
  {\bibinfo  {journal} {New Journal of Physics}\ }\textbf {\bibinfo {volume}
  {14}},\ \bibinfo {pages} {095024} (\bibinfo {year} {2012})}\BibitemShut
  {NoStop}%
\bibitem [{\citenamefont {Jurcevic}\ \emph {et~al.}(2017)\citenamefont
  {Jurcevic}, \citenamefont {Shen}, \citenamefont {Hauke}, \citenamefont
  {Maier}, \citenamefont {Brydges}, \citenamefont {Hempel}, \citenamefont
  {Lanyon}, \citenamefont {Heyl}, \citenamefont {Blatt},\ and\ \citenamefont
  {Roos}}]{expt_trapped_ions2}%
  \BibitemOpen
  \bibfield  {author} {\bibinfo {author} {\bibfnamefont {P.}~\bibnamefont
  {Jurcevic}}, \bibinfo {author} {\bibfnamefont {H.}~\bibnamefont {Shen}},
  \bibinfo {author} {\bibfnamefont {P.}~\bibnamefont {Hauke}}, \bibinfo
  {author} {\bibfnamefont {C.}~\bibnamefont {Maier}}, \bibinfo {author}
  {\bibfnamefont {T.}~\bibnamefont {Brydges}}, \bibinfo {author} {\bibfnamefont
  {C.}~\bibnamefont {Hempel}}, \bibinfo {author} {\bibfnamefont {B.~P.}\
  \bibnamefont {Lanyon}}, \bibinfo {author} {\bibfnamefont {M.}~\bibnamefont
  {Heyl}}, \bibinfo {author} {\bibfnamefont {R.}~\bibnamefont {Blatt}}, \ and\
  \bibinfo {author} {\bibfnamefont {C.~F.}\ \bibnamefont {Roos}},\ }\href
  {\doibase 10.1103/PhysRevLett.119.080501} {\bibfield  {journal} {\bibinfo
  {journal} {Phys. Rev. Lett.}\ }\textbf {\bibinfo {volume} {119}},\ \bibinfo
  {pages} {080501} (\bibinfo {year} {2017})}\BibitemShut {NoStop}%
\bibitem [{\citenamefont {Neyenhuis}\ \emph {et~al.}(2017)\citenamefont
  {Neyenhuis}, \citenamefont {Zhang}, \citenamefont {Hess}, \citenamefont
  {Smith}, \citenamefont {Lee}, \citenamefont {Richerme}, \citenamefont {Gong},
  \citenamefont {Gorshkov},\ and\ \citenamefont {Monroe}}]{expt_trapped_ions3}%
  \BibitemOpen
  \bibfield  {author} {\bibinfo {author} {\bibfnamefont {B.}~\bibnamefont
  {Neyenhuis}}, \bibinfo {author} {\bibfnamefont {J.}~\bibnamefont {Zhang}},
  \bibinfo {author} {\bibfnamefont {P.~W.}\ \bibnamefont {Hess}}, \bibinfo
  {author} {\bibfnamefont {J.}~\bibnamefont {Smith}}, \bibinfo {author}
  {\bibfnamefont {A.~C.}\ \bibnamefont {Lee}}, \bibinfo {author} {\bibfnamefont
  {P.}~\bibnamefont {Richerme}}, \bibinfo {author} {\bibfnamefont {Z.-X.}\
  \bibnamefont {Gong}}, \bibinfo {author} {\bibfnamefont {A.~V.}\ \bibnamefont
  {Gorshkov}}, \ and\ \bibinfo {author} {\bibfnamefont {C.}~\bibnamefont
  {Monroe}},\ }\href {\doibase 10.1126/sciadv.1700672} {\bibfield  {journal}
  {\bibinfo  {journal} {Science Advances}\ }\textbf {\bibinfo {volume} {3}}
  (\bibinfo {year} {2017}),\ 10.1126/sciadv.1700672}\BibitemShut {NoStop}%
\bibitem [{\citenamefont {Britton}\ \emph {et~al.}(2012)\citenamefont
  {Britton}, \citenamefont {Sawyer}, \citenamefont {Keith}, \citenamefont
  {Wang}, \citenamefont {Freericks}, \citenamefont {Uys}, \citenamefont
  {Biercuk},\ and\ \citenamefont {Bollinger}}]{expt_trapped_ions4}%
  \BibitemOpen
  \bibfield  {author} {\bibinfo {author} {\bibfnamefont {J.~W.}\ \bibnamefont
  {Britton}}, \bibinfo {author} {\bibfnamefont {B.~C.}\ \bibnamefont {Sawyer}},
  \bibinfo {author} {\bibfnamefont {A.~C.}\ \bibnamefont {Keith}}, \bibinfo
  {author} {\bibfnamefont {C.~C.~J.}\ \bibnamefont {Wang}}, \bibinfo {author}
  {\bibfnamefont {J.~K.}\ \bibnamefont {Freericks}}, \bibinfo {author}
  {\bibfnamefont {H.}~\bibnamefont {Uys}}, \bibinfo {author} {\bibfnamefont
  {M.~J.}\ \bibnamefont {Biercuk}}, \ and\ \bibinfo {author} {\bibfnamefont
  {J.~J.}\ \bibnamefont {Bollinger}},\ }\href
  {https://doi.org/10.1038/nature10981} {\bibfield  {journal} {\bibinfo
  {journal} {Nature}\ }\textbf {\bibinfo {volume} {484}},\ \bibinfo {pages}
  {489 EP } (\bibinfo {year} {2012})}\BibitemShut {NoStop}%
\bibitem [{\citenamefont {Richerme}\ \emph {et~al.}(2014)\citenamefont
  {Richerme}, \citenamefont {Gong}, \citenamefont {Lee}, \citenamefont {Senko},
  \citenamefont {Smith}, \citenamefont {Foss-Feig}, \citenamefont {Michalakis},
  \citenamefont {Gorshkov},\ and\ \citenamefont {Monroe}}]{expt_trapped_ions5}%
  \BibitemOpen
  \bibfield  {author} {\bibinfo {author} {\bibfnamefont {P.}~\bibnamefont
  {Richerme}}, \bibinfo {author} {\bibfnamefont {Z.-X.}\ \bibnamefont {Gong}},
  \bibinfo {author} {\bibfnamefont {A.}~\bibnamefont {Lee}}, \bibinfo {author}
  {\bibfnamefont {C.}~\bibnamefont {Senko}}, \bibinfo {author} {\bibfnamefont
  {J.}~\bibnamefont {Smith}}, \bibinfo {author} {\bibfnamefont
  {M.}~\bibnamefont {Foss-Feig}}, \bibinfo {author} {\bibfnamefont
  {S.}~\bibnamefont {Michalakis}}, \bibinfo {author} {\bibfnamefont {A.~V.}\
  \bibnamefont {Gorshkov}}, \ and\ \bibinfo {author} {\bibfnamefont
  {C.}~\bibnamefont {Monroe}},\ }\href {https://doi.org/10.1038/nature13450}
  {\bibfield  {journal} {\bibinfo  {journal} {Nature}\ }\textbf {\bibinfo
  {volume} {511}},\ \bibinfo {pages} {198 EP } (\bibinfo {year}
  {2014})}\BibitemShut {NoStop}%
\bibitem [{\citenamefont {Zhang}\ \emph
  {et~al.}(2017{\natexlab{a}})\citenamefont {Zhang}, \citenamefont {Pagano},
  \citenamefont {Hess}, \citenamefont {Kyprianidis}, \citenamefont {Becker},
  \citenamefont {Kaplan}, \citenamefont {Gorshkov}, \citenamefont {Gong},\ and\
  \citenamefont {Monroe}}]{expt_trapped_ions7}%
  \BibitemOpen
  \bibfield  {author} {\bibinfo {author} {\bibfnamefont {J.}~\bibnamefont
  {Zhang}}, \bibinfo {author} {\bibfnamefont {G.}~\bibnamefont {Pagano}},
  \bibinfo {author} {\bibfnamefont {P.~W.}\ \bibnamefont {Hess}}, \bibinfo
  {author} {\bibfnamefont {A.}~\bibnamefont {Kyprianidis}}, \bibinfo {author}
  {\bibfnamefont {P.}~\bibnamefont {Becker}}, \bibinfo {author} {\bibfnamefont
  {H.}~\bibnamefont {Kaplan}}, \bibinfo {author} {\bibfnamefont {A.~V.}\
  \bibnamefont {Gorshkov}}, \bibinfo {author} {\bibfnamefont {Z.-X.}\
  \bibnamefont {Gong}}, \ and\ \bibinfo {author} {\bibfnamefont
  {C.}~\bibnamefont {Monroe}},\ }\href {https://doi.org/10.1038/nature24654}
  {\bibfield  {journal} {\bibinfo  {journal} {Nature}\ }\textbf {\bibinfo
  {volume} {551}},\ \bibinfo {pages} {601 EP } (\bibinfo {year}
  {2017}{\natexlab{a}})}\BibitemShut {NoStop}%
\bibitem [{\citenamefont {Zhang}\ \emph
  {et~al.}(2017{\natexlab{b}})\citenamefont {Zhang}, \citenamefont {Hess},
  \citenamefont {Kyprianidis}, \citenamefont {Becker}, \citenamefont {Lee},
  \citenamefont {Smith}, \citenamefont {Pagano}, \citenamefont {Potirniche},
  \citenamefont {Potter}, \citenamefont {Vishwanath}, \citenamefont {Yao},\
  and\ \citenamefont {Monroe}}]{expt_time_crystal2}%
  \BibitemOpen
  \bibfield  {author} {\bibinfo {author} {\bibfnamefont {J.}~\bibnamefont
  {Zhang}}, \bibinfo {author} {\bibfnamefont {P.~W.}\ \bibnamefont {Hess}},
  \bibinfo {author} {\bibfnamefont {A.}~\bibnamefont {Kyprianidis}}, \bibinfo
  {author} {\bibfnamefont {P.}~\bibnamefont {Becker}}, \bibinfo {author}
  {\bibfnamefont {A.}~\bibnamefont {Lee}}, \bibinfo {author} {\bibfnamefont
  {J.}~\bibnamefont {Smith}}, \bibinfo {author} {\bibfnamefont
  {G.}~\bibnamefont {Pagano}}, \bibinfo {author} {\bibfnamefont {I.-D.}\
  \bibnamefont {Potirniche}}, \bibinfo {author} {\bibfnamefont {A.~C.}\
  \bibnamefont {Potter}}, \bibinfo {author} {\bibfnamefont {A.}~\bibnamefont
  {Vishwanath}}, \bibinfo {author} {\bibfnamefont {N.~Y.}\ \bibnamefont {Yao}},
  \ and\ \bibinfo {author} {\bibfnamefont {C.}~\bibnamefont {Monroe}},\ }\href
  {https://doi.org/10.1038/nature21413} {\bibfield  {journal} {\bibinfo
  {journal} {Nature}\ }\textbf {\bibinfo {volume} {543}},\ \bibinfo {pages}
  {217 EP } (\bibinfo {year} {2017}{\natexlab{b}})}\BibitemShut {NoStop}%
\bibitem [{\citenamefont {Jurcevic}\ \emph {et~al.}(2014)\citenamefont
  {Jurcevic}, \citenamefont {Lanyon}, \citenamefont {Hauke}, \citenamefont
  {Hempel}, \citenamefont {Zoller}, \citenamefont {Blatt},\ and\ \citenamefont
  {Roos}}]{expt_trapped_ions6}%
  \BibitemOpen
  \bibfield  {author} {\bibinfo {author} {\bibfnamefont {P.}~\bibnamefont
  {Jurcevic}}, \bibinfo {author} {\bibfnamefont {B.~P.}\ \bibnamefont
  {Lanyon}}, \bibinfo {author} {\bibfnamefont {P.}~\bibnamefont {Hauke}},
  \bibinfo {author} {\bibfnamefont {C.}~\bibnamefont {Hempel}}, \bibinfo
  {author} {\bibfnamefont {P.}~\bibnamefont {Zoller}}, \bibinfo {author}
  {\bibfnamefont {R.}~\bibnamefont {Blatt}}, \ and\ \bibinfo {author}
  {\bibfnamefont {C.~F.}\ \bibnamefont {Roos}},\ }\href
  {https://doi.org/10.1038/nature13461} {\bibfield  {journal} {\bibinfo
  {journal} {Nature}\ }\textbf {\bibinfo {volume} {511}},\ \bibinfo {pages}
  {202 EP } (\bibinfo {year} {2014})}\BibitemShut {NoStop}%
\bibitem [{\citenamefont {Maier}\ \emph {et~al.}(2019)\citenamefont {Maier},
  \citenamefont {Brydges}, \citenamefont {Jurcevic}, \citenamefont {Trautmann},
  \citenamefont {Hempel}, \citenamefont {Lanyon}, \citenamefont {Hauke},
  \citenamefont {Blatt},\ and\ \citenamefont {Roos}}]{Experiment_transport}%
  \BibitemOpen
  \bibfield  {author} {\bibinfo {author} {\bibfnamefont {C.}~\bibnamefont
  {Maier}}, \bibinfo {author} {\bibfnamefont {T.}~\bibnamefont {Brydges}},
  \bibinfo {author} {\bibfnamefont {P.}~\bibnamefont {Jurcevic}}, \bibinfo
  {author} {\bibfnamefont {N.}~\bibnamefont {Trautmann}}, \bibinfo {author}
  {\bibfnamefont {C.}~\bibnamefont {Hempel}}, \bibinfo {author} {\bibfnamefont
  {B.~P.}\ \bibnamefont {Lanyon}}, \bibinfo {author} {\bibfnamefont
  {P.}~\bibnamefont {Hauke}}, \bibinfo {author} {\bibfnamefont
  {R.}~\bibnamefont {Blatt}}, \ and\ \bibinfo {author} {\bibfnamefont {C.~F.}\
  \bibnamefont {Roos}},\ }\href {\doibase 10.1103/PhysRevLett.122.050501}
  {\bibfield  {journal} {\bibinfo  {journal} {Phys. Rev. Lett.}\ }\textbf
  {\bibinfo {volume} {122}},\ \bibinfo {pages} {050501} (\bibinfo {year}
  {2019})}\BibitemShut {NoStop}%
\bibitem [{\citenamefont {Yan}\ \emph {et~al.}(2013)\citenamefont {Yan},
  \citenamefont {Moses}, \citenamefont {Gadway}, \citenamefont {Covey},
  \citenamefont {Hazzard}, \citenamefont {Rey}, \citenamefont {Jin},\ and\
  \citenamefont {Ye}}]{expt_polar_molecules1}%
  \BibitemOpen
  \bibfield  {author} {\bibinfo {author} {\bibfnamefont {B.}~\bibnamefont
  {Yan}}, \bibinfo {author} {\bibfnamefont {S.~A.}\ \bibnamefont {Moses}},
  \bibinfo {author} {\bibfnamefont {B.}~\bibnamefont {Gadway}}, \bibinfo
  {author} {\bibfnamefont {J.~P.}\ \bibnamefont {Covey}}, \bibinfo {author}
  {\bibfnamefont {K.~R.~A.}\ \bibnamefont {Hazzard}}, \bibinfo {author}
  {\bibfnamefont {A.~M.}\ \bibnamefont {Rey}}, \bibinfo {author} {\bibfnamefont
  {D.~S.}\ \bibnamefont {Jin}}, \ and\ \bibinfo {author} {\bibfnamefont
  {J.}~\bibnamefont {Ye}},\ }\href {https://doi.org/10.1038/nature12483}
  {\bibfield  {journal} {\bibinfo  {journal} {Nature}\ }\textbf {\bibinfo
  {volume} {501}},\ \bibinfo {pages} {521 EP } (\bibinfo {year}
  {2013})}\BibitemShut {NoStop}%
\bibitem [{\citenamefont {Moses}\ \emph {et~al.}(2016)\citenamefont {Moses},
  \citenamefont {Covey}, \citenamefont {Miecnikowski}, \citenamefont {Jin},\
  and\ \citenamefont {Ye}}]{expt_polar_molecules3}%
  \BibitemOpen
  \bibfield  {author} {\bibinfo {author} {\bibfnamefont {S.~A.}\ \bibnamefont
  {Moses}}, \bibinfo {author} {\bibfnamefont {J.~P.}\ \bibnamefont {Covey}},
  \bibinfo {author} {\bibfnamefont {M.~T.}\ \bibnamefont {Miecnikowski}},
  \bibinfo {author} {\bibfnamefont {D.~S.}\ \bibnamefont {Jin}}, \ and\
  \bibinfo {author} {\bibfnamefont {J.}~\bibnamefont {Ye}},\ }\href
  {https://doi.org/10.1038/nphys3985} {\bibfield  {journal} {\bibinfo
  {journal} {Nature Physics}\ }\textbf {\bibinfo {volume} {13}},\ \bibinfo
  {pages} {13 EP } (\bibinfo {year} {2016})}\BibitemShut {NoStop}%
\bibitem [{\citenamefont {de~Paz}\ \emph {et~al.}(2013)\citenamefont {de~Paz},
  \citenamefont {Sharma}, \citenamefont {Chotia}, \citenamefont {Mar\'echal},
  \citenamefont {Huckans}, \citenamefont {Pedri}, \citenamefont {Santos},
  \citenamefont {Gorceix}, \citenamefont {Vernac},\ and\ \citenamefont
  {Laburthe-Tolra}}]{expt_dipolar_gas1}%
  \BibitemOpen
  \bibfield  {author} {\bibinfo {author} {\bibfnamefont {A.}~\bibnamefont
  {de~Paz}}, \bibinfo {author} {\bibfnamefont {A.}~\bibnamefont {Sharma}},
  \bibinfo {author} {\bibfnamefont {A.}~\bibnamefont {Chotia}}, \bibinfo
  {author} {\bibfnamefont {E.}~\bibnamefont {Mar\'echal}}, \bibinfo {author}
  {\bibfnamefont {J.~H.}\ \bibnamefont {Huckans}}, \bibinfo {author}
  {\bibfnamefont {P.}~\bibnamefont {Pedri}}, \bibinfo {author} {\bibfnamefont
  {L.}~\bibnamefont {Santos}}, \bibinfo {author} {\bibfnamefont
  {O.}~\bibnamefont {Gorceix}}, \bibinfo {author} {\bibfnamefont
  {L.}~\bibnamefont {Vernac}}, \ and\ \bibinfo {author} {\bibfnamefont
  {B.}~\bibnamefont {Laburthe-Tolra}},\ }\href {\doibase
  10.1103/PhysRevLett.111.185305} {\bibfield  {journal} {\bibinfo  {journal}
  {Phys. Rev. Lett.}\ }\textbf {\bibinfo {volume} {111}},\ \bibinfo {pages}
  {185305} (\bibinfo {year} {2013})}\BibitemShut {NoStop}%
\bibitem [{\citenamefont {{\'A}lvarez}\ \emph {et~al.}(2015)\citenamefont
  {{\'A}lvarez}, \citenamefont {Suter},\ and\ \citenamefont
  {Kaiser}}]{expt_nuclear_spins}%
  \BibitemOpen
  \bibfield  {author} {\bibinfo {author} {\bibfnamefont {G.~A.}\ \bibnamefont
  {{\'A}lvarez}}, \bibinfo {author} {\bibfnamefont {D.}~\bibnamefont {Suter}},
  \ and\ \bibinfo {author} {\bibfnamefont {R.}~\bibnamefont {Kaiser}},\ }\href
  {\doibase 10.1126/science.1261160} {\bibfield  {journal} {\bibinfo  {journal}
  {Science}\ }\textbf {\bibinfo {volume} {349}},\ \bibinfo {pages} {846}
  (\bibinfo {year} {2015})}\BibitemShut {NoStop}%
\bibitem [{\citenamefont {Ni}\ \emph {et~al.}(2008)\citenamefont {Ni},
  \citenamefont {Ospelkaus}, \citenamefont {de~Miranda}, \citenamefont
  {Pe{\textquoteright}er}, \citenamefont {Neyenhuis}, \citenamefont {Zirbel},
  \citenamefont {Kotochigova}, \citenamefont {Julienne}, \citenamefont {Jin},\
  and\ \citenamefont {Ye}}]{expt_polar_molecules2}%
  \BibitemOpen
  \bibfield  {author} {\bibinfo {author} {\bibfnamefont {K.-K.}\ \bibnamefont
  {Ni}}, \bibinfo {author} {\bibfnamefont {S.}~\bibnamefont {Ospelkaus}},
  \bibinfo {author} {\bibfnamefont {M.~H.~G.}\ \bibnamefont {de~Miranda}},
  \bibinfo {author} {\bibfnamefont {A.}~\bibnamefont {Pe{\textquoteright}er}},
  \bibinfo {author} {\bibfnamefont {B.}~\bibnamefont {Neyenhuis}}, \bibinfo
  {author} {\bibfnamefont {J.~J.}\ \bibnamefont {Zirbel}}, \bibinfo {author}
  {\bibfnamefont {S.}~\bibnamefont {Kotochigova}}, \bibinfo {author}
  {\bibfnamefont {P.~S.}\ \bibnamefont {Julienne}}, \bibinfo {author}
  {\bibfnamefont {D.~S.}\ \bibnamefont {Jin}}, \ and\ \bibinfo {author}
  {\bibfnamefont {J.}~\bibnamefont {Ye}},\ }\href {\doibase
  10.1126/science.1163861} {\bibfield  {journal} {\bibinfo  {journal}
  {Science}\ }\textbf {\bibinfo {volume} {322}},\ \bibinfo {pages} {231}
  (\bibinfo {year} {2008})}\BibitemShut {NoStop}%
\bibitem [{\citenamefont {Choi}\ \emph {et~al.}(2017)\citenamefont {Choi},
  \citenamefont {Choi}, \citenamefont {Landig}, \citenamefont {Kucsko},
  \citenamefont {Zhou}, \citenamefont {Isoya}, \citenamefont {Jelezko},
  \citenamefont {Onoda}, \citenamefont {Sumiya}, \citenamefont {Khemani},
  \citenamefont {von Keyserlingk}, \citenamefont {Yao}, \citenamefont
  {Demler},\ and\ \citenamefont {Lukin}}]{expt_time_crystal1}%
  \BibitemOpen
  \bibfield  {author} {\bibinfo {author} {\bibfnamefont {S.}~\bibnamefont
  {Choi}}, \bibinfo {author} {\bibfnamefont {J.}~\bibnamefont {Choi}}, \bibinfo
  {author} {\bibfnamefont {R.}~\bibnamefont {Landig}}, \bibinfo {author}
  {\bibfnamefont {G.}~\bibnamefont {Kucsko}}, \bibinfo {author} {\bibfnamefont
  {H.}~\bibnamefont {Zhou}}, \bibinfo {author} {\bibfnamefont {J.}~\bibnamefont
  {Isoya}}, \bibinfo {author} {\bibfnamefont {F.}~\bibnamefont {Jelezko}},
  \bibinfo {author} {\bibfnamefont {S.}~\bibnamefont {Onoda}}, \bibinfo
  {author} {\bibfnamefont {H.}~\bibnamefont {Sumiya}}, \bibinfo {author}
  {\bibfnamefont {V.}~\bibnamefont {Khemani}}, \bibinfo {author} {\bibfnamefont
  {C.}~\bibnamefont {von Keyserlingk}}, \bibinfo {author} {\bibfnamefont
  {N.~Y.}\ \bibnamefont {Yao}}, \bibinfo {author} {\bibfnamefont
  {E.}~\bibnamefont {Demler}}, \ and\ \bibinfo {author} {\bibfnamefont {M.~D.}\
  \bibnamefont {Lukin}},\ }\href {https://doi.org/10.1038/nature21426}
  {\bibfield  {journal} {\bibinfo  {journal} {Nature}\ }\textbf {\bibinfo
  {volume} {543}},\ \bibinfo {pages} {221 EP } (\bibinfo {year}
  {2017})}\BibitemShut {NoStop}%
\bibitem [{\citenamefont {Roses}\ \emph {et~al.}(2021)\citenamefont {Roses},
  \citenamefont {Landa},\ and\ \citenamefont
  {Dalla~Torre}}]{superconducting_qubit_long_range_experiment}%
  \BibitemOpen
  \bibfield  {author} {\bibinfo {author} {\bibfnamefont {M.~M.}\ \bibnamefont
  {Roses}}, \bibinfo {author} {\bibfnamefont {H.}~\bibnamefont {Landa}}, \ and\
  \bibinfo {author} {\bibfnamefont {E.~G.}\ \bibnamefont {Dalla~Torre}},\
  }\href {\doibase 10.1103/PhysRevResearch.3.033288} {\bibfield  {journal}
  {\bibinfo  {journal} {Phys. Rev. Research}\ }\textbf {\bibinfo {volume}
  {3}},\ \bibinfo {pages} {033288} (\bibinfo {year} {2021})}\BibitemShut
  {NoStop}%
\bibitem [{\citenamefont {Smale}\ \emph {et~al.}(2019)\citenamefont {Smale},
  \citenamefont {He}, \citenamefont {Olsen}, \citenamefont {Jackson},
  \citenamefont {Sharum}, \citenamefont {Trotzky}, \citenamefont {Marino},
  \citenamefont {Rey},\ and\ \citenamefont {Thywissen}}]{expt_atoms_in_trap}%
  \BibitemOpen
  \bibfield  {author} {\bibinfo {author} {\bibfnamefont {S.}~\bibnamefont
  {Smale}}, \bibinfo {author} {\bibfnamefont {P.}~\bibnamefont {He}}, \bibinfo
  {author} {\bibfnamefont {B.~A.}\ \bibnamefont {Olsen}}, \bibinfo {author}
  {\bibfnamefont {K.~G.}\ \bibnamefont {Jackson}}, \bibinfo {author}
  {\bibfnamefont {H.}~\bibnamefont {Sharum}}, \bibinfo {author} {\bibfnamefont
  {S.}~\bibnamefont {Trotzky}}, \bibinfo {author} {\bibfnamefont
  {J.}~\bibnamefont {Marino}}, \bibinfo {author} {\bibfnamefont {A.~M.}\
  \bibnamefont {Rey}}, \ and\ \bibinfo {author} {\bibfnamefont {J.~H.}\
  \bibnamefont {Thywissen}},\ }\href
  {https://advances.sciencemag.org/content/5/8/eaax1568} {\bibfield  {journal}
  {\bibinfo  {journal} {Science Advances}\ }\textbf {\bibinfo {volume} {5}}
  (\bibinfo {year} {2019})}\BibitemShut {NoStop}%
\bibitem [{\citenamefont {Saha}\ \emph
  {et~al.}(2019{\natexlab{a}})\citenamefont {Saha}, \citenamefont {Maiti},\
  and\ \citenamefont {Purkayastha}}]{Archak4}%
  \BibitemOpen
  \bibfield  {author} {\bibinfo {author} {\bibfnamefont {M.}~\bibnamefont
  {Saha}}, \bibinfo {author} {\bibfnamefont {S.~K.}\ \bibnamefont {Maiti}}, \
  and\ \bibinfo {author} {\bibfnamefont {A.}~\bibnamefont {Purkayastha}},\
  }\href {\doibase 10.1103/PhysRevB.100.174201} {\bibfield  {journal} {\bibinfo
   {journal} {Phys. Rev. B}\ }\textbf {\bibinfo {volume} {100}},\ \bibinfo
  {pages} {174201} (\bibinfo {year} {2019}{\natexlab{a}})}\BibitemShut
  {NoStop}%
\bibitem [{\citenamefont {Akhanjee}(2009)}]{long_range_transport}%
  \BibitemOpen
  \bibfield  {author} {\bibinfo {author} {\bibfnamefont {S.}~\bibnamefont
  {Akhanjee}},\ }\href {\doibase 10.1103/PhysRevB.79.205101} {\bibfield
  {journal} {\bibinfo  {journal} {Phys. Rev. B}\ }\textbf {\bibinfo {volume}
  {79}},\ \bibinfo {pages} {205101} (\bibinfo {year} {2009})}\BibitemShut
  {NoStop}%
\bibitem [{\citenamefont {Saha}\ \emph
  {et~al.}(2019{\natexlab{b}})\citenamefont {Saha}, \citenamefont
  {Purkayastha},\ and\ \citenamefont {Maiti}}]{Archak5}%
  \BibitemOpen
  \bibfield  {author} {\bibinfo {author} {\bibfnamefont {M.}~\bibnamefont
  {Saha}}, \bibinfo {author} {\bibfnamefont {A.}~\bibnamefont {Purkayastha}}, \
  and\ \bibinfo {author} {\bibfnamefont {S.~K.}\ \bibnamefont {Maiti}},\ }\href
  {\doibase 10.1088/1361-648x/ab4494} {\bibfield  {journal} {\bibinfo
  {journal} {Journal of Physics: Condensed Matter}\ }\textbf {\bibinfo {volume}
  {32}},\ \bibinfo {pages} {025303} (\bibinfo {year}
  {2019}{\natexlab{b}})}\BibitemShut {NoStop}%
\bibitem [{\citenamefont {Kloss}\ and\ \citenamefont
  {Bar~Lev}(2019)}]{Kloss_2019}%
  \BibitemOpen
  \bibfield  {author} {\bibinfo {author} {\bibfnamefont {B.}~\bibnamefont
  {Kloss}}\ and\ \bibinfo {author} {\bibfnamefont {Y.}~\bibnamefont
  {Bar~Lev}},\ }\href {\doibase 10.1103/PhysRevA.99.032114} {\bibfield
  {journal} {\bibinfo  {journal} {Phys. Rev. A}\ }\textbf {\bibinfo {volume}
  {99}},\ \bibinfo {pages} {032114} (\bibinfo {year} {2019})}\BibitemShut
  {NoStop}%
\bibitem [{\citenamefont {Kloss}\ and\ \citenamefont
  {Bar~Lev}(2020)}]{Kloss_2020}%
  \BibitemOpen
  \bibfield  {author} {\bibinfo {author} {\bibfnamefont {B.}~\bibnamefont
  {Kloss}}\ and\ \bibinfo {author} {\bibfnamefont {Y.}~\bibnamefont
  {Bar~Lev}},\ }\href {\doibase 10.1103/PhysRevB.102.060201} {\bibfield
  {journal} {\bibinfo  {journal} {Phys. Rev. B}\ }\textbf {\bibinfo {volume}
  {102}},\ \bibinfo {pages} {060201} (\bibinfo {year} {2020})}\BibitemShut
  {NoStop}%
\bibitem [{\citenamefont {Kawa}\ and\ \citenamefont
  {Machnikowski}(2020)}]{Kawa_2020}%
  \BibitemOpen
  \bibfield  {author} {\bibinfo {author} {\bibfnamefont {K.}~\bibnamefont
  {Kawa}}\ and\ \bibinfo {author} {\bibfnamefont {P.}~\bibnamefont
  {Machnikowski}},\ }\href {\doibase 10.1103/PhysRevB.102.174203} {\bibfield
  {journal} {\bibinfo  {journal} {Phys. Rev. B}\ }\textbf {\bibinfo {volume}
  {102}},\ \bibinfo {pages} {174203} (\bibinfo {year} {2020})}\BibitemShut
  {NoStop}%
\bibitem [{\citenamefont {Schneider}\ \emph {et~al.}(2021)\citenamefont
  {Schneider}, \citenamefont {Despres}, \citenamefont {Thomson}, \citenamefont
  {Tagliacozzo},\ and\ \citenamefont {Sanchez-Palencia}}]{Schneider_2021}%
  \BibitemOpen
  \bibfield  {author} {\bibinfo {author} {\bibfnamefont {J.~T.}\ \bibnamefont
  {Schneider}}, \bibinfo {author} {\bibfnamefont {J.}~\bibnamefont {Despres}},
  \bibinfo {author} {\bibfnamefont {S.~J.}\ \bibnamefont {Thomson}}, \bibinfo
  {author} {\bibfnamefont {L.}~\bibnamefont {Tagliacozzo}}, \ and\ \bibinfo
  {author} {\bibfnamefont {L.}~\bibnamefont {Sanchez-Palencia}},\ }\href
  {\doibase 10.1103/PhysRevResearch.3.L012022} {\bibfield  {journal} {\bibinfo
  {journal} {Phys. Rev. Research}\ }\textbf {\bibinfo {volume} {3}},\ \bibinfo
  {pages} {L012022} (\bibinfo {year} {2021})}\BibitemShut {NoStop}%
\bibitem [{\citenamefont {Prasad}\ and\ \citenamefont
  {Garg}(2021)}]{Prasad_2021}%
  \BibitemOpen
  \bibfield  {author} {\bibinfo {author} {\bibfnamefont {Y.}~\bibnamefont
  {Prasad}}\ and\ \bibinfo {author} {\bibfnamefont {A.}~\bibnamefont {Garg}},\
  }\href {\doibase 10.1103/PhysRevB.103.064203} {\bibfield  {journal} {\bibinfo
   {journal} {Phys. Rev. B}\ }\textbf {\bibinfo {volume} {103}},\ \bibinfo
  {pages} {064203} (\bibinfo {year} {2021})}\BibitemShut {NoStop}%
\bibitem [{\citenamefont {Modak}\ and\ \citenamefont
  {Nag}(2020)}]{Ranjan_Tanay}%
  \BibitemOpen
  \bibfield  {author} {\bibinfo {author} {\bibfnamefont {R.}~\bibnamefont
  {Modak}}\ and\ \bibinfo {author} {\bibfnamefont {T.}~\bibnamefont {Nag}},\
  }\href {\doibase 10.1103/PhysRevResearch.2.012074} {\bibfield  {journal}
  {\bibinfo  {journal} {Phys. Rev. Research}\ }\textbf {\bibinfo {volume}
  {2}},\ \bibinfo {pages} {012074} (\bibinfo {year} {2020})}\BibitemShut
  {NoStop}%
\bibitem [{\citenamefont {Katzer}\ \emph {et~al.}(2020)\citenamefont {Katzer},
  \citenamefont {Knorr}, \citenamefont {Finsterh\"olzl},\ and\ \citenamefont
  {Carmele}}]{Katzer_2020}%
  \BibitemOpen
  \bibfield  {author} {\bibinfo {author} {\bibfnamefont {M.}~\bibnamefont
  {Katzer}}, \bibinfo {author} {\bibfnamefont {W.}~\bibnamefont {Knorr}},
  \bibinfo {author} {\bibfnamefont {R.}~\bibnamefont {Finsterh\"olzl}}, \ and\
  \bibinfo {author} {\bibfnamefont {A.}~\bibnamefont {Carmele}},\ }\href
  {\doibase 10.1103/PhysRevB.102.125101} {\bibfield  {journal} {\bibinfo
  {journal} {Phys. Rev. B}\ }\textbf {\bibinfo {volume} {102}},\ \bibinfo
  {pages} {125101} (\bibinfo {year} {2020})}\BibitemShut {NoStop}%
\bibitem [{sup()}]{supp}%
  \BibitemOpen
  \href@noop {} {\bibinfo  {journal} {See supplemental material}\ }\BibitemShut
  {NoStop}%
\bibitem [{\citenamefont {Kosloff}\ and\ \citenamefont
  {Levy}(2014)}]{Kosloff_Levy_2014}%
  \BibitemOpen
\bibfield  {journal} {  }\bibfield  {author} {\bibinfo {author} {\bibfnamefont
  {R.}~\bibnamefont {Kosloff}}\ and\ \bibinfo {author} {\bibfnamefont
  {A.}~\bibnamefont {Levy}},\ }\href {\doibase
  10.1146/annurev-physchem-040513-103724} {\bibfield  {journal} {\bibinfo
  {journal} {Annual Review of Physical Chemistry}\ }\textbf {\bibinfo {volume}
  {65}},\ \bibinfo {pages} {365} (\bibinfo {year} {2014})},\ \bibinfo {note}
  {pMID: 24689798}\BibitemShut {NoStop}%
\bibitem [{\citenamefont {Benenti}\ \emph
  {et~al.}(2017{\natexlab{b}})\citenamefont {Benenti}, \citenamefont {Casati},
  \citenamefont {Saito},\ and\ \citenamefont {Whitney}}]{Benenti_2017}%
  \BibitemOpen
  \bibfield  {author} {\bibinfo {author} {\bibfnamefont {G.}~\bibnamefont
  {Benenti}}, \bibinfo {author} {\bibfnamefont {G.}~\bibnamefont {Casati}},
  \bibinfo {author} {\bibfnamefont {K.}~\bibnamefont {Saito}}, \ and\ \bibinfo
  {author} {\bibfnamefont {R.~S.}\ \bibnamefont {Whitney}},\ }\href {\doibase
  https://doi.org/10.1016/j.physrep.2017.05.008} {\bibfield  {journal}
  {\bibinfo  {journal} {Physics Reports}\ }\textbf {\bibinfo {volume} {694}},\
  \bibinfo {pages} {1} (\bibinfo {year} {2017}{\natexlab{b}})}\BibitemShut
  {NoStop}%
\bibitem [{\citenamefont {Haug}\ and\ \citenamefont
  {Jauho}(2008)}]{Jauho_book}%
  \BibitemOpen
  \bibfield  {author} {\bibinfo {author} {\bibfnamefont {H.}~\bibnamefont
  {Haug}}\ and\ \bibinfo {author} {\bibfnamefont {A.-P.}\ \bibnamefont
  {Jauho}},\ }\href {\doibase 10.1007/978-3-540-73564-9} {\emph {\bibinfo
  {title} {Quantum Kinetics in Transport and Optics of Semiconductors}}}\
  (\bibinfo  {publisher} {Springer-Verlag Berlin Heidelberg},\ \bibinfo {year}
  {2008})\BibitemShut {NoStop}%
\bibitem [{\citenamefont {Di~Ventra}(2008)}]{di_Ventra_book}%
  \BibitemOpen
  \bibfield  {author} {\bibinfo {author} {\bibfnamefont {M.}~\bibnamefont
  {Di~Ventra}},\ }\href {\doibase 10.1017/CBO9780511755606} {\emph {\bibinfo
  {title} {Electrical Transport in Nanoscale Systems}}}\ (\bibinfo  {publisher}
  {Cambridge University Press},\ \bibinfo {year} {2008})\BibitemShut {NoStop}%
\bibitem [{\citenamefont {Dhar}\ and\ \citenamefont {Sen}(2006)}]{Dhar_2006}%
  \BibitemOpen
  \bibfield  {author} {\bibinfo {author} {\bibfnamefont {A.}~\bibnamefont
  {Dhar}}\ and\ \bibinfo {author} {\bibfnamefont {D.}~\bibnamefont {Sen}},\
  }\href {\doibase 10.1103/PhysRevB.73.085119} {\bibfield  {journal} {\bibinfo
  {journal} {Phys. Rev. B}\ }\textbf {\bibinfo {volume} {73}},\ \bibinfo
  {pages} {085119} (\bibinfo {year} {2006})}\BibitemShut {NoStop}%
\bibitem [{\citenamefont {Rodriguez}\ \emph {et~al.}(2017)\citenamefont
  {Rodriguez}, \citenamefont {Casteels}, \citenamefont {Storme}, \citenamefont
  {Carlon~Zambon}, \citenamefont {Sagnes}, \citenamefont {Le~Gratiet},
  \citenamefont {Galopin}, \citenamefont {Lema\^{\i}tre}, \citenamefont {Amo},
  \citenamefont {Ciuti},\ and\ \citenamefont
  {Bloch}}]{non_eq_phase_transition_expt1}%
  \BibitemOpen
  \bibfield  {author} {\bibinfo {author} {\bibfnamefont {S.~R.~K.}\
  \bibnamefont {Rodriguez}}, \bibinfo {author} {\bibfnamefont {W.}~\bibnamefont
  {Casteels}}, \bibinfo {author} {\bibfnamefont {F.}~\bibnamefont {Storme}},
  \bibinfo {author} {\bibfnamefont {N.}~\bibnamefont {Carlon~Zambon}}, \bibinfo
  {author} {\bibfnamefont {I.}~\bibnamefont {Sagnes}}, \bibinfo {author}
  {\bibfnamefont {L.}~\bibnamefont {Le~Gratiet}}, \bibinfo {author}
  {\bibfnamefont {E.}~\bibnamefont {Galopin}}, \bibinfo {author} {\bibfnamefont
  {A.}~\bibnamefont {Lema\^{\i}tre}}, \bibinfo {author} {\bibfnamefont
  {A.}~\bibnamefont {Amo}}, \bibinfo {author} {\bibfnamefont {C.}~\bibnamefont
  {Ciuti}}, \ and\ \bibinfo {author} {\bibfnamefont {J.}~\bibnamefont
  {Bloch}},\ }\href {\doibase 10.1103/PhysRevLett.118.247402} {\bibfield
  {journal} {\bibinfo  {journal} {Phys. Rev. Lett.}\ }\textbf {\bibinfo
  {volume} {118}},\ \bibinfo {pages} {247402} (\bibinfo {year}
  {2017})}\BibitemShut {NoStop}%
\bibitem [{\citenamefont {Heugel}\ \emph {et~al.}(2019)\citenamefont {Heugel},
  \citenamefont {Biondi}, \citenamefont {Zilberberg},\ and\ \citenamefont
  {Chitra}}]{non_eq_phase_transition_expt2}%
  \BibitemOpen
  \bibfield  {author} {\bibinfo {author} {\bibfnamefont {T.~L.}\ \bibnamefont
  {Heugel}}, \bibinfo {author} {\bibfnamefont {M.}~\bibnamefont {Biondi}},
  \bibinfo {author} {\bibfnamefont {O.}~\bibnamefont {Zilberberg}}, \ and\
  \bibinfo {author} {\bibfnamefont {R.}~\bibnamefont {Chitra}},\ }\href
  {\doibase 10.1103/PhysRevLett.123.173601} {\bibfield  {journal} {\bibinfo
  {journal} {Phys. Rev. Lett.}\ }\textbf {\bibinfo {volume} {123}},\ \bibinfo
  {pages} {173601} (\bibinfo {year} {2019})}\BibitemShut {NoStop}%
\bibitem [{\citenamefont {Fink}\ \emph {et~al.}(2017)\citenamefont {Fink},
  \citenamefont {Dombi}, \citenamefont {Vukics}, \citenamefont {Wallraff},\
  and\ \citenamefont {Domokos}}]{non_eq_phase_transition_expt3}%
  \BibitemOpen
  \bibfield  {author} {\bibinfo {author} {\bibfnamefont {J.~M.}\ \bibnamefont
  {Fink}}, \bibinfo {author} {\bibfnamefont {A.}~\bibnamefont {Dombi}},
  \bibinfo {author} {\bibfnamefont {A.}~\bibnamefont {Vukics}}, \bibinfo
  {author} {\bibfnamefont {A.}~\bibnamefont {Wallraff}}, \ and\ \bibinfo
  {author} {\bibfnamefont {P.}~\bibnamefont {Domokos}},\ }\href {\doibase
  10.1103/PhysRevX.7.011012} {\bibfield  {journal} {\bibinfo  {journal} {Phys.
  Rev. X}\ }\textbf {\bibinfo {volume} {7}},\ \bibinfo {pages} {011012}
  (\bibinfo {year} {2017})}\BibitemShut {NoStop}%
\bibitem [{\citenamefont {Fitzpatrick}\ \emph {et~al.}(2017)\citenamefont
  {Fitzpatrick}, \citenamefont {Sundaresan}, \citenamefont {Li}, \citenamefont
  {Koch},\ and\ \citenamefont {Houck}}]{non_eq_phase_transition_expt4}%
  \BibitemOpen
  \bibfield  {author} {\bibinfo {author} {\bibfnamefont {M.}~\bibnamefont
  {Fitzpatrick}}, \bibinfo {author} {\bibfnamefont {N.~M.}\ \bibnamefont
  {Sundaresan}}, \bibinfo {author} {\bibfnamefont {A.~C.~Y.}\ \bibnamefont
  {Li}}, \bibinfo {author} {\bibfnamefont {J.}~\bibnamefont {Koch}}, \ and\
  \bibinfo {author} {\bibfnamefont {A.~A.}\ \bibnamefont {Houck}},\ }\href
  {\doibase 10.1103/PhysRevX.7.011016} {\bibfield  {journal} {\bibinfo
  {journal} {Phys. Rev. X}\ }\textbf {\bibinfo {volume} {7}},\ \bibinfo {pages}
  {011016} (\bibinfo {year} {2017})}\BibitemShut {NoStop}%
\bibitem [{\citenamefont {Jo}\ \emph {et~al.}(2021)\citenamefont {Jo},
  \citenamefont {Lee}, \citenamefont {Choi},\ and\ \citenamefont
  {Kahng}}]{Jo_2021}%
  \BibitemOpen
  \bibfield  {author} {\bibinfo {author} {\bibfnamefont {M.}~\bibnamefont
  {Jo}}, \bibinfo {author} {\bibfnamefont {J.}~\bibnamefont {Lee}}, \bibinfo
  {author} {\bibfnamefont {K.}~\bibnamefont {Choi}}, \ and\ \bibinfo {author}
  {\bibfnamefont {B.}~\bibnamefont {Kahng}},\ }\href {\doibase
  10.1103/PhysRevResearch.3.013238} {\bibfield  {journal} {\bibinfo  {journal}
  {Phys. Rev. Research}\ }\textbf {\bibinfo {volume} {3}},\ \bibinfo {pages}
  {013238} (\bibinfo {year} {2021})}\BibitemShut {NoStop}%
\bibitem [{\citenamefont {Gamayun}\ \emph {et~al.}(2021)\citenamefont
  {Gamayun}, \citenamefont {Slobodeniuk}, \citenamefont {Caux},\ and\
  \citenamefont {Lychkovskiy}}]{Gamayun_2021}%
  \BibitemOpen
  \bibfield  {author} {\bibinfo {author} {\bibfnamefont {O.}~\bibnamefont
  {Gamayun}}, \bibinfo {author} {\bibfnamefont {A.}~\bibnamefont
  {Slobodeniuk}}, \bibinfo {author} {\bibfnamefont {J.-S.}\ \bibnamefont
  {Caux}}, \ and\ \bibinfo {author} {\bibfnamefont {O.}~\bibnamefont
  {Lychkovskiy}},\ }\href {\doibase 10.1103/PhysRevB.103.L041405} {\bibfield
  {journal} {\bibinfo  {journal} {Phys. Rev. B}\ }\textbf {\bibinfo {volume}
  {103}},\ \bibinfo {pages} {L041405} (\bibinfo {year} {2021})}\BibitemShut
  {NoStop}%
\bibitem [{\citenamefont {Zamora}\ \emph {et~al.}(2020)\citenamefont {Zamora},
  \citenamefont {Dagvadorj}, \citenamefont {Comaron}, \citenamefont
  {Carusotto}, \citenamefont {Proukakis},\ and\ \citenamefont
  {Szyma\ifmmode~\acute{n}\else \'{n}\fi{}ska}}]{Zamora_2020}%
  \BibitemOpen
  \bibfield  {author} {\bibinfo {author} {\bibfnamefont {A.}~\bibnamefont
  {Zamora}}, \bibinfo {author} {\bibfnamefont {G.}~\bibnamefont {Dagvadorj}},
  \bibinfo {author} {\bibfnamefont {P.}~\bibnamefont {Comaron}}, \bibinfo
  {author} {\bibfnamefont {I.}~\bibnamefont {Carusotto}}, \bibinfo {author}
  {\bibfnamefont {N.~P.}\ \bibnamefont {Proukakis}}, \ and\ \bibinfo {author}
  {\bibfnamefont {M.~H.}\ \bibnamefont {Szyma\ifmmode~\acute{n}\else
  \'{n}\fi{}ska}},\ }\href {\doibase 10.1103/PhysRevLett.125.095301} {\bibfield
   {journal} {\bibinfo  {journal} {Phys. Rev. Lett.}\ }\textbf {\bibinfo
  {volume} {125}},\ \bibinfo {pages} {095301} (\bibinfo {year}
  {2020})}\BibitemShut {NoStop}%
\bibitem [{\citenamefont {Jo}\ \emph {et~al.}(2019)\citenamefont {Jo},
  \citenamefont {Um},\ and\ \citenamefont {Kahng}}]{Jo_2019}%
  \BibitemOpen
  \bibfield  {author} {\bibinfo {author} {\bibfnamefont {M.}~\bibnamefont
  {Jo}}, \bibinfo {author} {\bibfnamefont {J.}~\bibnamefont {Um}}, \ and\
  \bibinfo {author} {\bibfnamefont {B.}~\bibnamefont {Kahng}},\ }\href
  {\doibase 10.1103/PhysRevE.99.032131} {\bibfield  {journal} {\bibinfo
  {journal} {Phys. Rev. E}\ }\textbf {\bibinfo {volume} {99}},\ \bibinfo
  {pages} {032131} (\bibinfo {year} {2019})}\BibitemShut {NoStop}%
\bibitem [{\citenamefont {Carollo}\ \emph {et~al.}(2019)\citenamefont
  {Carollo}, \citenamefont {Gillman}, \citenamefont {Weimer},\ and\
  \citenamefont {Lesanovsky}}]{Carollo_2019}%
  \BibitemOpen
  \bibfield  {author} {\bibinfo {author} {\bibfnamefont {F.}~\bibnamefont
  {Carollo}}, \bibinfo {author} {\bibfnamefont {E.}~\bibnamefont {Gillman}},
  \bibinfo {author} {\bibfnamefont {H.}~\bibnamefont {Weimer}}, \ and\ \bibinfo
  {author} {\bibfnamefont {I.}~\bibnamefont {Lesanovsky}},\ }\href {\doibase
  10.1103/PhysRevLett.123.100604} {\bibfield  {journal} {\bibinfo  {journal}
  {Phys. Rev. Lett.}\ }\textbf {\bibinfo {volume} {123}},\ \bibinfo {pages}
  {100604} (\bibinfo {year} {2019})}\BibitemShut {NoStop}%
\bibitem [{\citenamefont {Minganti}\ \emph {et~al.}(2018)\citenamefont
  {Minganti}, \citenamefont {Biella}, \citenamefont {Bartolo},\ and\
  \citenamefont {Ciuti}}]{Minganti_2018}%
  \BibitemOpen
  \bibfield  {author} {\bibinfo {author} {\bibfnamefont {F.}~\bibnamefont
  {Minganti}}, \bibinfo {author} {\bibfnamefont {A.}~\bibnamefont {Biella}},
  \bibinfo {author} {\bibfnamefont {N.}~\bibnamefont {Bartolo}}, \ and\
  \bibinfo {author} {\bibfnamefont {C.}~\bibnamefont {Ciuti}},\ }\href
  {\doibase 10.1103/PhysRevA.98.042118} {\bibfield  {journal} {\bibinfo
  {journal} {Phys. Rev. A}\ }\textbf {\bibinfo {volume} {98}},\ \bibinfo
  {pages} {042118} (\bibinfo {year} {2018})}\BibitemShut {NoStop}%
\bibitem [{\citenamefont {Marcuzzi}\ \emph {et~al.}(2016)\citenamefont
  {Marcuzzi}, \citenamefont {Buchhold}, \citenamefont {Diehl},\ and\
  \citenamefont {Lesanovsky}}]{Marcuzzi_2016}%
  \BibitemOpen
  \bibfield  {author} {\bibinfo {author} {\bibfnamefont {M.}~\bibnamefont
  {Marcuzzi}}, \bibinfo {author} {\bibfnamefont {M.}~\bibnamefont {Buchhold}},
  \bibinfo {author} {\bibfnamefont {S.}~\bibnamefont {Diehl}}, \ and\ \bibinfo
  {author} {\bibfnamefont {I.}~\bibnamefont {Lesanovsky}},\ }\href {\doibase
  10.1103/PhysRevLett.116.245701} {\bibfield  {journal} {\bibinfo  {journal}
  {Phys. Rev. Lett.}\ }\textbf {\bibinfo {volume} {116}},\ \bibinfo {pages}
  {245701} (\bibinfo {year} {2016})}\BibitemShut {NoStop}%
\bibitem [{\citenamefont {Dagvadorj}\ \emph {et~al.}(2015)\citenamefont
  {Dagvadorj}, \citenamefont {Fellows}, \citenamefont
  {Matyja\ifmmode~\acute{s}\else \'{s}\fi{}kiewicz}, \citenamefont {Marchetti},
  \citenamefont {Carusotto},\ and\ \citenamefont {Szyma\ifmmode~\acute{n}\else
  \'{n}\fi{}ska}}]{Dagvadorj_2015}%
  \BibitemOpen
  \bibfield  {author} {\bibinfo {author} {\bibfnamefont {G.}~\bibnamefont
  {Dagvadorj}}, \bibinfo {author} {\bibfnamefont {J.~M.}\ \bibnamefont
  {Fellows}}, \bibinfo {author} {\bibfnamefont {S.}~\bibnamefont
  {Matyja\ifmmode~\acute{s}\else \'{s}\fi{}kiewicz}}, \bibinfo {author}
  {\bibfnamefont {F.~M.}\ \bibnamefont {Marchetti}}, \bibinfo {author}
  {\bibfnamefont {I.}~\bibnamefont {Carusotto}}, \ and\ \bibinfo {author}
  {\bibfnamefont {M.~H.}\ \bibnamefont {Szyma\ifmmode~\acute{n}\else
  \'{n}\fi{}ska}},\ }\href {\doibase 10.1103/PhysRevX.5.041028} {\bibfield
  {journal} {\bibinfo  {journal} {Phys. Rev. X}\ }\textbf {\bibinfo {volume}
  {5}},\ \bibinfo {pages} {041028} (\bibinfo {year} {2015})}\BibitemShut
  {NoStop}%
\bibitem [{\citenamefont {Nagy}\ and\ \citenamefont
  {Domokos}(2015)}]{Nagy_2015}%
  \BibitemOpen
  \bibfield  {author} {\bibinfo {author} {\bibfnamefont {D.}~\bibnamefont
  {Nagy}}\ and\ \bibinfo {author} {\bibfnamefont {P.}~\bibnamefont {Domokos}},\
  }\href {\doibase 10.1103/PhysRevLett.115.043601} {\bibfield  {journal}
  {\bibinfo  {journal} {Phys. Rev. Lett.}\ }\textbf {\bibinfo {volume} {115}},\
  \bibinfo {pages} {043601} (\bibinfo {year} {2015})}\BibitemShut {NoStop}%
\bibitem [{\citenamefont {Bastidas}\ \emph {et~al.}(2012)\citenamefont
  {Bastidas}, \citenamefont {Emary}, \citenamefont {Regler},\ and\
  \citenamefont {Brandes}}]{Bastidas_2012}%
  \BibitemOpen
  \bibfield  {author} {\bibinfo {author} {\bibfnamefont {V.~M.}\ \bibnamefont
  {Bastidas}}, \bibinfo {author} {\bibfnamefont {C.}~\bibnamefont {Emary}},
  \bibinfo {author} {\bibfnamefont {B.}~\bibnamefont {Regler}}, \ and\ \bibinfo
  {author} {\bibfnamefont {T.}~\bibnamefont {Brandes}},\ }\href {\doibase
  10.1103/PhysRevLett.108.043003} {\bibfield  {journal} {\bibinfo  {journal}
  {Phys. Rev. Lett.}\ }\textbf {\bibinfo {volume} {108}},\ \bibinfo {pages}
  {043003} (\bibinfo {year} {2012})}\BibitemShut {NoStop}%
\bibitem [{\citenamefont {Prosen}\ and\ \citenamefont
  {Pi\ifmmode~\check{z}\else \v{z}\fi{}orn}(2008)}]{Prosen_2008}%
  \BibitemOpen
  \bibfield  {author} {\bibinfo {author} {\bibfnamefont {T.}~\bibnamefont
  {Prosen}}\ and\ \bibinfo {author} {\bibfnamefont {I.}~\bibnamefont
  {Pi\ifmmode~\check{z}\else \v{z}\fi{}orn}},\ }\href {\doibase
  10.1103/PhysRevLett.101.105701} {\bibfield  {journal} {\bibinfo  {journal}
  {Phys. Rev. Lett.}\ }\textbf {\bibinfo {volume} {101}},\ \bibinfo {pages}
  {105701} (\bibinfo {year} {2008})}\BibitemShut {NoStop}%
\bibitem [{\citenamefont {Breuer}\ and\ \citenamefont
  {Petruccione}(2007)}]{Breuer_book}%
  \BibitemOpen
  \bibfield  {author} {\bibinfo {author} {\bibfnamefont {H.-P.}\ \bibnamefont
  {Breuer}}\ and\ \bibinfo {author} {\bibfnamefont {F.}~\bibnamefont
  {Petruccione}},\ }\href {\doibase 10.1093/acprof:oso/9780199213900.001.0001}
  {\emph {\bibinfo {title} {The Theory of Open Quantum Systems}}}\ (\bibinfo
  {publisher} {Oxford University Press},\ \bibinfo {year} {2007})\BibitemShut
  {NoStop}%
\bibitem [{\citenamefont {Walls}(1970)}]{Walls1970}%
  \BibitemOpen
  \bibfield  {author} {\bibinfo {author} {\bibfnamefont {D.~F.}\ \bibnamefont
  {Walls}},\ }\href {\doibase 10.1007/bf01396784} {\bibfield  {journal}
  {\bibinfo  {journal} {Zeitschrift f\"{u}r Physik A Hadrons and nuclei}\
  }\textbf {\bibinfo {volume} {234}},\ \bibinfo {pages} {231} (\bibinfo {year}
  {1970})}\BibitemShut {NoStop}%
\bibitem [{\citenamefont {Wichterich}\ \emph {et~al.}(2007)\citenamefont
  {Wichterich}, \citenamefont {Henrich}, \citenamefont {Breuer}, \citenamefont
  {Gemmer},\ and\ \citenamefont {Michel}}]{Wichterich_2007}%
  \BibitemOpen
  \bibfield  {author} {\bibinfo {author} {\bibfnamefont {H.}~\bibnamefont
  {Wichterich}}, \bibinfo {author} {\bibfnamefont {M.~J.}\ \bibnamefont
  {Henrich}}, \bibinfo {author} {\bibfnamefont {H.-P.}\ \bibnamefont {Breuer}},
  \bibinfo {author} {\bibfnamefont {J.}~\bibnamefont {Gemmer}}, \ and\ \bibinfo
  {author} {\bibfnamefont {M.}~\bibnamefont {Michel}},\ }\href {\doibase
  10.1103/PhysRevE.76.031115} {\bibfield  {journal} {\bibinfo  {journal} {Phys.
  Rev. E}\ }\textbf {\bibinfo {volume} {76}},\ \bibinfo {pages} {031115}
  (\bibinfo {year} {2007})}\BibitemShut {NoStop}%
\bibitem [{\citenamefont {Rivas}\ \emph {et~al.}(2010)\citenamefont {Rivas},
  \citenamefont {Plato}, \citenamefont {Huelga},\ and\ \citenamefont
  {Plenio}}]{Rivas_2010}%
  \BibitemOpen
  \bibfield  {author} {\bibinfo {author} {\bibfnamefont {{\'{A}}.}~\bibnamefont
  {Rivas}}, \bibinfo {author} {\bibfnamefont {A.~D.~K.}\ \bibnamefont {Plato}},
  \bibinfo {author} {\bibfnamefont {S.~F.}\ \bibnamefont {Huelga}}, \ and\
  \bibinfo {author} {\bibfnamefont {M.~B.}\ \bibnamefont {Plenio}},\ }\href
  {\doibase 10.1088/1367-2630/12/11/113032} {\bibfield  {journal} {\bibinfo
  {journal} {New Journal of Physics}\ }\textbf {\bibinfo {volume} {12}},\
  \bibinfo {pages} {113032} (\bibinfo {year} {2010})}\BibitemShut {NoStop}%
\bibitem [{\citenamefont {Deçordi}\ and\ \citenamefont
  {Vidiella-Barranco}(2017)}]{barranco_2014}%
  \BibitemOpen
  \bibfield  {author} {\bibinfo {author} {\bibfnamefont {G.}~\bibnamefont
  {Deçordi}}\ and\ \bibinfo {author} {\bibfnamefont {A.}~\bibnamefont
  {Vidiella-Barranco}},\ }\href {\doibase
  https://doi.org/10.1016/j.optcom.2016.10.017} {\bibfield  {journal} {\bibinfo
   {journal} {Optics Communications}\ }\textbf {\bibinfo {volume} {387}},\
  \bibinfo {pages} {366} (\bibinfo {year} {2017})}\BibitemShut {NoStop}%
\bibitem [{\citenamefont {Levy}\ and\ \citenamefont
  {Kosloff}(2014)}]{Levy2014}%
  \BibitemOpen
  \bibfield  {author} {\bibinfo {author} {\bibfnamefont {A.}~\bibnamefont
  {Levy}}\ and\ \bibinfo {author} {\bibfnamefont {R.}~\bibnamefont {Kosloff}},\
  }\href {\doibase 10.1209/0295-5075/107/20004} {\bibfield  {journal} {\bibinfo
   {journal} {{EPL} (Europhysics Letters)}\ }\textbf {\bibinfo {volume}
  {107}},\ \bibinfo {pages} {20004} (\bibinfo {year} {2014})}\BibitemShut
  {NoStop}%
\bibitem [{\citenamefont {Purkayastha}\ \emph {et~al.}(2016)\citenamefont
  {Purkayastha}, \citenamefont {Dhar},\ and\ \citenamefont
  {Kulkarni}}]{Archak6}%
  \BibitemOpen
  \bibfield  {author} {\bibinfo {author} {\bibfnamefont {A.}~\bibnamefont
  {Purkayastha}}, \bibinfo {author} {\bibfnamefont {A.}~\bibnamefont {Dhar}}, \
  and\ \bibinfo {author} {\bibfnamefont {M.}~\bibnamefont {Kulkarni}},\ }\href
  {\doibase 10.1103/PhysRevA.93.062114} {\bibfield  {journal} {\bibinfo
  {journal} {Phys. Rev. A}\ }\textbf {\bibinfo {volume} {93}},\ \bibinfo
  {pages} {062114} (\bibinfo {year} {2016})}\BibitemShut {NoStop}%
\bibitem [{\citenamefont {Trushechkin}\ and\ \citenamefont
  {Volovich}(2016)}]{Trushechkin_2016}%
  \BibitemOpen
  \bibfield  {author} {\bibinfo {author} {\bibfnamefont {A.~S.}\ \bibnamefont
  {Trushechkin}}\ and\ \bibinfo {author} {\bibfnamefont {I.~V.}\ \bibnamefont
  {Volovich}},\ }\href {\doibase 10.1209/0295-5075/113/30005} {\bibfield
  {journal} {\bibinfo  {journal} {{EPL} (Europhysics Letters)}\ }\textbf
  {\bibinfo {volume} {113}},\ \bibinfo {pages} {30005} (\bibinfo {year}
  {2016})}\BibitemShut {NoStop}%
\bibitem [{\citenamefont {Eastham}\ \emph {et~al.}(2016)\citenamefont
  {Eastham}, \citenamefont {Kirton}, \citenamefont {Cammack}, \citenamefont
  {Lovett},\ and\ \citenamefont {Keeling}}]{Eastham_2016}%
  \BibitemOpen
  \bibfield  {author} {\bibinfo {author} {\bibfnamefont {P.~R.}\ \bibnamefont
  {Eastham}}, \bibinfo {author} {\bibfnamefont {P.}~\bibnamefont {Kirton}},
  \bibinfo {author} {\bibfnamefont {H.~M.}\ \bibnamefont {Cammack}}, \bibinfo
  {author} {\bibfnamefont {B.~W.}\ \bibnamefont {Lovett}}, \ and\ \bibinfo
  {author} {\bibfnamefont {J.}~\bibnamefont {Keeling}},\ }\href
  {http://dx.doi.org/10.1103/PhysRevA.94.012110} {\bibfield  {journal}
  {\bibinfo  {journal} {Physical Review A}\ }\textbf {\bibinfo {volume} {94}}
  (\bibinfo {year} {2016})}\BibitemShut {NoStop}%
\bibitem [{\citenamefont {Hofer}\ \emph {et~al.}(2017)\citenamefont {Hofer},
  \citenamefont {Perarnau-Llobet}, \citenamefont {Miranda}, \citenamefont
  {Haack}, \citenamefont {Silva}, \citenamefont {Brask},\ and\ \citenamefont
  {Brunner}}]{Hofer_2017}%
  \BibitemOpen
  \bibfield  {author} {\bibinfo {author} {\bibfnamefont {P.~P.}\ \bibnamefont
  {Hofer}}, \bibinfo {author} {\bibfnamefont {M.}~\bibnamefont
  {Perarnau-Llobet}}, \bibinfo {author} {\bibfnamefont {L.~D.~M.}\ \bibnamefont
  {Miranda}}, \bibinfo {author} {\bibfnamefont {G.}~\bibnamefont {Haack}},
  \bibinfo {author} {\bibfnamefont {R.}~\bibnamefont {Silva}}, \bibinfo
  {author} {\bibfnamefont {J.~B.}\ \bibnamefont {Brask}}, \ and\ \bibinfo
  {author} {\bibfnamefont {N.}~\bibnamefont {Brunner}},\ }\href {\doibase
  10.1088/1367-2630/aa964f} {\bibfield  {journal} {\bibinfo  {journal} {New
  Journal of Physics}\ }\textbf {\bibinfo {volume} {19}},\ \bibinfo {pages}
  {123037} (\bibinfo {year} {2017})}\BibitemShut {NoStop}%
\bibitem [{\citenamefont {González}\ \emph {et~al.}(2017)\citenamefont
  {González}, \citenamefont {Correa}, \citenamefont {Nocerino}, \citenamefont
  {Palao}, \citenamefont {Alonso},\ and\ \citenamefont
  {Adesso}}]{Gonzalez_2017}%
  \BibitemOpen
  \bibfield  {author} {\bibinfo {author} {\bibfnamefont {J.~O.}\ \bibnamefont
  {González}}, \bibinfo {author} {\bibfnamefont {L.~A.}\ \bibnamefont
  {Correa}}, \bibinfo {author} {\bibfnamefont {G.}~\bibnamefont {Nocerino}},
  \bibinfo {author} {\bibfnamefont {J.~P.}\ \bibnamefont {Palao}}, \bibinfo
  {author} {\bibfnamefont {D.}~\bibnamefont {Alonso}}, \ and\ \bibinfo {author}
  {\bibfnamefont {G.}~\bibnamefont {Adesso}},\ }\href {\doibase
  10.1142/S1230161217400108} {\bibfield  {journal} {\bibinfo  {journal} {Open
  Systems \& Information Dynamics}\ }\textbf {\bibinfo {volume} {24}},\
  \bibinfo {pages} {1740010} (\bibinfo {year} {2017})}\BibitemShut {NoStop}%
\bibitem [{\citenamefont {Mitchison}\ and\ \citenamefont
  {Plenio}(2018)}]{Mitchison_2018}%
  \BibitemOpen
  \bibfield  {author} {\bibinfo {author} {\bibfnamefont {M.~T.}\ \bibnamefont
  {Mitchison}}\ and\ \bibinfo {author} {\bibfnamefont {M.~B.}\ \bibnamefont
  {Plenio}},\ }\href {\doibase 10.1088/1367-2630/aa9f70} {\bibfield  {journal}
  {\bibinfo  {journal} {New Journal of Physics}\ }\textbf {\bibinfo {volume}
  {20}},\ \bibinfo {pages} {033005} (\bibinfo {year} {2018})}\BibitemShut
  {NoStop}%
\bibitem [{\citenamefont {Cattaneo}\ \emph {et~al.}(2019)\citenamefont
  {Cattaneo}, \citenamefont {Giorgi}, \citenamefont {Maniscalco},\ and\
  \citenamefont {Zambrini}}]{Cattaneo_2019}%
  \BibitemOpen
  \bibfield  {author} {\bibinfo {author} {\bibfnamefont {M.}~\bibnamefont
  {Cattaneo}}, \bibinfo {author} {\bibfnamefont {G.~L.}\ \bibnamefont
  {Giorgi}}, \bibinfo {author} {\bibfnamefont {S.}~\bibnamefont {Maniscalco}},
  \ and\ \bibinfo {author} {\bibfnamefont {R.}~\bibnamefont {Zambrini}},\
  }\href {\doibase 10.1088/1367-2630/ab54ac} {\bibfield  {journal} {\bibinfo
  {journal} {New Journal of Physics}\ }\textbf {\bibinfo {volume} {21}},\
  \bibinfo {pages} {113045} (\bibinfo {year} {2019})}\BibitemShut {NoStop}%
\bibitem [{\citenamefont {Hartmann}\ and\ \citenamefont
  {Strunz}(2020)}]{Hartmann_2020_1}%
  \BibitemOpen
  \bibfield  {author} {\bibinfo {author} {\bibfnamefont {R.}~\bibnamefont
  {Hartmann}}\ and\ \bibinfo {author} {\bibfnamefont {W.~T.}\ \bibnamefont
  {Strunz}},\ }\href {\doibase 10.1103/PhysRevA.101.012103} {\bibfield
  {journal} {\bibinfo  {journal} {Phys. Rev. A}\ }\textbf {\bibinfo {volume}
  {101}},\ \bibinfo {pages} {012103} (\bibinfo {year} {2020})}\BibitemShut
  {NoStop}%
\bibitem [{\citenamefont {Konopik}\ and\ \citenamefont
  {Lutz}(2020)}]{konopik_2020local}%
  \BibitemOpen
  \bibfield  {author} {\bibinfo {author} {\bibfnamefont {M.}~\bibnamefont
  {Konopik}}\ and\ \bibinfo {author} {\bibfnamefont {E.}~\bibnamefont {Lutz}},\
  }\href@noop {} {\  (\bibinfo {year} {2020})},\ \Eprint
  {http://arxiv.org/abs/2012.09907} {arXiv:2012.09907 [quant-ph]} \BibitemShut
  {NoStop}%
\bibitem [{\citenamefont {Scali}\ \emph {et~al.}(2021)\citenamefont {Scali},
  \citenamefont {Anders},\ and\ \citenamefont {Correa}}]{Scali_2021}%
  \BibitemOpen
  \bibfield  {author} {\bibinfo {author} {\bibfnamefont {S.}~\bibnamefont
  {Scali}}, \bibinfo {author} {\bibfnamefont {J.}~\bibnamefont {Anders}}, \
  and\ \bibinfo {author} {\bibfnamefont {L.~A.}\ \bibnamefont {Correa}},\
  }\href {http://dx.doi.org/10.22331/q-2021-05-01-451} {\bibfield  {journal}
  {\bibinfo  {journal} {Quantum}\ }\textbf {\bibinfo {volume} {5}},\ \bibinfo
  {pages} {451} (\bibinfo {year} {2021})}\BibitemShut {NoStop}%
\bibitem [{\citenamefont {Nathan}\ and\ \citenamefont {Rudner}(2020)}]{ule}%
  \BibitemOpen
  \bibfield  {author} {\bibinfo {author} {\bibfnamefont {F.}~\bibnamefont
  {Nathan}}\ and\ \bibinfo {author} {\bibfnamefont {M.~S.}\ \bibnamefont
  {Rudner}},\ }\href {\doibase 10.1103/PhysRevB.102.115109} {\bibfield
  {journal} {\bibinfo  {journal} {Phys. Rev. B}\ }\textbf {\bibinfo {volume}
  {102}},\ \bibinfo {pages} {115109} (\bibinfo {year} {2020})}\BibitemShut
  {NoStop}%
\bibitem [{\citenamefont {Kleinherbers}\ \emph {et~al.}(2020)\citenamefont
  {Kleinherbers}, \citenamefont {Szpak}, \citenamefont {K\"onig},\ and\
  \citenamefont {Sch\"utzhold}}]{Kleinherbers_2020}%
  \BibitemOpen
  \bibfield  {author} {\bibinfo {author} {\bibfnamefont {E.}~\bibnamefont
  {Kleinherbers}}, \bibinfo {author} {\bibfnamefont {N.}~\bibnamefont {Szpak}},
  \bibinfo {author} {\bibfnamefont {J.}~\bibnamefont {K\"onig}}, \ and\
  \bibinfo {author} {\bibfnamefont {R.}~\bibnamefont {Sch\"utzhold}},\ }\href
  {\doibase 10.1103/PhysRevB.101.125131} {\bibfield  {journal} {\bibinfo
  {journal} {Phys. Rev. B}\ }\textbf {\bibinfo {volume} {101}},\ \bibinfo
  {pages} {125131} (\bibinfo {year} {2020})}\BibitemShut {NoStop}%
\bibitem [{\citenamefont {Davidovi{\'c}}(2020)}]{Davidovic_2020}%
  \BibitemOpen
  \bibfield  {author} {\bibinfo {author} {\bibfnamefont {D.}~\bibnamefont
  {Davidovi{\'c}}},\ }\href {\doibase 10.22331/q-2020-09-21-326} {\bibfield
  {journal} {\bibinfo  {journal} {Quantum}\ }\textbf {\bibinfo {volume} {4}},\
  \bibinfo {pages} {326} (\bibinfo {year} {2020})}\BibitemShut {NoStop}%
\bibitem [{\citenamefont {Mozgunov}\ and\ \citenamefont
  {Lidar}(2020)}]{mozgunov2020}%
  \BibitemOpen
  \bibfield  {author} {\bibinfo {author} {\bibfnamefont {E.}~\bibnamefont
  {Mozgunov}}\ and\ \bibinfo {author} {\bibfnamefont {D.}~\bibnamefont
  {Lidar}},\ }\href {http://dx.doi.org/10.22331/q-2020-02-06-227} {\bibfield
  {journal} {\bibinfo  {journal} {Quantum}\ }\textbf {\bibinfo {volume} {4}},\
  \bibinfo {pages} {227} (\bibinfo {year} {2020})}\BibitemShut {NoStop}%
\bibitem [{\citenamefont {McCauley}\ \emph {et~al.}(2020)\citenamefont
  {McCauley}, \citenamefont {Cruikshank}, \citenamefont {Bondar},\ and\
  \citenamefont {Jacobs}}]{mccauley2020}%
  \BibitemOpen
  \bibfield  {author} {\bibinfo {author} {\bibfnamefont {G.}~\bibnamefont
  {McCauley}}, \bibinfo {author} {\bibfnamefont {B.}~\bibnamefont
  {Cruikshank}}, \bibinfo {author} {\bibfnamefont {D.~I.}\ \bibnamefont
  {Bondar}}, \ and\ \bibinfo {author} {\bibfnamefont {K.}~\bibnamefont
  {Jacobs}},\ }\href {http://dx.doi.org/10.1038/s41534-020-00299-6} {\bibfield
  {journal} {\bibinfo  {journal} {npj Quantum Information}\ }\textbf {\bibinfo
  {volume} {6}} (\bibinfo {year} {2020})}\BibitemShut {NoStop}%
\bibitem [{\citenamefont {Kir\ifmmode~\check{s}\else \v{s}\fi{}anskas}\ \emph
  {et~al.}(2018)\citenamefont {Kir\ifmmode~\check{s}\else \v{s}\fi{}anskas},
  \citenamefont {Francki\'e},\ and\ \citenamefont {Wacker}}]{kirvsanskas2018}%
  \BibitemOpen
  \bibfield  {author} {\bibinfo {author} {\bibfnamefont {G.}~\bibnamefont
  {Kir\ifmmode~\check{s}\else \v{s}\fi{}anskas}}, \bibinfo {author}
  {\bibfnamefont {M.}~\bibnamefont {Francki\'e}}, \ and\ \bibinfo {author}
  {\bibfnamefont {A.}~\bibnamefont {Wacker}},\ }\href {\doibase
  10.1103/PhysRevB.97.035432} {\bibfield  {journal} {\bibinfo  {journal} {Phys.
  Rev. B}\ }\textbf {\bibinfo {volume} {97}},\ \bibinfo {pages} {035432}
  (\bibinfo {year} {2018})}\BibitemShut {NoStop}%
\end{thebibliography}%

\widetext{
\clearpage
\begin{center}
\textbf{\large Supplemental Material: Sub-diffusive phases in open clean long-range systems}\\
\end{center}
}

\vspace{5mm}
\setcounter{equation}{0}
\setcounter{figure}{0}
\renewcommand{\theequation}{S\arabic{equation}}
\renewcommand{\thefigure}{S\arabic{figure}}
\renewcommand{\thesection}{S\arabic{section}}

\section{Exact analytical dispersion relation for isolated system in the thermodynamic limit}

In the derivation of our main results, we have used the bare retarded Green's function of the isolated system in the thermodynamic limit. In this section we derive the dispersion relation for the clean long-range isolated system. First, we take our system Hamiltonian in Eq.(7) and re-label the sites to $r\rightarrow r-\frac{N}{2}-1$, assuming, for simplicity, that $N$ is even,
\begin{align}
\label{Hs2}
\hat{\mathcal{H}}_S = -\sum_{m=1}^N \left(\sum_{r=-N/2}^{N/2-1-m} \frac{1}{m^\alpha} \left(\hat{c}_r^\dagger \hat{c}_{r+m} + \hat{c}_{r+m}^\dagger \hat{c}_{r}\right)\right).
\end{align}
Next, we take $N\rightarrow \infty$ in the above equation. This gives,
\begin{align}
\label{Hs3}
\hat{\mathcal{H}}_S = -\sum_{r=-\infty}^{\infty} \sum_{m=1}^\infty  \frac{1}{m^\alpha} \left(\hat{c}_r^\dagger \hat{c}_{r+m} + \hat{c}_{r+m}^\dagger \hat{c}_{r}\right),
\end{align}
where we have neglected some boundary terms. Now the system has translational invariance. We can diagonalize the above Hamiltonian by going to momentum space via a Fourier transform,
\begin{align}
& \hat{\tilde{c}}(k) = \sum_{r=-\infty}^\infty \hat{c}_r e^{i rk},~~k \in [-\pi, \pi] \nonumber \\
& \hat{c}_r = \frac{1}{2\pi} \int_{-\pi}^{\pi} dk  e^{-i rk} \hat{\tilde{c}}(k).
\end{align}
This gives
\begin{align}
\hat{\mathcal{H}}_S = \int_{-\pi}^{\pi} dk~\varepsilon(k,\alpha) \hat{\tilde{c}}^\dagger(k) \hat{\tilde{c}}(k),
\end{align}
with the dispersion relation
\begin{align}
\varepsilon(k,\alpha) = -2\sum_{m=1}^\infty \frac{\cos(m k)}{m^\alpha} .
\end{align}
The retarded Green's function of the isolated system in the thermodynamic limit in the momentum-frequency space is then given by
\begin{align}
g^+(k,\omega) = \lim_{\epsilon\rightarrow 0}\frac{1}{\omega - \varepsilon(k,\alpha)-i\epsilon}.
\end{align}
Returning to the site basis $(p,q=\{1,N\})$, we get the bare retarded Green's function as 
\begin{align}
\mathbf{g}_{pq}^+(\omega) = \lim_{\epsilon\rightarrow 0} \frac{1}{2\pi} \int_{-\pi}^{\pi} dk \frac{e^{-ik |p-q|}}{\omega - \varepsilon(k,\alpha)-i\epsilon}.
\end{align}

\section{Properties of the dispersion relation}
\begin{figure}
\includegraphics[width=\textwidth]{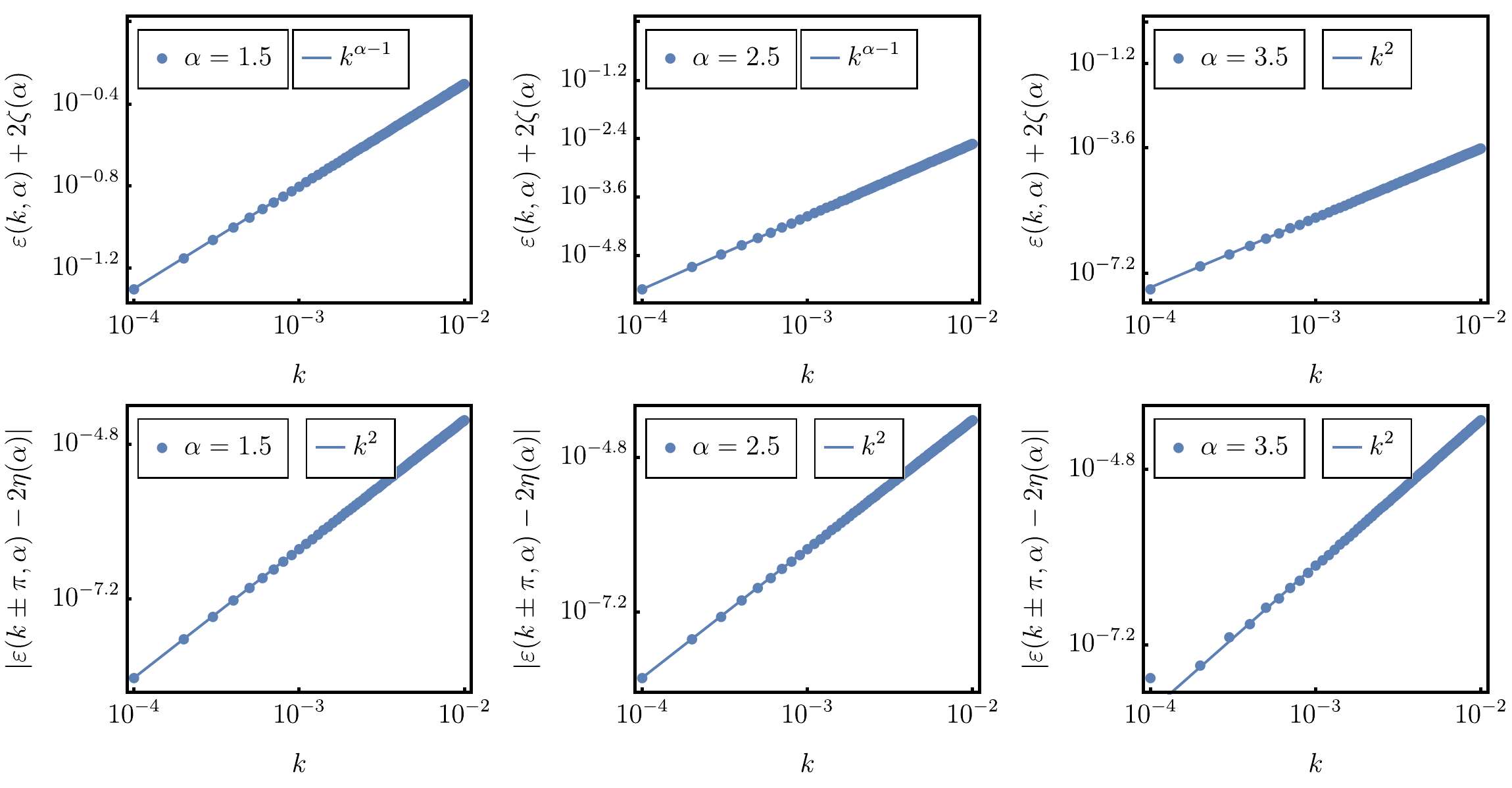} 
\caption{(Color online) In the first row, numerically we have plotted $\varepsilon(k,\alpha)+2\zeta(\alpha)$ in small $k$ regime for $\alpha=1.5, 2.5$ and $3.5$ respectively. For $\alpha<3$, $\varepsilon(k,\alpha)+2\zeta(\alpha)\sim k^{\alpha-1}$ and for $\alpha>3$, $\varepsilon(k,\alpha)+2\zeta(\alpha)\sim k^2$. This also matches with our analytical result eq.~\ref{small_k_dispersion}. Similarly, in the second row, we have plotted $\vert \varepsilon(k\pm \pi)-2\eta(\alpha)\vert$ in small $k$ regime for three different values of $\alpha$. Here we can see $\vert \varepsilon(k\pm \pi)-2\eta(\alpha)\vert$ always goes as $k^2$. This also matches with our analytical result eq.~\ref{k_pi_dispersion}.}
\label{fig:dispersion} 
\end{figure}
In this section, we state the relevant properties of the dispersion relation which are crucial for the proof of the scaling in conductance. The infinite series giving the dispersion relation is absolutely convergent for all $k$ when $\alpha>1$. The following relation holds connecting $\varepsilon(k,\alpha)$ with $\varepsilon(k \pm \pi,\alpha)= 2\sum_{m=1}^\infty (-1)^{m-1} \frac{\cos(m k)}{m^\alpha}$,
\begin{align}
\label{dispersion_property}
\varepsilon(k \pm \pi,\alpha) =  - \varepsilon(k,\alpha) + 2^{1-\alpha} \varepsilon (2k, \alpha)
\end{align}
The minimum of the dispersion relation is given at $k=0$, 
\begin{align}
\varepsilon_{min}(\alpha)=\varepsilon(0,\alpha)=-2 \, \zeta(\alpha),
\end{align}
where $\zeta(\alpha)=\sum_{m=1}^\infty \frac{1}{m^\alpha}$ is the Reimann-zeta function. Similarly, the maximum of the dispersion relation is at $k=\pm \pi$, 
\begin{align}
\varepsilon_{max}(\alpha)=\varepsilon(\pm \pi,\alpha)=2 \, \eta(\alpha),
\end{align}
 where $\eta(\alpha)$ is the Dirichlet eta function, $\eta(\alpha)=\sum_{m=1}^\infty \frac{(-1)^{m-1}}{m^\alpha}=(1-2^{1-\alpha})\zeta(\alpha)$, also known as the alternating zeta function. Importantly, the dispersion relation is non-analytic at $k=0$ for all values of $\alpha$. This is because, 
\begin{align}
\lim_{k\rightarrow 0} \frac{\partial^p \varepsilon(k,\alpha)}{\partial k^p} \rightarrow \infty,~~\forall~p>\alpha-1.
\end{align}
Thus a Taylor expansion around $k=0$ does not exist. However, we still need to find the small $k$ behavior of $\varepsilon(k,\alpha)$. To do this we define
\begin{align}
f(s k,\alpha)=\zeta(\alpha)+ \frac{\varepsilon(s k,\alpha)}{2}=\sum_{m=1}^\infty \frac{1-\cos(s m k)}{m^\alpha}.
\end{align}
Here, $s$ is an integer. Next, we divide the summation in the right-hand-side into two parts,
\begin{align}
& f(s k,\alpha)=\sum_{m=1}^{1/s |k|} \frac{1-\cos(s m k)}{m^\alpha} + |k|^\alpha B, \nonumber \\
& B=\sum_{y\geq 1/s} \frac{1-\cos(s y)}{y^\alpha} \simeq s^{\alpha -1} \int_1^\infty dy^{\prime} \frac{1-\cos(y^{\prime})}{y^{\prime \alpha}}  \nonumber \\
&\simeq s^{\alpha -1} B^{\prime}, ~~B^{\prime}=\int_1^\infty dy^{\prime} \frac{1-\cos(y^{\prime})}{y^{\prime \alpha}} \nonumber
\end{align}
where we have used $y=m|k|$ and $y^{\prime}=s y$. The expression in the definition of $B^{\prime}$ converges. Thus $B^{\prime}$ is a real number which depends on $\alpha$. Now we expand the cosine to obtain
\begin{align}
f(sk,\alpha)=s^{\alpha -1}|k|^\alpha B^{\prime} - \sum_{p=1}^\infty  \frac{(-1)^p s^{2p} k^{2p}}{(2p)!} \sum_{m=1}^{1/s|k|}  \frac{1}{m^{\alpha-2p}}.    
\end{align}
Till now, the expression is exact. After this we make some approximations and assumptions. We replace the summation over $m$ by an integration and further assume that $\alpha$ is not an integer. This then gives,
\begin{align}
\label{general}
& f(sk,\alpha)\simeq s^{\alpha -1}|k|^\alpha B^{\prime} - \sum_{p=1}^\infty  \frac{(-1)^p s^{2p} k^{2p}}{(2p)!} \frac{(1/s|k|)^{2p-\alpha+1} -1}{2p-\alpha+1} \nonumber \\
& = s^{\alpha -1}|k|^\alpha B^{\prime} + s^{\alpha-1}|k|^{\alpha-1} a_1 - \sum_{p=1}^\infty \frac{(-1)^{p+1} s^{2p}k^{2p}}{(2p)!~(2p-\alpha+1)} \nonumber ,\\
& a_1 = \sum_{p=1}^\infty  \frac{(-1)^{p+1} }{(2p)!~(2p-\alpha+1)},~~\alpha>1, \alpha \notin \mathbb{Z},
\end{align} 
where $\mathbb{Z}$ is the set of all integers.
It can be checked by ratio test that the infinite series in the definition of $a_1$ converges. So, $a_1$ is a real number which depends on $\alpha$. Now considering $s=1$, we have an approximate series expansion for $\varepsilon(k,\alpha)$ around $k=0$,
\begin{align}
\label{kequation}
\varepsilon(k,\alpha) & \simeq -2\Big[\zeta(\alpha)-|k|^{\alpha-1} a_1-|k|^\alpha B^{\prime} \nonumber \\
&+\sum_{p=1}^\infty \frac{(-1)^{p+1} k^{2p}}{(2p)!~(2p-\alpha+1)} \Big]
\end{align}
It is interesting to note that $a_1$, which is the coefficient of $|k|^{\alpha-1}$, has contribution from all terms coming from the expansion of the cosine. This is consistent with the fact that Taylor series expansion around $k=0$ is invalid. The presence of absolute values and the terms raised to non-integer powers, both of which make $k=0$ non-analytic, clearly distinguishing the above series expansion from a Taylor series expansion. Armed with the series expansion, we obtain the small $k$ behavior of the dispersion relation by keeping the lowest order terms with non-integer and integer powers,
\begin{align}
\label{small_k_dispersion}
& \varepsilon(k,\alpha)\simeq-2\left[ \zeta(\alpha)- a_1 |k|^{\alpha-1}- a_2 k^2 \right],~~|k| \ll 1, \\
& a_2 = \frac{1}{2(\alpha-3)}.  \nonumber 
\end{align}
Similarly putting $s=2$ in Eq.(\ref{general}) one can compute,
\begin{align}
\label{2kequation}
\varepsilon(2k,\alpha) & \simeq -2\Big[\zeta(\alpha)-2^{\alpha-1}|k|^{\alpha-1} a_1- 2^{\alpha -1}|k|^\alpha B^{\prime} \nonumber \\
&+\sum_{p=1}^\infty \frac{(-1)^{p+1} 2^{2p}k^{2p}}{(2p)!~(2p-\alpha+1)} \Big]
\end{align}
Further, using Eq.(\ref{kequation}) and Eq.(\ref{2kequation}) in Eq.(\ref{dispersion_property}) and considering the leading order term $p=1$ we can also obtain an equivalent expansion around $k=\pm \pi$,
\begin{align}
\label{k_pi_dispersion}
& \varepsilon(k\pm \pi,\alpha)\simeq 2\,\eta(\alpha)- 2 a_2 (1-2^{3-\alpha}) k^2,\nonumber \\
&|k| \ll 1.
\end{align}
which is analytic as the expansion contains only integer powers. The above two expansions are used to obtain the scaling of current with system size. We have also checked this two expansions eq.~\ref{small_k_dispersion} and eq.~\ref{k_pi_dispersion} numerically in fig.\ref{fig:dispersion}. Though the above expansions are obtained for non-integer values of $\alpha$, the results can be analytically continued to include integer values of $\alpha>1$ by making the fractional part arbitrarily small.

{
\section{The NEGF formalism}
In the main text we have used the expression of conductance at zero temperature as obtained from the non-equilibrium Green's function (NEGF) approach. Here we give details of this approach. This pedagogical section follows standard texts and references \cite{Jauho_book,di_Ventra_book,Dhar_2006}.

We want to describe an open system set-up of the form $\hat{\mathcal{H}}=\hat{\mathcal{H}}_S+\hat{\mathcal{H}}_{SB}+\hat{\mathcal{H}}_{B_1}+\hat{\mathcal{H}}_{B_N}$,
\begin{align}
\hat{\mathcal{H}}_S = \sum_{\ell m=1}^N \mathbf{H}_{\ell m} \hat{c}_\ell^\dagger \hat{c}_m, ~~\hat{\mathcal{H}}_{B_1}=\sum_{r=1}^{N_B} \Omega_{r1} \hat{B}_{r1}^\dagger\hat{B}_{r1} ,~~\hat{\mathcal{H}}_{B_N}=\sum_{r=1}^{N_B} \Omega_{rN} \hat{B}_{rN}^\dagger\hat{B}_{rN},~~\hat{\mathcal{H}}_{SB}=\sum_{\ell=1,N}\sum_{r=1}^{N_B} (\kappa_{r\ell} \hat{c}_\ell^\dagger\hat{B}_{r\ell} +~\kappa_{r\ell}^* \hat{B}_{r\ell}^\dagger\hat{c}_\ell), 
\end{align}
where $\hat{c}_r$ is the fermionic annihilation operator at the $r$th site of the system, $\hat{B}_{r1}$ ($\hat{B}_{rN}$) is the fermionic annhilation operator of the $r$th mode of the left (right) bath and $N_B$ is the number of modes in the baths, which will shortly be taken to infinity. We will assume that $\mathbf{H}$ is a real symmetric matrix. The Hamiltonian of the entire set-up can be written in the form 
$
\hat{\mathcal{H}}= \sum_{p,q=1}^{N+2N_B} \mathbf{H}_{p,q}^{\rm tot} \hat{d}_p^\dagger \hat{d}_q,
$
where $\hat{d}_p$ is the fermionic annihilation operator of either a system or a bath site. The retarded single-particle Green's function of the entire set-up in frequency space is given by the $(N+2N_B) \times (N+2N_B)$ matrix,
\begin{align}
\mathbf{G}^{{\rm tot} +}(\omega) = \big[(\omega-i\epsilon) \mathbb{I}-\mathbf{H}^{\rm tot}\big]^{-1} \Rightarrow \left[(\omega-i\epsilon) \mathbb{I}-\mathbf{H}^{\rm tot}\right]\mathbf{G}^{{\rm tot} +}(\omega)=\mathbb{I},
\end{align}
where $\mathbb{I}$ is the identity matrix of the corresponding dimension and $\epsilon$ is a  small positive number that takes care of the causality condition of the retarded Green's function. Breaking $\mathbf{G}^{{\rm tot} +}(\omega)$ into various blocks, the above expression can be re-written in the following form
\begin{align}
\left(
\begin{array}{ccc}
(\omega-i\epsilon)\mathbb{I}-\mathbf{H} & -\mathbf{\kappa}_1 &-\mathbf{\kappa}_N \\
\mathbf{\kappa}_1^\dagger & (\omega-i\epsilon)\mathbb{I}-\mathbf{\Omega}_1 & 0 \\
\mathbf{\kappa}_N^\dagger & 0 & (\omega-i\epsilon)\mathbb{I}-\mathbf{\Omega}_N \\
\end{array}
\right)
\left(
\begin{array}{ccc}
\mathbf{G}^+(\omega) & \mathbf{G}^{SL+}(\omega) & \mathbf{G}^{SR+}(\omega) \\
\mathbf{G}^{LS+}(\omega) & \mathbf{G}^{L+}(\omega) & \mathbf{G}^{LR+}(\omega) \\
\mathbf{G}^{RL+}(\omega) & \mathbf{G}^{RS+}(\omega) & \mathbf{G}^{R+}(\omega) \\
\end{array}
\right)=\mathbb{I},
\end{align}
where $\mathbf{\Omega_1}$ ($\mathbf{\Omega_N}$) is a $N_B$ dimensional diagonal matrix whose elements are the mode frequencies of the left (right) bath, and $\mathbf{\kappa}_1$ ($\mathbf{\kappa}_N$) is a matrix whose elements are hopping between the system sites and the various modes of the left (right) bath. Since only the first (last) site is attached to the left (right) bath, the only the first (last) row of $\mathbf{\kappa}_1$ ($\mathbf{\kappa}_N$) is non-zero. In the above equation, the $N \times N$ matrix $\mathbf{G}^+(\omega)$ is the retarded non-equilibrium Green's function (NEGF) of the system. By solving the above equation for  $\mathbf{G}^+(\omega)$, one obtains
\begin{align}
\mathbf{G}^+(\omega)=\big[(\omega -i\epsilon)\mathbb{I}-\mathbf{H}-\Sigma^{(1)}(\omega)-\Sigma^{(N)}(\omega)\big]^{-1}, \textrm{ with } \Sigma^{(1)}(\omega)=\mathbf{\kappa}_1^\dagger \mathbf{g}^{L+}(\omega)\mathbf{\kappa}_1,~~\Sigma^{(N)}(\omega)=\mathbf{\kappa}_N^\dagger \mathbf{g}^{R+}(\omega)\mathbf{\kappa}_N,
\end{align}
where $\mathbf{g}^{L+}(\omega)=\left[(\omega-i\epsilon)\mathbb{I}-\mathbf{\Omega}_L\right]^{-1}$ ($\mathbf{g}^{R+}(\omega)=\left[(\omega-i\epsilon)\mathbb{I}-\mathbf{\Omega}_R\right]^{-1}$) is the bare retarded Green's function of the left (right) bath is absence of coupling with the system. Here the $N \times N$ matrix  $\Sigma^{(1)}(\omega)$ ($\Sigma^{(N)}(\omega)$) is the self-energy of the left (right) bath. Since only the first (last) site of the system is coupled to the left (right) bath, the form of $\kappa_1$ ($\kappa_N$) enforces that only the top left (bottom right) corner element of $\Sigma^{(1)}(\omega)$ ($\Sigma^{(N)}(\omega)$) is non-zero. Now upon taking the number of bath modes to infinity ($N_B \rightarrow \infty$) such that the bath spectral functions become continuous, we can obtain the following expressions for the only non-zero elements of the self-energy matrices as
\begin{align}
\label{self_energy_elements}
\Sigma^{(\ell)}_{\ell \ell}(\omega)= -i\frac{\mathfrak{J}_\ell(\omega)}{2}-\mathcal{P}\int \frac{d\omega^\prime}{2\pi}\frac{\mathfrak{J}_\ell(\omega^\prime)}{\omega-\omega^\prime},~~\mathfrak{J}_\ell(\omega)=\sum_{r=1}^\infty |\kappa_{r \ell}|^2 \delta(\omega- \Omega_\alpha),~~\ell=\{1,N\}, 
\end{align}
$\delta(\omega)$ being the Dirac delta function.
We are interested in the non-equilibrium steady state (NESS) of the system, starting from an arbitrary initial state of the system and thermal states of the baths. The correlation functions involving system operators can be expressed in terms of the NEGF as \cite{Jauho_book,di_Ventra_book,Dhar_2006}
\begin{align}
\langle \hat{c}_p^\dagger \hat{c}_q \rangle_{\rm NESS}=& \int \frac{d\omega}{2\pi} \Big[ \mathbf{G}_{p 1}^*(\omega) \mathbf{G}_{q 1}(\omega) \mathfrak{J}_1(\omega) \mathfrak{n}_1(\omega)+ \mathbf{G}_{p N}^*(\omega) \mathbf{G}_{q N}(\omega) \mathfrak{J}_{N}(\omega) \mathfrak{n}_{N}(\omega) \Big],
\end{align}
where $\mathfrak{n}_1(\omega)=[e^{\beta_1(\omega-\mu_1)}+1]^{-1}$ ($\mathfrak{n}_N(\omega)=[e^{\beta_N(\omega-\mu_N)}+1]^{-1}$) is the Fermi distribution corresponding to the initial temperatures and chemical potentials of the left (right) bath. In NESS, the particle current from the left bath is the same as the particle current into the right bath and its general expression can be written as \cite{Jauho_book,di_Ventra_book,Dhar_2006}
\begin{align}
\label{current}
I = \int \frac{d\omega}{2\pi} \mathcal{T}(\omega) \left(\mathfrak{n}_1(\omega) - \mathfrak{n}_N(\omega)\right),~~\mathcal{T}(\omega)={\rm Tr}\left(\mathbf{\Gamma}^{(1)}(\omega) \mathbf{G}^{+*}(\omega)\mathbf{\Gamma}^{(N)}(\omega)\mathbf{G}^{+}(\omega)\right)
\end{align}
where $\mathbf{\Gamma}^{(\ell)}(\omega)=  {\rm Im}\left(\Sigma^{(\ell)}(\omega)\right)$, $\ell=\{1,N\}$. The above expression has the form of a Landauer-Buttiker formula for current, with $\mathcal{T}(\omega)$ being the transmission function.  Since the only non-zero elements of the self-energy matrices are as given in Eq.(\ref{self_energy_elements}), the transmission function simplifies to
\begin{align}
\mathcal{T}(\omega) = \mathfrak{J}_1(\omega)\mathfrak{J}_N(\omega) |\mathbf{G}^{+}_{1N}(\omega)|^2.
\end{align}
Going to zero temperature limit, $\beta_1,\beta_N \rightarrow \infty$, the expression for current reduces to 
\begin{align}
I = \int_{\mu_1}^{\mu_N} \frac{d\omega}{2\pi} \mathcal{T}(\omega) 
\end{align}
Writing $\mu_1=\mu$ and $\mu_N = \mu - \Delta \mu$, the conductance at zero temperature is given by
\begin{align}
G(\mu) = \lim_{\Delta \mu \rightarrow 0} \frac{I}{\Delta \mu}= \frac{\mathcal{T}(\mu)}{2\pi} = \frac{\mathfrak{J}_1(\mu)\mathfrak{J}_N(\mu) |\mathbf{G}^{+}_{1N}(\mu)|^2}{2\pi}.
\end{align}
This is the expression used in the main text to calculate the system-size scaling of conductance.

}

\section{Analytical Scaling of $\mathbf{G}^{+}_{1N}(\mu)$ with system size}
Our main conjecture is that for large $N$ the system size scaling of $\mathbf{G}^{+}_{1N}(\mu)$ will be same as that of the bare retarded Green's function $\mathbf{g}^{+}_{1N}(\mu)$, i.e, $\mathbf{G}^{+}_{1N}(\mu) \propto \mathbf{g}_{1N}(\mu)$, with proportionality constant being independent of $N$. If we further assume that $\mathbf{g}^{+}_{1N}(\mu)$ is evaluated in the thermodynamic limit, we get
\begin{align}
\label{conjecture2}
\mathbf{G}^{+}_{1N}(\mu) \sim \lim_{\epsilon\rightarrow 0} \frac{1}{2\pi} \int_{-\pi}^{\pi} dk \frac{e^{-ik N}}{\mu - \varepsilon(k,\alpha)-i\epsilon}=\mathbf{\mathcal{G}}^{+}_{1N}(\mu).
\end{align}
The major contribution to the integral on the right-hand-side comes from the singularities of the integrand.

{\it Case 1:   $-2\,\zeta(\alpha)<\mu<2 \, \eta(\alpha)$} : Clearly, if $\mu$ lies within the bandwidth of the system, $-2\zeta(\alpha)<\mu<2 \eta(\alpha)$, then the integrand will have poles on the real line.  Poles on the real line can, at best, generate terms which oscillate with $N$, and not any scaling behavior with $N$. So, as far as system size scaling is concerned, we can infer, 
\begin{align}
\mathbf{G}^{+}_{1N}(\mu) \sim N^0~~\forall~~ -2\zeta(\alpha)<\mu<2\eta(\alpha).
\end{align}
This is what leads to the ballistic behavior of current. 

\begin{figure}
\includegraphics[scale=0.5]{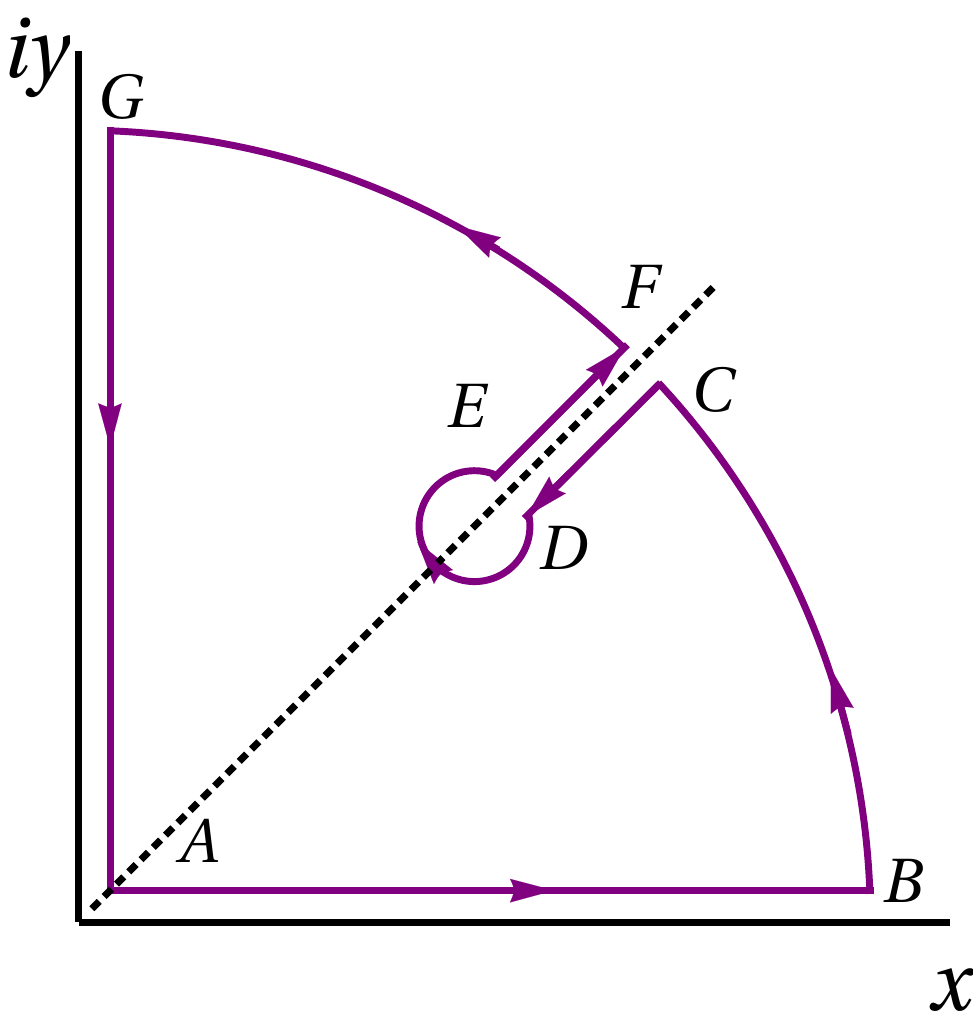} 
\caption{The contour chosen to carry out the integration in Eq.(\ref{contour_int}). }
\label{fig:contour} 
\end{figure}

{\it Case 2: $\mu < -2\zeta(\alpha)$:} Next, we consider the case when $\mu$ lies below the lower band edge i.e., $\mu \leq-2\,\zeta(\alpha)$ which is our main regime of interest. In this case, the maximum contribution to the integral comes from small values of $k$. Therefore we use Eq.(\ref{small_k_dispersion}) to obtain
\begin{align}
& \frac{1}{\mu - \varepsilon(k,\alpha)-i\epsilon}\simeq -\frac{1}{a_0(\omega)+2 a_1 |k|^{\alpha-1}+2 a_2 k^2 + i \epsilon }, \\
& a_0(\mu) = -2\zeta(\alpha)-\mu \geq 0. \nonumber
\end{align}
Then we have,
\begin{align}
\mathbf{G}^{+}_{1N}(\mu) \sim -\frac{1}{2 \pi} \, \int_{-\infty}^\infty dk \frac{e^{-ikN}}{a_0(\mu)+2 a_1 |k|^{\alpha-1}+2 a_2 k^2 + i \epsilon} = -\frac{1}{2 \pi} \big(A_{+} + A_{-}\big) ,
\end{align} 
where we have extended both the upper and the lower limit of the integral to infinity since we have already assumed that large values of $k$ gives a negligible contribution.  Here
\begin{align}
A_{\pm} =  \int_{0}^\infty dk \frac{e^{\mp ikN}}{a_0(\mu)+2 a_1 |k|^{\alpha-1}+2 a_2 k^2 + i \epsilon}
\label{contour_int}
\end{align}
The integration in Eq.(\ref{contour_int}) can be carried out using contour integration techniques, by choosing a proper contour as shown in Fig.(\ref{fig:contour}). This is a valid contour for computing $A_{-}$, whereas for $A_{+}$ a valid contour is the one enclosing the lower half of the complex plane.   Let us first focus on computing $A_{-}$. Depending on the value of $\alpha$, the integrand may or may not have branch point singularities in the right upper half plane.  In Fig.(\ref{fig:contour}), we assume there is one such singularity. For $\alpha>3$, it can be argued that this will be case, since the $k^{\alpha-1}$ term in the denominator will be sub-leading. By carrying out the integration along the curves CD, DE and EF in the contour, it can be checked that the contribution from them scales with system size as $e^{-a N}$, $(a>0)$. So, the contribution from any branch point singularity in the right upper half plane is exponentially decaying with system size.  The contribution from BC and FG is zero, as is standard. The leading contribution then comes from the line GA of the contour. The integral along the line GA, after some simplification is
\begin{align}
&A_{-}|_{\rm GA} = \frac{i}{N} \int_0^\infty dy \, \frac{e^{-y}} { x_R + i x_I}, & A_{+}|_{\rm GA} = - \frac{i}{N} \int_0^\infty dy \, \frac{e^{-y}} { x_R + i \tilde{x}_I},
\end{align}
where 
\begin{eqnarray}
x_R &=& a_0 \!+\!2\, a_1 (\frac{y}{N})^{\alpha-1} \cos\Big[{\frac{\pi (\alpha\!-\!1)}{2}}\Big]\!-\!2 a_2 \left(\frac{y}{N}\right)^2, \nonumber \\
x_I &=& \epsilon + 2 \, a_1 (\frac{y}{N})^{\alpha-1} \sin\Big[{\frac{\pi (\alpha\!-\!1)}{2}}\Big],\nonumber \\
\tilde{x}_I &=& \epsilon - 2 \, a_1 (\frac{y}{N})^{\alpha-1} \sin\Big[{\frac{\pi (\alpha\!-\!1)}{2}}\Big],\nonumber \\
\end{eqnarray}
One can finally write 
\begin{align}
\mathbf{G}^{+}_{1N}(\mu) \sim - \frac{2 \, a_1  \sin\Big[{\frac{\pi (\alpha\!-\!1)}{2}\Big]}}{N^{\alpha} \, \pi } \, \int_0^\infty dy \, \frac{e^{-y} \, y^{ \alpha-1}}{(x_R + i x_I)\, (x_R + i \tilde{x}_I)} 
\end{align}
Given that $a_0$ is finite, one can ignore $N$ dependent terms in $x_R, x_I, \tilde{x}_I$ in the thermodynamics limit ($N \to \infty$) leading to $1/N^{\alpha}$ dependence for the NEGF and thus conductance $G(\mu) $ scaling as $\sim N^{-2\alpha}$. 


\begin{figure}
\includegraphics[width=\textwidth]{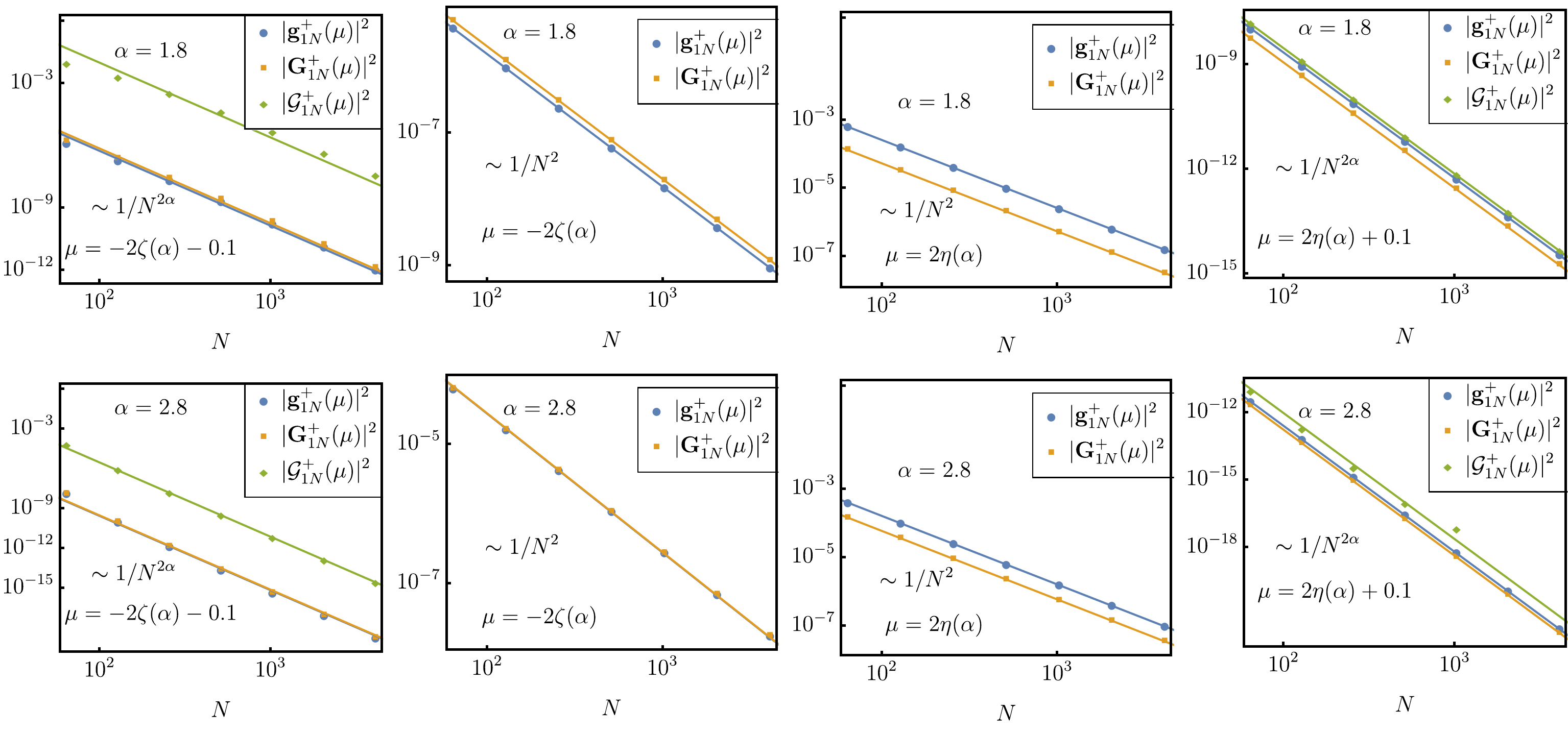} 
\caption{(Color online). Here numerically we have shown system size scaling of three different retarded Green's functions (i) central system's bare retarded Green's function $\mathbf{g}^{+}_{1N}(\mu)$, (ii) actual retarded Green's function $\mathbf{G}^{+}_{1N}(\mu)$ that appears in the conductance formula (iii) further approximated bare retarded Green's function $\mathbf{\mathcal{G}}^{+}_{1N}(\mu)$. The first row is for $\alpha=1.8$ and the second row is for $\alpha=2.8$. For the calculation of $\mathbf{G}^{+}_{1N}(\mu)$, the bath spectral functions are chosen to be $\mathfrak{J}_1(\omega)=\mathfrak{J}_N(\omega)=\Gamma \sqrt{1-\left(\frac{\omega}{\Lambda}\right)^2}$, with $\Lambda=10$ and $\Gamma=1.6$. We can see that all the retarded Green's functions can capture the $1/N^{2\alpha}$ scaling outside the band edge for both $\alpha=1.8$ and $2.8$. At the edges $1/N^2$ scaling can not be captured by $\mathbf{\mathcal{G}}^{+}_{1N}(\mu)$ but it is captured by systems's bare retarded Green's function $\mathbf{g}^{+}_{1N}(\mu)$. It also ensures that all these sub-diffusive scalings are actually the property of the central system.}
\label{fig:scaling} 
\end{figure}

{\it Case 3: $\mu > 2\,\eta(\alpha)$:} 
A similar analysis like above can be done also in the case when $\mu$ lies above the maximum band energy $2 \,\eta(\alpha)$ corresponding to $k=\pm \pi$. Interestingly, as $\varepsilon(k\pm \pi)$ is analytic around $k = \pm \pi$, (Eq.~(\ref{k_pi_dispersion})), it is easy to show that this leads to an exponential contribution in the system size, i.e., $\mathbf{G}^{+}_{1N}(\mu) \sim e^{-b N}, b>0$. The leading order contribution in $N$ once again arises from the non-analyticity behavior of the dispersion relation at $k=0$ and following similar contour integration steps as above one obtains exactly the same scaling
\begin{align}
& \mathbf{G}^{+}_{1N}(\mu) \sim N^{-\alpha},~~\forall~\mu<-2\zeta(\alpha), \mu>2\eta(\alpha)
\end{align}
{\it Case 4: $\mu = 2\,\eta(\alpha)$, and $\mu=-2\zeta(\alpha)$:} At any finite $N$, these values of $\mu$ do not correspond to any eigenvalue of $\mathbf{H}$, but the minimum and the maximum eigenvalues of $\mathbf{H}$ tend to these values with increase in $N$.  We find that this case is difficult to obtain from scaling of $\mathcal{G}_{1N}(\omega)$ defined in Eq.(\ref{conjecture2}). In other words, we cannot use the expression for $\mathbf{g}_{1N}(\mu)$ in the thermodynamic limit. However, direct numerical evaluation gives $\mathbf{G}^{+}_{1N}(\mu) \propto \mathbf{g}_{1N}(\mu)$, confirming the original conjecture, as we show in the next section.

\section{Numerical Scaling of $\mathbf{g}^{+}_{1N}(\mu),\mathbf{G}^{+}_{1N}(\mu), \mathbf{\mathcal{G}}^{+}_{1N}(\mu)$ with system size}
In the previous section, we have analytically calculated the approximated bare retarded Green's function $\mathbf{\mathcal{G}}^{+}_{1N}(\mu)$ for different cases like inside the band and outside the band. The analytical results give clear understanding of sub-diffusive behaviour $(1/N^{2\alpha})$ outside the band and ballistic behaviour $N^0$ inside the band. But, this approximated retarded Green's function $\mathbf{\mathcal{G}}^{+}_{1N}(\mu)$ can not capture the sub-diffusive scaling $1/N^2$ at the two band edges. Thus, numerically we have plotted system size scaling of all three different retarded Green's functions (i) central system's bare retarded Green's function $\mathbf{g}^{+}_{1N}(\mu)$, (ii) actual retarded Green's function $\mathbf{G}^{+}_{1N}(\mu)$ that appears in the conductance formula (iii)  approximated bare retarded Green's function $\mathbf{\mathcal{G}}^{+}_{1N}(\mu)$ in Fig.~\ref{fig:scaling} for $\alpha=1.8$ and $\alpha=2.8$. Here, we can numerically also see that all the retarded Green's function can capture the sub-diffusive scaling ($1/N^{2\alpha}$) outside the band. The scaling at band-edge can not be captured by $\mathbf{\mathcal{G}}^{+}_{1N}(\mu)$ but can be captured by system's bare retarded Green's function $\mathbf{g}^{+}_{1N}(\mu)$.

\section{Effect of system-bath coupling}
\begin{figure}
\includegraphics[width=\textwidth]{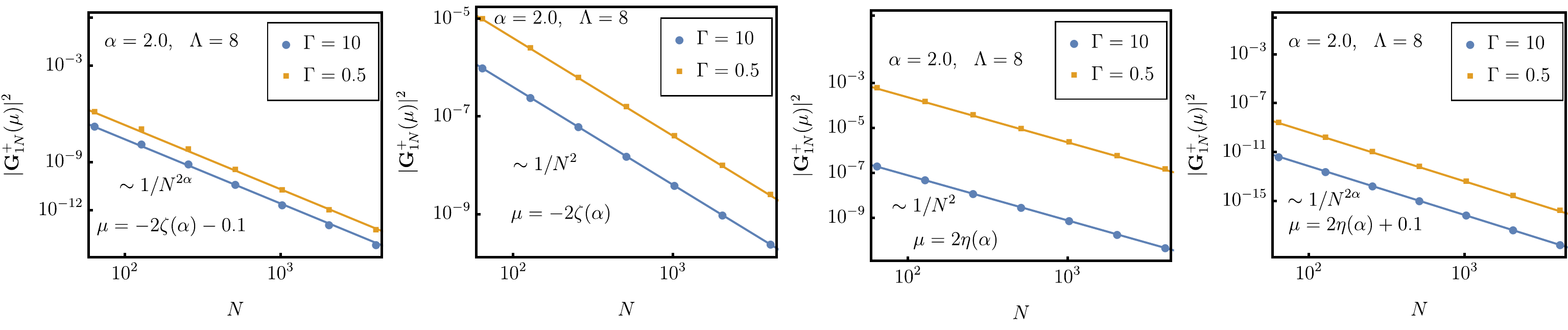} 
\caption{(Color online). We have plotted the system size scaling of exact retarded Green's function $\mathbf{G}^{+}_{1N}(\mu)$ for different system-bath couplings for $\alpha=2$. Here, we can see all the scaling properties are unaffected by the strength of system-bath coupling. For the plots, the bath spectral functions are chosen to be $\mathfrak{J}_1(\omega)=\mathfrak{J}_N(\omega)=\Gamma \sqrt{1-\left(\frac{\omega}{\Lambda}\right)^2}$, with $\Lambda=8$.}
\label{fig:coupling} 
\end{figure}

In the main text, we have said that the system-size scaling of conductance is independent of the strength of system-bath coupling.
In this section, we explicitly check this numerically. In Fig.~\ref{fig:coupling}, we have shown the system size scaling of $\mathbf{G}^{+}_{1N}(\mu)$ for two widely different strengths of system-bath coupling. We clearly see that all the scaling properties are unaffected by the system-bath coupling strength.

{

\section{Inability of local and global Lindblad approaches to describe the sub-diffusive phases}

Even though the system size scaling of conductance is independent of the strength of system-bath coupling, standard quantum master equations like Lindblad equations in local and global forms, which are often used to describe weak-system-bath coupling situations,  are unable to capture the sub-diffusive behavior. In this section, we explicitly discuss this.

\subsubsection{The local Lindblad approach}

The commonly used local Lindblad approach corresponds to the following quantum master equation \cite{Breuer_book,Landi_2021,Archak6}
\begin{align}
\label{LLE}
\frac{\partial \hat{\rho}}{\partial t}=i\left[\hat{\rho},\hat{\mathcal{H}}_S + \hat{\mathcal{H}}_{LS}  \right] + \sum_{\ell=1,N}\left[\mathfrak{J}_\ell(\varepsilon_\ell)\Big(1-\mathfrak{n}_\ell(\varepsilon_\ell)\Big) \left(\hat{c}_\ell \hat{\rho} \hat{c}_\ell^\dagger - \frac{1}{2}\{\hat{c}_\ell^\dagger\hat{c}_\ell, \hat{\rho}\}\right)+\mathfrak{J}_\ell(\varepsilon_\ell)\mathfrak{n}_\ell(\varepsilon_\ell) \left(\hat{c}_\ell^\dagger \hat{\rho} \hat{c}_\ell - \frac{1}{2}\{\hat{c}_\ell\hat{c}_\ell^\dagger, \hat{\rho}\}\right) \right],
\end{align}
where the so-called Lamb-Shift Hamiltonian $\hat{\mathcal{H}}_{LS}$ is given by
\begin{align}
\hat{\mathcal{H}}_{LS}=\sum_{\ell=1,N} \mathfrak{J}_\ell^H(\varepsilon_\ell) \hat{c}_\ell^\dagger\hat{c}_\ell, \textrm{ where } \mathfrak{J}_\ell^H(\omega) = \frac{1}{\pi}\int d\omega^\prime \frac{\mathfrak{J}_\ell(\omega^\prime)}{\omega-\omega^\prime}, 
\end{align}
is the Hilbert transform of $\mathfrak{J}_\ell(\omega)$. Here $\varepsilon_\ell$ is the on-site energy of the site attached to the bath, which, for our set-up is $\varepsilon_1=\varepsilon_N=0$. The first dissipative term in Eq.(\ref{LLE}) is called often called the loss Lindblad term. It describes the process that results in loss of particle and energy due to coupling with bath. The second dissipative term is often called the gain Linblad term. It describes the process that results in gain of particle and energy. At zero temperature, and for chemical potentials $<0$, from Eq.(\ref{LLE}), we see that only the loss Lindblad term survives. As a consequence, the system looses all its particles and its steady state is empty. Exactly similarly, for chemical potentials $>0$, from Eq.(\ref{LLE}) we see that only the gain Lindblad term survives. As a consequence, the steady state is completely filled. There can be no transport in either of these cases and hence current is zero. Thus, not only is the local Lindblad equation unable to describe the sub-diffusive phases, but also it cannot describe the ballistic transport at zero temperature when both chemical potentials are either positive or negative. These observations can be also checked by direct calculation.

Such limitations of local Lindblad approach are known \cite{Walls1970,Wichterich_2007,Rivas_2010,barranco_2014,Levy2014,Archak6,
Trushechkin_2016,
Eastham_2016,Hofer_2017,Gonzalez_2017,Mitchison_2018,
Cattaneo_2019,Hartmann_2020_1,konopik_2020local,
Scali_2021}. Microscopic derivations suggest that it can only describe situations either at infinite temperature or at infinite voltage bias or when the connections within the system are small enough that the sites do not hybridize well with one another. The physics we are describing in this paper are far from all these regimes. So this physics is beyond the regime that can be described by a local Lindblad equation.

\subsubsection{Global Lindblad approach}

To circumvent some of the drawbacks of the local Lindblad approach, a different approach often advocated is the global Lindblad or the eigenbasis Lindblad approach \cite{Breuer_book,Landi_2021,Archak6}. In deriving this quantum master equation, we first need to diagonalize the system Hamiltonian. For non-interacting systems (quadratic Hamiltonians) that we are discussing, this can be done by diagonalizing the single-particle Hamiltonian,
\begin{align}
\Phi^T \mathbf{H} \Phi = \mathbf{D},~~\mathbf{D}=diag\{\omega_\alpha\}.
\end{align}
Here $\{\omega_\alpha\}$ are the single particle eigenvalues, and the columns of $\Phi$ are the single particle eigenvectors. The system Hamiltonian can be written in the form \cite{Archak6},
\begin{align}
\hat{\mathcal{H}}_S = \sum_{\ell,m=1}^N \mathbf{H}_{\ell m} \hat{c}^\dagger_\ell \hat{c}_m = \sum_{\alpha=1}^N \omega_\alpha \hat{A}_\alpha^\dagger \hat{A}_\alpha,~~\hat{A}_\alpha = \sum_{\ell=1}^N \Phi_{\ell \alpha } \hat{c}_\ell.
\end{align}
The eigenbasis Lindblad equation is given by
\begin{align}
\label{ELE}
\frac{\partial \hat{\rho}}{\partial t}=i\left[\hat{\rho},\hat{\mathcal{H}}_S + \hat{\mathcal{H}}_{LS}  \right] + \sum_{\alpha=1}^N\sum_{\ell=1,N}|\Phi_\ell \alpha|^2\Big[ & \mathfrak{J}_\ell(\omega_\alpha)\Big(1-\mathfrak{n}_\ell(\omega_\alpha)\Big) \left(\hat{A}_\alpha \hat{\rho} \hat{A}_\alpha^\dagger - \frac{1}{2}\{\hat{A}_\alpha^\dagger\hat{A}_\alpha, \hat{\rho}\}\right)\nonumber \\
&+\mathfrak{J}_\ell(\omega_\alpha)\mathfrak{n}_\ell(\omega_\alpha) \left(\hat{A}_\alpha^\dagger \hat{\rho} \hat{A}_\alpha - \frac{1}{2}\{\hat{A}_\alpha\hat{A}_\alpha^\dagger, \hat{\rho}\}\right) \Big],
\end{align}
where the Lamb-shift Hamiltonian is given by $\hat{\mathcal{H}}_{LS}=\sum_{\alpha=1}^N\sum_{\ell=1,N} |\Phi_\ell \alpha|^2 \mathfrak{J}_\ell^H(\omega_\alpha) \hat{A}_\alpha^\dagger\hat{A}_\alpha$. As in the local Lindblad case, the first dissipative term in above equation is a loss Lindblad term, and describes loss of particles from the system eigenmodes due to coupling with the baths. Likewise, the second dissipative term in above equation is a gain Lindblad term and describes the gain of particles into the system eigenmodes due to coupling with the baths. At zero temperature and for $\mu_1,\mu_N \leq -2\zeta(\alpha)$, the Fermi distributions appearing the above equation dictate that only the loss term survives and the gain term is zero. Thus, the system in steady state is completely empty. Likewise, for $\mu_1,\mu_N \geq 2\eta(\alpha)$, the only the gain term survives and the loss term is zero. Thus, in this case, the state state is completely full. Either of these cases cannot have any transport so currents from the baths will be zero. For $-2\zeta(\alpha)<\mu_1,\mu_N<2\eta(\alpha)$, however, the currents from the baths will capture the ballistic behavior. These statements can also be verified via direct calculation. Therefore, we see that the global Lindblad approach also fails to capture the sub-diffusive phases and the critical points.

There are a variety of more refined quantum master equation approaches \cite{ule,Kleinherbers_2020,Davidovic_2020,mozgunov2020,mccauley2020,
kirvsanskas2018}, including the Redfield equation. Whether such quantum master equations can capture the sub-diffusive behavior remains to be seen and requires further investigation.

}

\begin{figure}
\includegraphics[width=0.8\textwidth]{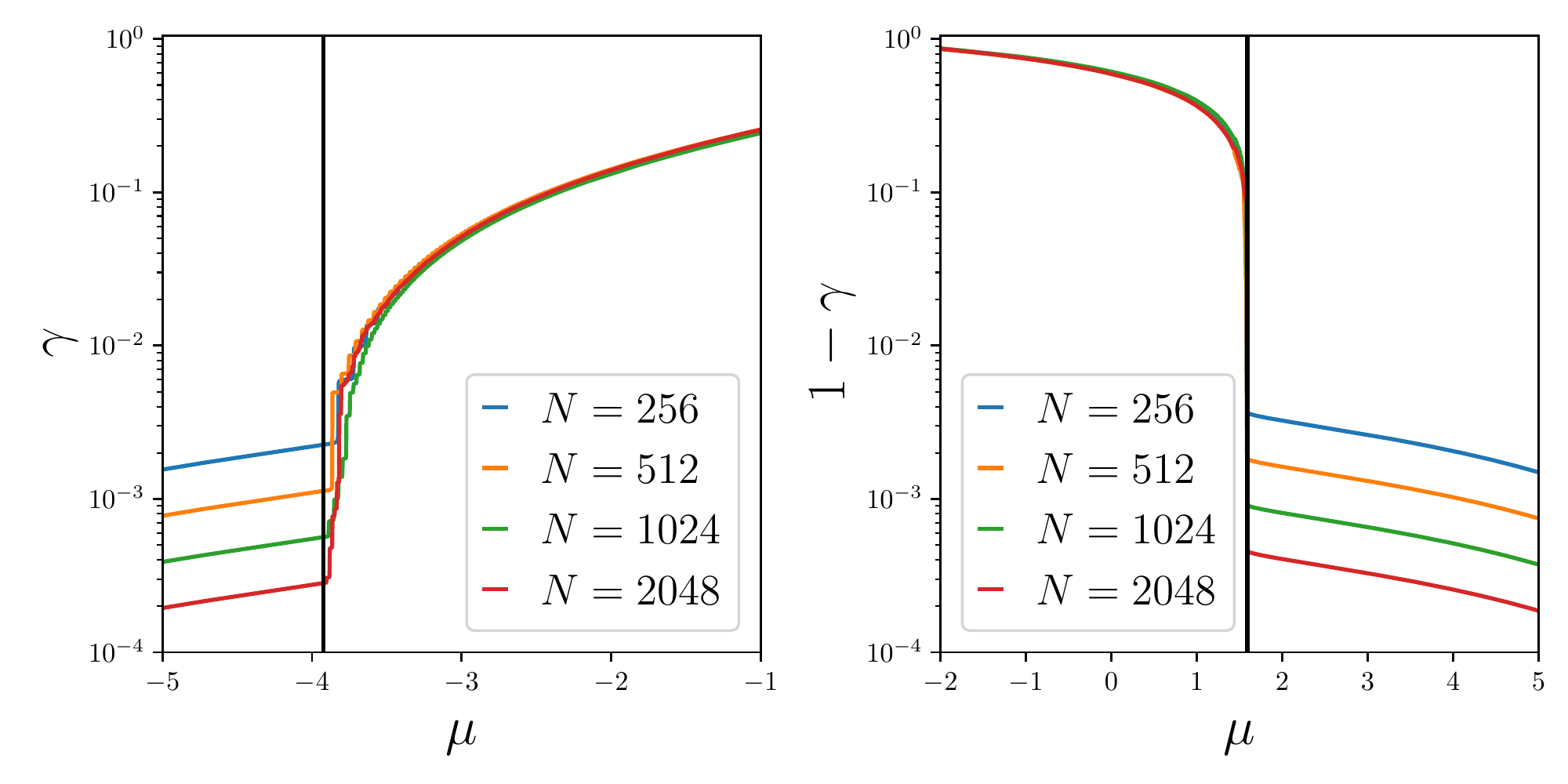}
\caption{{\bf (a)} The particle density $\gamma$ in the system is plotted as a function of chemical potential at zero temperature. The vertical black line corresponds to $\mu=-2\zeta(\alpha)$. For $\mu\leq-2\zeta(\alpha)$, the particle density decays with system size. {\bf (b)} The hole density in the system, given by $1-\gamma$, is plotted as function of chemical potential at zero temperature. The vertical line corresponds to $\mu=2\eta(\alpha)$. For $\mu\geq 2\eta(\alpha)$, the hole density decays with system size. For the plots, the bath spectral functions are chosen to be $\mathfrak{J}_1(\omega)=\mathfrak{J}_N(\omega)=\Gamma \sqrt{1-\left(\frac{\omega}{\Lambda}\right)^2}$, with $\Lambda=8$, $\Gamma=10$. \label{fig:particle_density}   }
\end{figure}

\section{Particle density in the system}
In the main text, we have remarked that there is a sub-extensive number of particle in the system for $\mu\leq-2\zeta(\alpha)$, while there is a sub-extensive number of holes for $\mu\geq 2\eta(\alpha)$. For $-2\zeta(\alpha)<\mu<2\eta(\alpha)$, there is an extensive number of both particles and holes. Here we explicitly check this. The particle density in the system is defined as
\begin{align}
\gamma=\frac{1}{N}\sum_{\ell=1}^N \langle \hat{c}_\ell^\dagger \hat{c}_\ell \rangle.
\end{align}
The occupation at $\ell$th site in NESS is given in terms of the NEGF as
\begin{align}
\langle \hat{c}_\ell^\dagger \hat{c}_\ell \rangle = \int_{-\Lambda}^\mu \frac{d\omega}{2\pi} \Big[\left|\mathbf{G}_{1\ell}^+(\omega)\right|^2 \mathfrak{J}_1(\omega) +
\left|\mathbf{G}_{N\ell}^+(\omega)\right|^2 \mathfrak{J}_N(\omega) \Big],
\end{align}
where $-\Lambda$ is the minimum energy of the band of the bath. We numerically calculate $\gamma$ and check its behavior with $\mu$ and $N$, as shown in Fig.~\ref{fig:particle_density}.
When there is a sub-extensive number of particles in the system, $\gamma$ decays with $N$, which happens for $\mu\leq-2\zeta(\alpha)$. When there is an extensive number of particles in the system, $\gamma$ is independent of $N$, which happens for $\mu>-2\zeta(\alpha)$ . When there is a sub-extensive number of holes in the system, $1-\gamma$ decays with $N$, which happens for $\mu\geq 2\eta(\alpha)$, while if there is an extensive number of holes in the system, $1-\gamma$ is independent of system-size, which happens for $\mu<2\eta(\alpha)$. 

{
\section{Relation to experiments}

In a number of experiments in various platforms like trapped ions \cite{expt_trapped_ions6,
Experiment_transport}, polar molecules \cite{expt_polar_molecules1,
expt_polar_molecules3}, dipolar gas \cite{expt_dipolar_gas1}, nuclear spins \cite{expt_nuclear_spins} the spin Hamiltonians of the following form has been realized,
\begin{align}
\label{spin_hamiltonian}
\hat{\mathcal{H}}_S = -\sum_{m=1}^N \left(\sum_{r=1}^{N-m} \frac{1}{m^\alpha} \left(\hat{\sigma}_r^+ \hat{\sigma}_{r+m}^- + \hat{\sigma}_{r+m}^+ \hat{\sigma}_{r}^- + \Delta \hat{\sigma}_{r+m}^z \hat{\sigma}_{r}^z  \right)\right),
\end{align}
where $\hat{\sigma}_r^{\pm}=\left(\hat{\sigma}_r^x \pm i\hat{\sigma}_r^y \right)/2$, and $\hat{\sigma}_r^{x,y,z}$ are the Pauli spin operators at site $r$. This Hamiltonian conserves the total magnetization, $\hat{M}_z= \sum_{r=1}^N \hat{\sigma}_{r}^z$, i.e, $[\hat{M}_z,\hat{\mathcal{H}}]=0$. Let us perform Jordan-Wigner transformation to convert this Hamiltonian into a fermionic one. The Jordan-Wigner transformation is given by
\begin{align}
\hat{\sigma}_r^{+} = \hat{c}_r^\dagger e^{-i \pi \sum_{p=1}^{r-1} \hat{n}_{p}},~~ \hat{\sigma}_r^{-} = e^{i \pi \sum_{p=1}^{r-1} \hat{n}_{p}}\hat{c}_r,~~\hat{n}_r = \hat{c}_r^\dagger\hat{c}_r = \frac{\hat{\sigma}_r^z +1}{2}.
\end{align}
Using these, we receive,
\begin{align}
\hat{\sigma}_r^+ \hat{\sigma}_{r+m}^- = \hat{c}_r^\dagger e^{i\pi \sum_{p=r}^{r+m-1} \hat{n}_p}  \hat{c}_{r+m} = \hat{c}_r^\dagger \prod_{p=r}^{r+m-1} \left(1-2\hat{n}_p\right) \hat{c}_{r+m}, \textrm{ and } \hat{\sigma}_r^z = 2\hat{n}_r -1,
\end{align}
where we have also used the result $e^{i\pi \hat{n}_p} = 1-2\hat{n}_p$, which can be proven by expanding the exponential. Substituting these into Eq.(\ref{spin_hamiltonian}) we see that the resulting fermionic Hamiltonian is of the form 
\begin{align}
\hat{\mathcal{H}}_S=-\sum_{m=1}^N \left(\sum_{r=1}^{N-m} \frac{1}{m^\alpha} \left(\hat{c}_r^\dagger \hat{c}_{r+m} + \hat{c}_{r+m}^\dagger \hat{c}_{r}\right)\right) + \hat{H}_{\rm int},
\end{align}
 where $\hat{H}_{\rm int}$ contains the many-body interacting terms, i.e, the higher-than-quadratic terms. The total magnetization operator $\hat{M}_z = \sum_{r=1}^N (2\hat{n}_r -1)$. Therefore, the conservation of net magnetization guarantees that in the fermionic picture the Hamiltonian is particle number conserving. Thus, upon Jordan-Wigner transformation, the spin Hamiltonians realized in several controlled experimental platforms can be mapped into number conserving fermionic Hamiltonians with power-law-decaying hopping and many-body interactions. Since, as argued in the main text, the sub-diffusive phases and critical points are expected to be robust against arbitrary number conserving many-body interaction terms, the physics described here is relevant to these experimental set-ups at low temperatures.
}

\end{document}